\documentclass[english,12pt]{article} 
\usepackage[T1]{fontenc} 
\usepackage[utf8]{inputenc} 
\usepackage{comment} 
\usepackage[english]{babel}
\usepackage{a4wide} 
\usepackage[pdftex]{graphicx}
\usepackage{verbatim} 
\usepackage{booktabs} 
\usepackage{multirow} 
\usepackage{subcaption}
\usepackage{amsmath,amssymb,amsopn,dsfont,marvosym,stmaryrd} 
\usepackage{threeparttable,multirow}
\usepackage{float} 
\usepackage[]{placeins} 
\usepackage{flafter} 
\usepackage{booktabs} 
\usepackage{multirow} 
\usepackage[]{longtable} 
\usepackage{verbatim} 
\usepackage{array} 
\usepackage{natbib} 
\usepackage{xr} 
\usepackage{tikz} 
\usepackage{eurosym} 
\usepackage{adjustbox} 
\usepackage{tabularx} 
\usepackage{dcolumn} 
\usepackage{setspace}

\newcolumntype{d}[1]{D{.}{.}{#1}} 
\newtheorem{thm}{Theorem}[section]

\newtheorem{prop}[thm]{Proposition}

\def\E{{\mathbb E}}

\def\BL{\begin{list}{}{}}
\def\ENDL{\end{list}}

\def\indic{\mathds{1}}

\usepackage{layout}
\usepackage[colorlinks=true,	
linkcolor=red,
urlcolor=blue,
citecolor=blue]{hyperref}
\usepackage[top=1in, bottom=1in, left=1in, right=1in]{geometry}

\title{\vspace{-2cm}  Biases-Informed Job Search Guidance:\\ \large Characterization, Implications, and Targeting Support\thanks{\footnotesize{Cr\'epon: CREST, crepon@ensae.fr; Frot: France-Travail, CREST, aurelien.frot@ensae.fr; Gaillac: University of Geneva, GSEM: christophe.gaillac@unige.ch. This paper is the result of a partnership with France Travail, the Public Employment Service in France, and does not represent their views. We thank Anita Bonnet, Cyril Nouveau, and Chantal Vessereau, for their support and feedback throughout this project. This research was supported by the Chaire Sécurisation des Parcours Professionels. Authors retained full intellectual freedom throughout this process, all errors are our own.  We thank Yves Le Yaouanq, Roland Rathelot, Arne Uhlendorff, as well as seminar participants at Oxford, CREST, INSEE, Paris School of Economics (PSE), the chaire travail seminar of PSE, the Chaire Sécurisation seminar at DARES, the 2025 Workshop on Subjective Expectations in Lisbon, and the IAAE 2025 in Turin for useful comments and suggestions. We thank Alexis Du for his useful participation.}}}
\author{Bruno Crépon, Aurélien Frot, Christophe Gaillac}

\usepackage{natbib}
\usepackage{graphicx}

\begin{document}

\maketitle

\vspace{-1cm}
\begin{abstract} 
 Job seekers' expectations about reemployment are increasingly used to study job search, but what their biases reveal about underlying beliefs and preferences is ambiguous.  We combine new survey data, structural modeling, and machine learning to uncover the informational content of these expectations and show how they can be used to improve the targeting of employment support. Using a new panel of French job seekers' subjective expectations linked to administrative records, we show that reemployment expectation biases are strongly associated with, and summarize, biases in beliefs about the two fundamentals of search, job offer arrival rates and the wage distribution. These underlying biases are heterogeneous but strongly positively correlated, so their effects on search compound. We then estimate a structural job search model with multiple sources of biased beliefs and show that correcting them helps pessimistic job seekers but can demotivate and hurt optimistic ones, providing a rationale for targeting. Finally, we develop a machine-learning stratification that recovers policy-relevant groups, with distinct patterns of biased beliefs and behaviors, from easily elicited reemployment expectations alone. This gives employment services a simple tool to target informational interventions.\\

\end{abstract}

\newpage


\section{Introduction}

In job search models, beliefs about the labor market and expectations about search outcomes are important determinants of search strategies and, consequently, the quality of reemployment. Following \cite{manski2004measuring} and, more recently, \cite{almaas2023economics}, researchers have increasingly emphasized the need to collect richer measures of beliefs and expectations to inform individual search behavior and to estimate more credible and realistic models. So far, mainly for practical reasons, most job search surveys elicit only job seekers' expectations about their hiring prospects \cite[see, \emph{e.g.}, the Survey of Consumer Expectations (SCE),][]{armantier2017overview}, and the literature has focused on the average differences between these expectations and realized reemployment for subgroups of the population \citep[see, \emph{e.g.},][]{mueller2021job,caliendo2023accuracy}.

\medskip

Yet what these reemployment expectation biases reveal about job seekers is ambiguous. In standard job-search models, reemployment expectations are determined by more fundamental beliefs about offer arrival rates, the distribution of posted wages, the returns to search effort, and individual preferences. A given bias in reemployment expectations can therefore arise from many different combinations of biased beliefs and preferences, which is why subjective expectations have remained difficult to use for tailoring support to individual job seekers \citep[see][]{mueller2023expectations}.

\medskip

In this paper, we address this ambiguity by combining new subjective-expectation data, a structural model of job search, and machine-learning tools. Together, these elements allow us to uncover the informational content of reemployment expectations, quantify the consequences of biased beliefs, and develop an operational framework for targeting employment support. In partnership with the French public employment service (France Travail), we designed a quarterly rotating panel survey linked to rich administrative records on job seekers' characteristics, search behavior, and reemployment outcomes. This unique longitudinal dataset offers an exceptionally rich picture of the job search process, jointly observing for the same individuals over time beliefs about labor market fundamentals, reemployment expectations, search behavior, and realized outcomes. It allows us to follow nearly 25,000 job seekers throughout their unemployment spell: we elicit beliefs about job offer arrival rates, posted wage distributions, reemployment probabilities, and the perceived returns to search effort, while tracking reservation wages, mobility, declared search effort, applications submitted through the PES platform, and realized reemployment.

\medskip

Our first contribution is to characterize the information contained in reemployment expectation biases. Using new measures of beliefs about the two key labor market fundamentals, i.e., job offer arrival rates and the wage distribution, we show that these beliefs are strong predictors of job search behavior: job seekers who expect more attractive opportunities choose different reservation wages and levels of search effort. These underlying beliefs are highly heterogeneous across job seekers, yet strongly positively correlated: individuals who are optimistic about the wage distribution also tend to be optimistic about arrival rates, so their behavioral effects reinforce rather than offset one another. Crucially, reemployment expectation biases are strongly associated with these underlying beliefs, demonstrating that a simple, easily elicited expectation contains valuable information about the sources of a job seeker's misperceptions. Along the way, we establish several facts of independent interest. We document how reemployment expectations are updated over the spell, showing that the upward revision emphasized in the literature is confined to short-term expectations, whereas long-term expectations are revised downward. We confirm that reemployment expectations strongly predict realized reemployment and find an average short-term optimism of 24 percentage points, similar to that documented for the United States. Finally, we provide new measures of the perceived arrival rate of offers and show that they are related to psychological traits, in particular the locus of control, and strongly predict search effort.

\medskip

Our second contribution is to develop and estimate a structural model of job search that quantifies the consequences of these biases and  highlights that correcting need not benefit all job seekers equally. In a standard McCall model, more accurate beliefs lead to better search decisions and higher welfare. This conclusion becomes less clear once beliefs also affect motivation: a growing literature emphasizes that optimism may sustain search effort while pessimism may discourage it (\citealp{spinnewijn2015unemployed}; \citealp{cooper2020behavioral}; \citealp{giustinelli2022expectations}). We therefore develop a subjective version of the McCall model \citep{mccall1970economics} with endogenous search intensity that accommodates \emph{multiple} sources of biased beliefs and lets them affect both decisions and motivation; we estimate some parameters and calibrate others using our rich subjective measures, thereby relaxing assumptions commonly made in the literature. Correcting beliefs about the wage distribution primarily affects job seekers with over-optimistic reemployment expectations, who revise their reservation wages downward, while having little effect on pessimists, whose reservation wages are already near the minimum wage. Correcting beliefs about arrival rates has a sharply different effect: for pessimists it substantially raises search effort (by 34\%) and the three-month reemployment probability (by 1.3 percentage points), whereas for optimists the effect is ambiguous, because lower perceived employability both reduces the perceived return to effort (a demotivation effect) and lowers reservation wages (increasing acceptance). The net effect depends on true employability, which is unobservable in practice. Once motivational effects are taken into account, the value of correcting a given belief thus depends on the job seeker's profile, providing a rationale for \emph{targeted} informational interventions rather than uniform information provision.

\medskip

Our third contribution is to show how the relevant profiles can be identified in practice using only reemployment expectations. While the structural model underscores the importance of the underlying sources of bias, directly eliciting these beliefs requires rich survey instruments that are unlikely to be feasible in routine employment services; reemployment expectations, by contrast, can be collected with a single question. We develop a flexible, data-driven stratification that uses these expectations to form groups of job seekers with maximally different predicted reemployment biases, adapting the Generic Machine Learning framework of \cite{chernozhukov2018generic} while remaining agnostic about the underlying prediction algorithm (random forests, boosting, neural networks, lasso, etc.). The classification uncovers four economically meaningful groups, namely pessimistic, rational, optimistic, and over-optimistic. They represent substantial shares of the unemployed, with between 16\% and 20\% of job seekers classified as pessimistic and around 30\% as over-optimistic. These groups differ markedly in their underlying labor market beliefs, search behavior, and informational needs: biases about wage distributions and arrival rates increase monotonically with reemployment optimism, and reservation wages, search effort, and application behavior differ substantially across groups. Reemployment expectations thus provide a simple and operational basis for tailoring informational interventions to heterogeneous job seekers.

\medskip

These results have direct implications for the design of informational interventions within public employment services. Whereas existing interventions typically provide the same information to all job seekers, our findings suggest that reemployment expectations can be used to diagnose heterogeneous patterns of biased beliefs and to tailor interventions accordingly, targeting both the content and the recipients of information. More broadly, by using reemployment expectations as signals of job seekers' underlying beliefs, employment services can design targeted informational or behavioral interventions in the spirit of Bayesian persuasion \citep{kamenica2011bayesian} and, more generally, information design. Our approach illustrates how subjective expectations can become part of the standard toolkit of public employment services.

\paragraph{Organization of the paper.} Section~\ref{sec:lit} reviews the related literature. Section~\ref{sec:data} describes the survey and administrative data. Section~\ref{sec:information} provides a statistical characterization of the biases in beliefs and expectations. Section~\ref{sec:pred_outcome} develops and estimates the structural job search model and quantifies the costs of biased beliefs. Section~\ref{sec:heterogeneity} presents the stratification methodology and results. Section~\ref{sec:conclude} concludes.

\section*{Related literature}\label{sec:lit}

We contribute to several strands of the literature. First, our paper expands the documentation of biases in beliefs and subjective expectations about the labor market \cite[see earlier work of][]{manski1999worker}. Several papers, \emph{e.g.},
\cite{mueller2021job,spinnewijn2015unemployed,van2024predicting}, document job seekers' optimism about their reemployment expectations, and in particular an upward revision with duration at a 3 months horizon. Using our panel data, we find a similar result for short-term expectations. However, by eliciting reemployment expectations for different time horizons, we also demonstrate that job seekers actually update their long-term expectations downward \citep[see the related discussion in][]{he2024understanding}. Our measures of the bias with respect to the wage distribution, leveraging administrative data, complement the insights from \cite{conlon2018labor,caliendo2023accuracy,altmann2025,christensen2026perceived}, who document deviations from rationality in perceptions of the wage distribution of job offers or postings. \cite{adams2023perceived} examines the perceived returns to job search effort and shows that, while there is overoptimism about the probability of receiving offers conditional on effort, the perceived marginal returns to additional search are low on average. We enrich this description by relating the marginal returns to effort to both psychological traits and other subjective expectations \citep[see also][]{caliendo2015locus}.  Finally, important behavioral mechanisms have been documented and related to job search expectations \citep{katz1990unemployment,cooper2020behavioral,giustinelli2022expectations}. Our new measurements allow to extend this knowledge with a comprehensive picture of the evolution of beliefs and expectations.  

\medskip

Our stratification methodology builds on the recent ML methods for heterogeneous treatment effects, in particular \cite{chernozhukov2018generic}, and establishes the link between this literature and the flexible nonparametric estimation of the biases. Similarly to the treatment effect literature, our outcome of interest (the job search biases) is not directly observed. This methodology can be used more generally in such situations. In a complementary exercise and finance context, \cite{von2021heterogeneity} rather consider building groups based on $k$-means of beliefs data then estimating panel data models of expectations. This approach is also possible with our data but the important cost of collecting subjective beliefs would prevent designing subsequent targeting based on them. Importantly, our methodology provides new tools to stratify the population only leveraging information about their reemployment biases, identifying heterogeneous intervention needs. %
Thus, our paper importantly relates these methods to the large existing literature about these types of experiments, which mostly considers the effect of \textit{uniform} informational interventions \citep{bandiera2025search,belot2019providing,altmann2018learning,altmann2025,jones2022can,tekleselassie2025feedback,harmon2026job}.   
Interestingly, we also characterize the heterogeneity of such biases with respect to certain demographics. Previous descriptions were limited in this regard \citep[see the challenges underlined in][]{mueller2023expectations}.

\medskip

Finally, a growing literature  uses subjective expectation data to consider the consequences of deviations from rationality within standard job search models.  Job search models used in the literature \citep[see, \emph{e.g},][for a survey]{mueller2023expectations} evaluate the costs of the documented unidirectional deviations from rationality separately: associated with beliefs about the arrival rate of job offers in \cite{mueller2021job} or the wage distribution in \cite{hall2018wage,conlon2018labor,altmann2025}. They show that biases might have important costs in particular by increasing the risk of long-term unemployment. In this respect, our job search model innovates by allowing for several sources of biased expectations.

\section{General context and Data}\label{sec:data}

To characterize job search biases, we use two sources of data. The first is a new panel survey that we designed in collaboration with the PES, focusing on expectations, beliefs, preferences, psychology, and search behaviors. The second is the administrative data available at the PES, which includes demographics, initial search parameters and reemployment outcomes.  Both can be perfectly matched, and the second concerns the universe of job seekers in France. We explain how we use the administrative data to reweight and control for sample selection when using variables from the survey. 

\subsection{Panel data on Beliefs}\label{ssec:panel_data}

We have designed a panel of job seekers' expectations and beliefs that shares and enriches the structure of the pioneering labor market module of the Survey of Consumer Expectations \citep[SCE hereafter, see][for an overview]{armantier2017overview}. Every three months since October 2021, a new sample from the pool of all job seekers registered with the PES in metropolitan France enters the panel. We stratify only by unemployment status and duration, and oversample job seekers newly registered with the PES.\footnote{We also exclude nursery assistants and entertainment workers who are registered with the PES in France, who usually work under specific type of contracts.} This is a rotating panel, so in addition to this refreshing sample, job seekers who report that they are actively looking for a job are interviewed about their expectations every 3 months. Importantly, the others are asked about their job search outcomes before they leave the panel. This survey was administered by email using the PES email address, but mentioning the research purpose of the survey, as well as guaranteeing anonymity with respect to their caseworker and complete separation from their case at the PES (see Figure \ref{fig:mail_de_contact}). This paper focuses on the first 8 waves of the panel, until July 2023, with sample sizes varying between 2,500 and 4,500. 

\paragraph{Expectations about return to work at different horizons.} An important part of the panel elicits the subjective expectations of the duration $T_i$ before finding a job conditional on their information set $\mathcal{I}$, denoted by $\psi_{i,t}:= \widetilde{P}(T_i \leq t | \mathcal{I})$, before $t=1, 3, 6, 12,$ or 24 months. The wording for these questions is: 

\medskip

\textit{``What are your chances of finding a job within t month(s), on a permanent or fixed-term contract of at least 1 month?''.}

\medskip

The end of the latter question controls what exactly is measured as an outcome in the PES administrative data. Respondents answer using a slider that can range from 0 to 100 percent chance (see figure \ref{fig:remployment_slider}). A small introductory text describes in words what these situations correspond to. In the following, we will use either ``probabilities'' to refer to these data converted to the $[0,1]$ range, or ``percent chances'' when using the original $[0,100]$ scale.\footnote{After several focus groups with job seekers, this seemed to be the best solution to make them aware of probabilistic reasoning in this PES-related survey. The fact that these expectations are predictive of behaviors and outcomes is reassuring about the information content of the former.} They are also asked what they consider to be a return to work in terms of duration and type of contract. 

Table \ref{tab:subj_exp} presents descriptive statistics on the elicited subjective expectations and some elicited job search behaviors. It is interesting to note that while the average expected chances of finding a job before 1, 3, 6, 12 or 24 months vary from 39 to 74.1\%, the standard deviation is also but slightly increasing (between 28.8 and 32.2, over more than 25,449 individuals). We find that, on average, subjective reemployment expectations increase with the horizon. However, this is not true for the entire population: around 15\% of our sample have decreasing reemployment beliefs with time horizon. Using this as a preliminary test of the understanding of probabilistic reasoning, we remove such individuals for the subsequent analyses, even though they are not related to reemployment expectations.

\paragraph{Representation of the wage offers distribution.} The first representation of market fundamentals that we elicit are three points about the perceived distribution of wage offers. We use data on all the wages posted in vacancies on the PES platform to compute the quartiles of the wage distribution over three months preceding each survey, conditional on occupation and location (at the French \textit{région} level).\footnote{According to \cite{le2017unemployment}, this represents about 60\% of the job offers in France in 2017.} As suggested by the literature \citep[see,][]{manski2004measuring}, we use these values to personalize the three thresholds at which we elicit the subjective wage distribution.\footnote{Specifically, if the upper and lower quartiles are less than 5\% close to the median, we multiply them by 1.1 and 0.9, respectively, to obtain significant and noticeable differences between the scores obtained.} The wording of the question is:
  
\medskip

\textit{``Imagine that you are offered a job. A job offer is not necessarily an offer you will accept.  What are the chances that it offers a salary of more than $x$ euros gross per month full-time?}

\medskip

Respondents answer using a slider that can range from 0 to 100 percent chance, and we refer to these associated subjective probabilities as $\widetilde{P}(W \geq x)$, where $x$ is the wage level being asked. The three questions are in ascending order and the methodology to decide on the thresholds are discussed in more details in section \ref{ssec:wage_bias}. More than 16,250 job seekers answered these questions (see Table \ref{tab:subj_exp}).

\paragraph{Expected arrival rate of job offers and realized offers.} The second component of the subjective representation of the labor market that we elicit are two points on the subjective distribution of the arrival rate of job offers, collected after asking about the actual search effort per week, denoted by $e^*$. We use a wording similar to that used in the SCE:  

\medskip

\textit{``What do you think is the percent chance that within the coming three months, you will receive
at least x job offer(s)? Remember that a job offer is not necessarily a job you will accept.''}

\medskip

Again, respondents answer with a slider that can range from 0 to 100 percent chance. We denote these associated probabilities by $\widetilde{P}(N \geq x | e = e^*)$, where $x \in \{1,3\}$ and $N$ is the number of offers. Table \ref{tab:subj_exp} shows that around 19,000 job seekers answered these questions, with an average subjective probability of receiving more than 1 offer (resp. 3 offers) of 45.2$\%$ (resp. 41.3$\%$). Starting with wave 6, we also ask the respondents about the probability of receiving at least one job offer under counterfactual search effort scenarios, \emph{i.e.}, $\widetilde{P}(N \geq x | e )$ for $e \in\{10,20,30\}$ (see \cite{adams2023perceived} for a similar strategy).\footnote{This is only done for waves 6 to 8, however.} On average, this probability varies from 35.4$\%$ for searching 10 hours per week to 48.8$\%$ for searching 30 hours per week, with considerable heterogeneity (the standard deviations are 27.3 and 31.4, respectively, for about 6,000 job seekers).

\medskip

In the subsequent waves of the survey, we collect data on the number of offers received in the last 3 months, which gives us the realization of this risk $N_i$ for these individuals.  Respondents can use a slider to answer up to five, after which they must enter a numeric value. This data is discussed in more detail in section \ref{sec:bias:arrival}.

\paragraph{Actual search parameters.} Since an important part of this work is the relationship between representations and the choice of search parameters, we also ask about several of these parameters at the time of the survey. First, we elicit the actual level of search effort $e^*$:

\medskip

\textit{``On average, how many hours do you spend looking for a job in a week ?``}

\medskip

Job seekers answer using a slider going from 0 to 35 hours and can directly type their answer if they spend more than 35 hours. Table \ref{tab:subj_exp} shows that, on average, job seekers report searching 13.1 h/week with significant heterogeneity (standard deviation of 9.2).

In order to be able to link the representations of the wage offer distribution and the offer arrival rates to the probability of accepting job offers, we also elicit their reservation wages in each period, denoted by $w^*_{i,t}$:

\medskip

\textit{``What is the minimum gross monthly salary you would accept for the job you are willing to take?''}.

This question has the same wording as the corresponding question that all job seekers have to answer at the PES registration. This provides us with an administrative benchmark, while allowing to collect this information along the unemployment spell.

\paragraph{Personality and locus of control.} Since individuals' subjective perceptions and search behaviors might vary according to one's personality, we also collect information on psychology and locus of control \citep[see, \emph{e.g.},][for an analysis of their links with economics' preferences]{heckman2021some}. Questions about personality are such that they cover the seminal ``big five'', \emph{i.e.}, Conscientiousness, Openness, Neuroticism, Extraversion, and Agreeableness \footnote{Questions about personality traits are included from the 6th wave of the survey onwards.}. Our measures of internal and external locus of control are built based on 6 statements job seekers must rate how much they agree with. The specific items used to elicit psychological traits and locus of control are detailed in Appendix \ref{ssec:app_psycho_items}.

\subsection{Administrative data}\label{ssec:admin_data}

In order to characterize the expectations and biases of job seekers, we need two types of administrative information: first about the relevant demographics for job search, as well as actual job search behavior, then about the outcomes of the current one.

\paragraph{Demographics and initial search parameters.} We use information about job seeker’s demographics such as declared at the PES. In particular, we have detailed information about the job seeker status during this unemployment spell: date of registration, date of hiring, the duration of benefits received, the reason of registration, and years of experience in the desired sector. We also know about job search parameters such as declared at the registration at the PES: reservation wage, maximum commute time, the desired job, the desired type of labour contract (fixed-term/permanent), and working time (full/part time). Using this data, tables \ref{tab:non_resp1} and \ref{tab:non_resp2} present descriptive statistics on the demographics, unemployment history, and search parameters of our sample, respectively. The distribution of education is close to that of the SCE (see Table 1 in \cite{mueller2021job}), with 28.1\% college degree (Bachelor or Master) or more and 67.6\% high school degree or less, as is the gender distribution (56.2\% women).

\paragraph{Return to work and other outcomes.} A critical information for our study is the actual search behavior and outcomes of it. First, we have access to the applications made through the PES platform and the characteristics of the job. 
Second, and more importantly, we can compute the date of their return to work. This is done combining three administrative databases and answers to the following waves of the survey.\footnote{We combine the ``déclarations préalables à l'embauche'' (DPAE, pre-hiring declaration), which is a statement made by the firm and compulsory in France before hiring, a re-treated monthly indicator from the PES called the ``indicateur de retour à l'emploi'' (IRE, return-to-employment indicator), administrative records exit (exit from the ``Fichier historique'') and reemployment spells declared by job seekers in subsequent waves of the survey.}

\subsection{Attrition} For this study, our dataset consists of waves 1 to 8 of the panel (October 2021-July 2023) with 25,846 usable responses and the following distribution: 2,731 (Panel 1), 4,581 (Panel 2), 2,854 (Panel 3), 3,393 (Panel 4), 3,854 (Panel 5), 3,145 (Panel 6), 2,514 (Panel 7), 2,774 (Panel 8).\footnote{We consider as usable any response in which at least the perceived probability of finding a job in 3 months is filled in.} Table \ref{tab:non_resp_waves} provides details on the response rates for each survey, which averages 15\%.\footnote{Due to reputational and ethical concerns in our partnership with the PES and the fact that we can rely on administrative data to reweight job seekers who answer the survey we did not incentivize the responses to the survey.}
Conditional on responding to the first wave, the response rates to a second wave vary from 31.7\% to 44.2\%. Of these, on average only 13.5\% have found a job and are satisfied, 18.4\% have found a job without being satisfied, 7.3\% have stopped looking, and 60.8\% are still looking. These figures are broadly similar for subsequent waves and other panels. Note that only individuals who report that they are still actively looking for a job are re-interviewed in subsequent waves. Focusing on respondents with multiple interviews about their job search, this results in 9,356 respondents with 2 interviews, 3,937 with 3 interviews, and 2,037 and 1,000 with 4 and 5 interviews respectively. An important point is that despite this attrition, the administrative database allows us to observe some job search behaviors as well as the objective outcomes of the search without attrition.

\subsection{Representativeness}  

Our study sample may not be representative of the entire population of French job seekers. First, the panel oversamples new job seekers and mainly focuses on job seekers actively looking for a job, which makes the reached group not representative of the entire stock of French job seekers. Second, the decision of job seekers to actually participate in the survey may be correlated with their characteristics, making the sample of survey respondents unrepresentative of the group that was invited to participate. To correct for these two selection biases, we compute two sets of individual weights: sampling weights and participation weights. The sampling weights adjust for the probability that job seekers appear in the group who received the survey invitation. The participation weights are computed as the inverse of the predicted probability of survey participation, based on job seekers' characteristics (age, gender, level of education, length of unemployment period, etc.).

\medskip

Tables \ref{tab:non_resp1} and \ref{tab:non_resp2} present average characteristics in the sample of survey respondents (column 1) and the whole group of job seekers invited to take the survey (column 2). Women, older job seekers, and those with higher qualifications and reservation wages are more likely to participate in the survey, as they are strongly overrepresented in the sample of respondents compared to the group invited to take the survey. Column 3 of Tables \ref{tab:non_resp1} and \ref{tab:non_resp2} shows the average characteristics of survey respondents using only the participation weights. This weighted sample properly matches the characteristics of the group invited to take the survey. Finally, column 4 displays the characteristics of survey respondents when both sets of weights are used. Comparing column 3 with column 4 shows that we oversampled new entrants when building the sample of job seekers invited to take the survey. Hereafter, unless otherwise noted, we use these weights in all regressions and analyses.

\section{Job search biases characterization}\label{sec:information}

In this section, we provide a more detailed statistical analysis of the new measures of bias that we introduce. Specifically, we first analyze the information content of the biases on the distribution of potential wages and the arrival rate of offers separately, and then consider the updating of these beliefs. We then turn to the analysis of reemployment biases, focusing on their decomposition using the latter two. Finally, we document that perceptions of labor market primitives are strong determinants of search behavior and search outcomes, as expected from theory.

\subsection{Some measures of informational biases}

\subsubsection{Biases about the distribution of potential wages}\label{ssec:wage_bias}

We use both our survey data on wage perceptions and our administrative data on posted wages in job advertisements to construct measures of bias on the distribution of potential wages that are comparable across individuals. 

\medskip

We assume that the distribution of potential wages depends on (at least) occupation and location. Based on the pool of vacancies matching the occupation and location of individual $i$ posted on the PES website, we compute her or his empirical objective potential wage cdf $F_i(w) = P_i(W \leq w)$. We then select the quartiles and median of the true potential wage distribution $F_i^{-1}(\alpha) $ for $\alpha = 0.25,0.5,0.75$ and use them as different thresholds to elicit the values of the individual's $i$ perceived probability of receiving a job offer proposing a wage at least equal to the corresponding values.\footnote{This is up to the following modification made to the questions to make them more meaningful to job seekers. When the true values $F_i^{-1}(\alpha)$ are too close together, we have multiplied the 25\% and 75\% percentiles by 0.9 and 1.1, respectively, when \textit{eliciting} these perceived probabilities. We then recompute them at the quartile $F_i^{-1}(\alpha)$ for $\alpha = 0.25,0.5,0.75$, imposing a linear interpolation of the perceived cdf.}  Denoting the cumulative distribution function of the perceived potential wage by $\widetilde{F}_i(w) = \widetilde{P}_i(W \leq w)$, we end up eliciting $\widetilde{P}_i(W > F_i^{-1}(\alpha))$ for $\alpha = 0.25,0.5,0.75$.\footnote{Here, we do not have information on bargaining, but most offers posted on the PES webside present a minimum and maximum wage, and we take the average of these two bounds.} Thus, we observe the individual wage biases at quantile of order $\alpha = 0.25,0.5,0.75$:
$$B_i(\alpha) :=  
 \widetilde{P}_i(W > F_i^{-1}(\alpha)) - (1-\alpha):=\alpha-\widetilde{P}_i(W < F_i^{-1}(\alpha)),$$
which is positive (resp. negative) if individuals over-estimate (resp. under-estimate) the probability for a job vacancy to propose a wage higher than the threshold $F_i^{-1}(\alpha)$. Note that the wage bias measured at quantile $\alpha$, namely $B_i(\alpha)$, of the objective potential wage distribution is bounded by  construction between $\alpha-1$ and $\alpha$.

\medskip

\begin{table}
\centering
\caption{Summary statistics of perception biases}
\label{tab:summary_statistics_perception_biases}
\begin{adjustbox}{width = \textwidth}
\begin{threeparttable}
\begin{tabular}{rlllllllll}
\hline
\hline
  & \textbf{Mean} & \textbf{SD} & \textbf{Min} & \textbf{Q25} & \textbf{Q50} & \textbf{Q75} & \textbf{Max} & \textbf{Nb. Obs} \\ 
    \midrule
$B_i(.25)$ & -0.28 & 0.31 & -0.75 & -0.55 & -0.29 & -0.04 & 0.25 & 16,261 \\
$B_i(.5)$ &  -0.09 & 0.31 & -0.5 & -0.36 & -0.12 & 0.14 & 0.5 & 16,259 \\
$B_i(.75)$ & 0.10 & 0.32 & -0.25 & -0.18 & 0.02 & 0.3 & 0.75 & 16,259 \\
\midrule
$\widetilde{P}(N\geq 1) - P(N\geq 1)$ & -0.22 & 0.24 & -0.83 & -0.39 & -0.23 & -0.04 & 0.53 & 19,183 \\
$\widetilde{P}(N\geq 3) - P(N\geq 3)$  & 0.11 & 0.23 & -0.49 & -0.07 & 0.08 & 0.27 & 0.86 & 18,872 \\
\hline
\hline
\end{tabular}
\begin{tablenotes}[flushleft]
    \small
    \item \textit{Notes:} Summary statistics on the biases about the distribution of wage offers (top panel, $B_i(\alpha)$) for 3 quartiles of the distribution (resp. $\alpha=$0.25,0.5, 0.75) as well as the biases on the arrival rate of job offers, $\widetilde{P}(N\geq x) - P(N\geq x)$ (bottom panel), for 2 points of the distribution (resp. $x=1,3$). We report the mean, standard deviation, the min, the quartiles, the maximum and the number of observations. The sample pools Waves 1--8 of the panel survey (October 2021--July 2023) and is restricted to the 25,846 individuals who reported their perceived 3-month reemployment probability and a main occupation.
\end{tablenotes}
\end{threeparttable}
\end{adjustbox}
\end{table}

\medskip

Table \ref{tab:summary_statistics_perception_biases} and Figure \ref{fig:wage_biases_ecdf} show the empirical cumulative distributions over all job seekers of the three potential wage biases elicited at quantiles $\alpha=0.25,0.5,0.75$.  Interestingly, the cumulative distributions of the biases at different points in the potential wage distribution exhibit first-order stochastic dominance across quantiles: job seekers are systematically more optimistic at higher points of the wage distribution (70\% of individuals display pessimistic biases at the first quartile, compared to only 40\% at the third quartile). On average, biases at the median are relatively small (-9~pp), whereas there is substantial pessimism at the first quartile (-29~pp) and moderate optimism at the third quartile (+10~pp). Taken together, these results indicate that, although posted wages are quite compressed, job seekers perceive the wage distribution as considerably more dispersed than it actually is.

\begin{figure}
    \centering
    \caption{Average shape of the true and perceived wage offer distribution}
    \includegraphics[width = 1\textwidth]{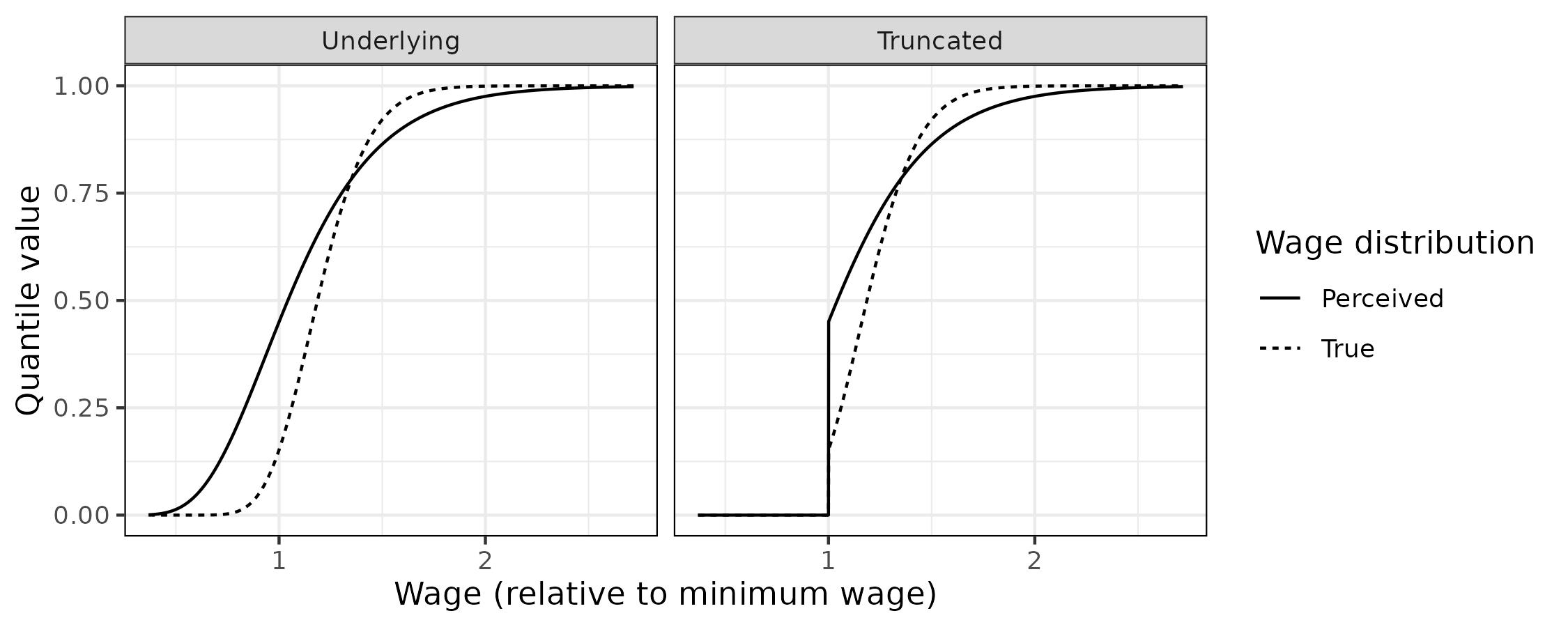}
    \caption*{\small \textit{Notes:} This figure plots the true and perceived average wage distribution. We first calibrate the true (and perceived) log normal parameters that best fit the true (and perceived) wage distribution faced by each job seeker. We then average these parameters on the whole sample and represent the cumulative distribution function and the truncated counterpart in term of wage levels relative to the legal gross minimum wage (\euro1766.92 in June 2023).}
    \label{fig:fitted_wage_distrib}
\end{figure}

To directly compare the shape of the perceived and objective wage distributions, we approximate both by log-normal distributions.  For each job seeker, we use the elicited three points of the true and perceived cumulative distribution to estimate the log-normal parameters that best fit each distribution.  Figure \ref{fig:fitted_wage_distrib} plots the resulting average log-normal distributions, expressed relative to the minimum wage, together with their minimum-wage-truncated counterparts. The figure shows that, on average across job seekers, the observed distribution dominates the perceived distribution at the second order. In standard models of job search, this would imply that, all things being equal, a rational population of job seekers would lower their reservation wage relative to this one.

\medskip

Turning to the determinants of these biases, Table \ref{tab:correlations_X_wages_biases} shows that women tend to be more pessimistic at all quantiles of the wage distribution, with an additional bias of -6.3pp at the median of the wage distribution. If these biases influence the job search strategy, they could explain part of the gender wage gap \cite[see][on the ask gap]{roussille2021central}. Similarly, we note that job seekers older than 55 years old also tend to be more pessimistic at all quantiles of the wage distribution, with an additional bias of -10.2pp for the median. Pessimism also increases with the duration of unemployment, from 0 to 4.3pp at the median after 1 year.  This could indicate either a selection effect, with those who remain unemployed being more pessimistic, or a shift away from labor market knowledge with increasing duration of unemployment.  

\subsubsection{Biases about the arrival rate of offers}\label{sec:bias:arrival}

Biases about the arrival rate of job offers can also be measured at different points in their distribution. However, when we elicit the perceived probabilities of receiving at least one and at least three job offers in the next three months, the objective values of these probabilities cannot be computed directly. Thus, we rely on the actual number of offers received reported by job seekers in their second interview to predict the objective counterparts of the perceived probabilities $\widetilde{P}(N\geq 1)$ and $\widetilde{P}(N\geq 3)$, denoted by $P(N\geq 1)$ and $P(N\geq 3)$, respectively. These objective points on the distribution of the job offer arrival rate may depend on observable characteristics as well as on the job seekers' private information contained in their perceived probabilities and their expected job-finding probability. Based on the sample of individuals who reported their true number of offers received in the second wave of the survey and who were still actively looking, we estimate a logit model to predict the objective moments of the number of offers received\footnote{We exclude individuals who were reemployed before their second interview. This ensures that we don't underestimate the number of applications received in the case they actually searched less than 3 months.}. Importantly, these predictive models include both observable characteristics, the level of effort, and associated subjective expectations.\footnote{We thus measure the bias on the arrival rate of job offers as $\tilde{P}_i(N\geq k)-\hat{P}(N\geq k|X_i)$}

\medskip

Table \ref{tab:summary_statistics_perception_biases} shows the distribution of these biases. On average job seekers underestimate the chances of receiving at least one offer by 22pp, and most are pessimistic about this event. In contrast, the average bias for receiving more than 3 offers is slightly positive (+11pp). Figure \ref{fig:ecdf_biases_lambda} presents the distribution over our sample of these biases. We note that 80\% underestimate the probability of receiving at least 3 offers, while only 40\% underestimate the probability of receiving at least 1 offer. 

\medskip

Columns (1)-(2) and (4)-(5) of table \ref{tab:determinants_arrival} show the determinants of the offer arrival rate and subjective expectations regarding the number of job offers in the next three months after the first survey date. The main results from columns (1) and (4) are that those aged 35--45 and 45--55 are more likely to receive offers (for the latter: 15.1 pp more at least one offer, 18.8 pp at least 3 offers). We also find a correlation between the duration of unemployment and the arrival rate of the offer, since the probabilities of receiving 3 offers are higher in the early months of unemployment. In columns (2) and (5) we also report estimates of probability linear models for the true indicators of offers received on observable characteristics but also including the subjective probabilities $\widetilde{P}(N\geq 1)$ and $\widetilde{P}(N\geq 3)$. An important finding is that expectations are strongly predictive of realizations (regression coefficients of 0.344 and 0.332, respectively), even after controlling for a large number of covariates.  However, if we were to test the rational expectations hypothesis of subjective expectations about the number of offers to be received, it would be rejected at all standard levels. Subjective expectations still provide relevant private information to predict the objective distribution of the number of offers received. Finally, columns (3) and (6) consider the determinants of subjective expectations. Focusing on the subjective probability of receiving 3 or more job offers (column 6), we observe that older job seekers, specifically those over 55, tend to have lower expectations about the number of offers they will receive (-13.1pp), while the opposite or no differences are observed for these groups in terms of realized offers. We also observe that these subjective expectations decrease with the duration of unemployment (from -4.9pp between 1-3 months to -12.5pp after 2 years compared with individuals with spell duration inferior to 1 month). 

\medskip

A caveat applies to the elicited offer probabilities. As we document in Appendix~\ref{app:perceived_arrival_rate}, the subjective probabilities of receiving at least one and at least three offers are empirically very close, which is hard to reconcile with the parsimonious Poisson benchmark and suggests that respondents do not interpret the ``at least one offer'' item literally. This concern is confined to \emph{cardinal} uses of these items, that is, to the mapping of an elicited probability into an arrival-rate parameter, which we perform only in the structural model of Section~\ref{sec:pred_outcome} and in the perceived-returns comparison of Figure~\ref{fig:perceived_return}. All of the descriptive and stratification results instead use these items nonparametrically---through subjective-minus-objective bias differences and as ordinal proxies for perceived employability---and rely primarily on the more reliable ``at least three offers'' item. They therefore do not depend on the Poisson assumption or on a literal reading of the ``at least one offer'' question.

\subsubsection{Beliefs updating}\label{sec:b_updating}

The panel dimension of our data also allows us to document the updating of beliefs with the duration of unemployment and close to the exhaustion of benefits. We first consider the following model, where the belief or bias of interest is generically denoted by $Y_{i,t} $: 
\begin{equation}\label{eq:update}
    Y_{i,t} = \sum_{k=1}^6 \gamma_k \indic\{ \tau_{i,t} = k\} + \alpha_i + \varepsilon_{i,t},
\end{equation}  
where $\tau_{i,t}$ is the elapsed duration of unemployment and $\alpha_i$ is an individual fixed effect. Table \ref{table_update_duration} shows the results of this updating for the probabilities of receiving job offers, the associated biases, the biases at the median of the wage distribution and the spread, as well as the bias at the reported reservation wage. The main results are that the bias in the perception of the wage distribution at the reservation wage is updated downward (-8pp/3m after 1 year), as is the bias at the spread (-6pp/3m after 1 year), while there is no adjustment at the median. Since job seekers tend to overestimate the dispersion of wages, this suggests some learning about the objective dispersion of the wage distribution they face. Job seekers tend to become increasingly, but slightly, optimistic about receiving a large number of offers (+3pp after 1 year) and pessimistic about receiving few offers (-4pp after 1 year). Similar to \cite{marinescu2021unemployment}, we checked for updating close to the end of benefits, but we do not find strong evidence of changes in these beliefs.

\medskip

\subsection{Decomposing reemployment biases using beliefs}

Using our panel data, we observe, for the same individual and at different points during the unemployment spell, subjective assessments of the probability of returning to employment at several horizons ($t = 1$ month, 3 months, 6 months, and 12 months), denoted
$\psi_t=\tilde{P}(T_i\geq t|\mathcal{I})$. Because we are also able to track job seekers through administrative records, we observe the corresponding realized outcomes through indicators of remaining unemployed at these horizons,
$\indic(T_i\geq t)$. Section~\ref{sec:heterogeneity} provides additional details on how these data allow us to identify reemployment biases and, in particular, their heterogeneity across individuals. Even at this stage, however, we can already document several important stylized facts, including the predictive power of subjective beliefs $\psi_t$ for subsequent unemployment durations $T$, the magnitude of re-employment biases at different horizons $t$, and their evolution over the course of the unemployment spell.

\subsubsection{Elicited expectations about reemployment: first facts}\label{ssec:mbias}

We present three main stylized facts about reemployment expectations and biases that either complement or extend the literature. 

\paragraph{Job-finding expectations have predictive power.} A first important fact is that these expectations contain important information for predicting unemployment duration that is not contained in the (very large) set of characteristics at our disposal \cite[see, \emph{e.g.}][in other countries and with different sets of characteristics used as controls]{mueller2021job,mueller2023expectations,van2024predicting}.\footnote{Formally, these beliefs contain useful information that differs from the usual observables $X$, if adding $\psi_t$ to the knowledge of $X$ helps to predict the duration of unemployment: $E(\indic\{ T\leq t\} |\psi_t,X ) \neq \psi_t$.
} Table \ref{tab:predictive} considers the linear regression of the dummy variable indicating return to work before $t$ months on the elicited probability of getting a job before $t=3$, $6$, or $12$ months, without controls (columns 2, 5 and 8), then additionnally controlling  for a large set of covariates including skills, demographics, past behavior, location, and search parameters, which we generically denote as $X$ (columns 3,6,9), and only on the latter controls (columns 1,4,7). First, we find that without or with $X$ and at all horizons, expectations are strong predictors of unemployment duration at the same horizon, highlighting their information content. For example, without controls, our point estimate for the coefficient of subjective job-finding expectations before 6 months is $0.410$, which is significant at the 1\% level. The explanatory power is substantial ($R^2=0.065$ at 6 months), close to that with controls ($R^2=0.076$).\footnote{Note that even if job seekers had perfect information and rational expectations $E(\indic\{ T\leq t\} |\psi_t,X ) =\psi_t$, then $R^2$ would be much less than 1 and equal to $\text{Var}(\psi_t)/(E(\psi_t)(1-E(\psi_t))$. For our data, this gives 0.36, 0.40, and 0.46 for the 3-month, 6-month, and 12-month horizons, respectively.} Our estimates are slightly reduced to $0.341$ for this 6-month horizon when controlling for $X$, which remains significant at the 1\% level. First, these results confirm that these elicited expectations do indeed contain useful information for learning about one's own search outcomes. Moreover, these results are broadly consistent with those observed for the SCE in \cite{mueller2021job}, i.e. a coefficient at the 3-month horizon of $0.414 \ (0.015)$. Finally, if without controls the predictive power seems stable with respect to the prediction horizon, it seems that the controls bring more explanatory power at longer horizons: $54\%$ gain at 3 months while $105\%$ gain at 1 year.

\paragraph{Average short-term optimism.} A natural next step is to focus on the average biases of these expectations. This can be assessed directly by comparing the averages of expectations with the averages of realizations. We perform an analysis of the heterogeneity of the determinants of these biases in section \ref{sec:heterogeneity}. 

\begin{figure}[H]
\caption{Average short-term optimism on the eight first waves of the panel}
\includegraphics[width = \textwidth]{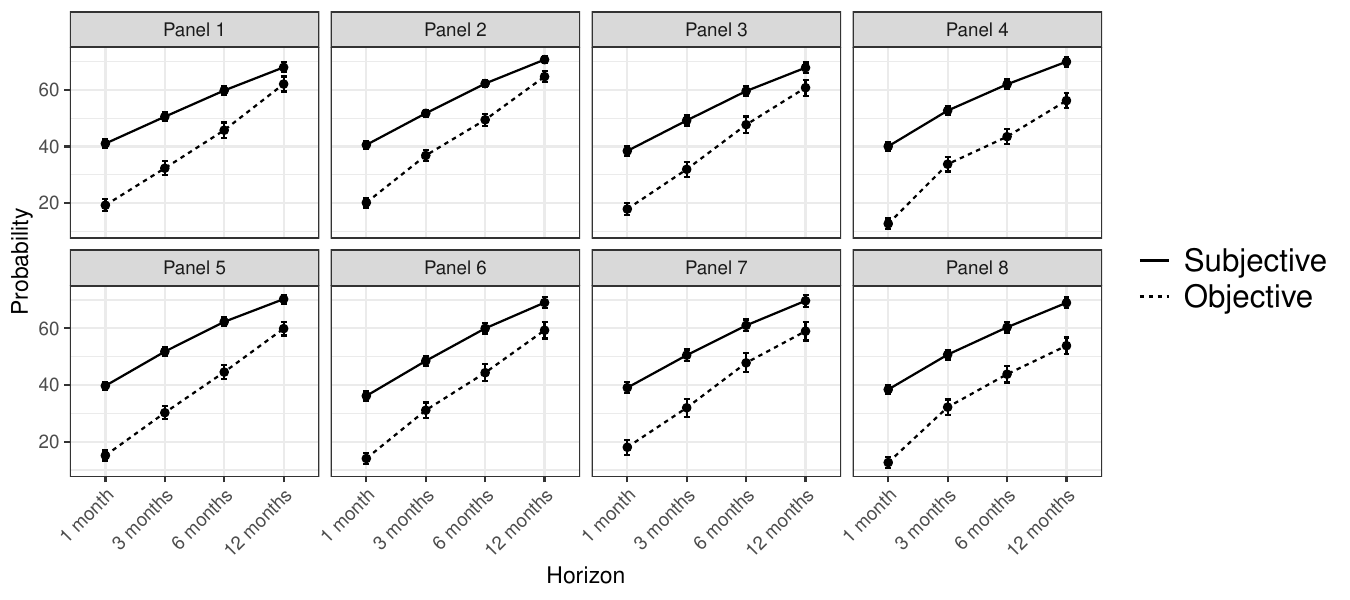}
  \caption*{\small{\textit{Notes:} Dark points represent the average subjective (in plain) or objective (in dots) probability of finding a job at the horizons specified on the $x$-axis. They are associated with 95\% confidence intervals. We consider the 8 first waves of the panel (every quarter from 10-2022 to 07-2023). We also use sampling weights calibrated using administrative data to be representative of the population of job seekers at this time.}}
	
\label{fig:avg_bias_by_panel}
\end{figure}

First, job seekers are on average optimistic about their unemployment spell at all horizons. Figure \ref{fig:avg_bias_by_panel} makes this comparison for the 8 panels we consider at the 1, 3, 6, and 12 month horizons. It shows that, on average, job seekers' expectations are significantly higher than the actual duration of their unemployment. This average bias is significant in the short run (1 to 3 months) and is reduced for long-term forecasts (24 pp at 1 month versus 10 pp at 1 year when pooling all panels). Importantly, this average optimism seems to be an empirical fact that is constant across the different waves of the panels in Figure \ref{fig:avg_bias_by_panel}. This adds to the picture of short-term (3-month) evidence documented for the US \citep[see, \emph{e.g.}][]{mueller2021job}. 

\paragraph{Updating of reemployment biases.}

Columns (1) to (4) of table \ref{table_update_duration_psi} show the results of updating subjective job-finding probabilities at different horizons (1, 3, 6, 12 months) using model \eqref{eq:update}. The main result of this table is that, on average, job-finding expectations are updated differently depending on whether they are short-term or long-term during the unemployment spell. Indeed, 1 and 3 month job-finding expectations are updated upward, by about 7pp/3m for the 1m probability and 3pp/3m for the 3m probability. However, these expectations are not updated or even downward for the relatively long horizons: no update at 6m after 1y, -3pp/3m at a 12m horizon. The latter result is consistent with theoretical predictions in a model with rational expectations, since the objective job-finding probability should hardly be updated at all or even slowly downward (see, \emph{e.g.}, \cite{mueller2021job}) due to human capital depreciation. Thus, this description of horizon-dependent updating importantly complements the previous puzzling upward updating described in \cite{mueller2021job} observed using the SCE data, which focuses only on 3 months expectations. This suggests that we should model the term structure of these expectations. 


\subsubsection{Reemployment biases and informational biases}

We now examine the influence of biased beliefs about the two key labor market parameters on job-finding expectations. The (perceived) re-employment probability can be decomposed as the product of the (perceived) acceptance probability conditional on receiving a job offer times the (perceived) probability of receiving a job offer. In a standard McCall job search model, these probabilities can be related to the (perceived) distribution of wage offers and the (perceived) offer arrival rate, respectively. This decomposition implies that, according to job search models, biases in reemployment expectations would potentially arise from the (linear) combination of these two information biases. 

\medskip

Figure \ref{fig:correlation_wage_arrival_biases} provides an initial empirical test of this theoretical relationship. We partition the set of observed job seekers into 100 cells based on their bias with respect to their offer arrival rate. Figure \ref{fig:correlation_wage_arrival_biases} then represents the associated group mean bias with respect to the median of the wage distribution as a function of the group mean bias with respect to offer arrival rate. The two main information biases appear to be strongly positively correlated with each other. As discussed in more detail in section \ref{sec:heterogeneity}, taking for each group the difference between the average subjective reemployment rate and the realized reemployment indicator provides an unbiased estimator of this group reemployment bias. The second important takeaway from figure \ref{fig:correlation_wage_arrival_biases}, is that the reemployment bias appears to be strongly correlated with both information biases. These biases appear to explain a significant portion of the heterogeneity in the reemployment biases, which range from -0.1 to 0.3, depending on the group's mean arrival rate of the offer bias.

\medskip

To sharpen this empirical evidence, table \ref{tab:reduced_form_jobfindingbias_perceptionbias} shows the estimates of a reduced-form regression of reemployment bias on informational biases.  
The results confirm the theoretical relationship between job search bias and information bias. Even after controlling for factors that may influence beliefs, we find that misperceptions about labor market parameters are strong predictors of biases in reemployment beliefs. Adding these information biases also significantly improves the explanatory power of the model, nearly doubling the $R^2$ compared to the model in which we include key observable characteristics and fixed effects for target occupation and region (from 0.049 to 0.082). Biases regarding the arrival rate of offers appear to have a larger impact: a 10 pp increase in this bias is associated with a 4 pp increase in the three-month reemployment bias. Wage biases have a more moderate effect, but this effect increases when considering biases with respect to higher quantiles of the wage distribution. Consistent with theoretical intuitions, we find that optimism about the arrival rate of offers and the wage distribution is associated with optimism about the probability of finding a job. Indeed, when the reservation wage is fixed, a perceived first-order dominance of the objective wage distribution implies a higher acceptance rate, while a higher offer arrival rate increases the probability of finding a job.  Overall, informational biases appear as important determinants of reemployment biases.

\subsection{Informational biases are determinants of search behaviors.}

In theory, job seekers' beliefs about fundamental parameters of the labor market should determine their search parameters. This section provides empirical evidence on these relationships, with a special focus on identification. We focus first on the reservation wage and then on search effort. We take care to restrict the samples to individuals with consistent beliefs about reemployment, wage distributions, and offer arrival rates.\footnote{More precisely, we restrict the sample to job seekers who perceive a higher reemployment rate as the horizon increases, lower probabilities of receiving more job offers, and lower probabilities that a job offer will propose a higher wage.}

\paragraph{Information biases and reservation wage.}

According to theory, a job seeker's reservation wage should be an increasing function of his beliefs about both the median of the wage offer distribution and the offer arrival rate. We leverage our measures of the two related informational biases to test this hypothesis. Table \ref{tab:search_behaviors_wage} shows, in column (1), the cross-sectional regression of the log reservation wage on biases regarding labor market parameters. As expected, a 10 pp more optimistic belief about the median wage is associated with a 2.3\% increase in the reservation wage. A more dispersed wage distribution also leads to a higher reservation wage (+1.2\% for a 10 pp higher bias regarding the probability of a wage offer falling between the 25th and 75th percentiles). 
An important concern with this first specification is the potential simultaneity problem. Indeed, the degree of bias in the offer arrival rate may depend on the actual level of search effort. In theory, job seekers jointly set their reservation wage and their level of effort, implying that the biases on the arrival rate of offers and the reservation wage are determined by the same process.

\medskip

\begin{table}[!htbp]\centering
\caption{Regression of log reservation wage on labor market representations}
\label{tab:search_behaviors_wage}

\begingroup

\begin{adjustbox}{width = \textwidth}
\begin{threeparttable}
\begin{tabular}{l*{6}{c}}
\hline
\hline
& \multicolumn{6}{c}{\textbf{Log reservation wage}} \\
\cline{2-7}
Sample & Waves 1--8 & \multicolumn{3}{c}{Waves 6--8} & \multicolumn{2}{c}{Waves 6--8} \\
Specification & \multicolumn{4}{c}{Cross section} & \multicolumn{2}{c}{Panel} \\
& (1) & (2) & (3) & (4) & (5) & (6) \\
\midrule

$B_{.5}$ 
& 0.231$^{***}$ & 0.230$^{***}$ & 0.232$^{***}$ & 0.257$^{***}$ & 0.094$^{**}$ & 0.119$^{***}$ \\
& (0.017)  & (0.037)  & (0.036)  & (0.045)  & (0.036) & (0.038) \\ 

$B_{.75}-B_{.25}$ 
& 0.118$^{***}$ & 0.129$^{*}$ & 0.123$^{*}$ & 0.080 & 0.027 & 0.013 \\ 
& (0.025) & (0.070) & (0.070) & (0.078)  & (0.042) & (0.042) \\ 

$\widetilde{p_N}(3)-p_N(3)$
& -0.023 &  &  &  &  &  \\
& (0.022) &  &  &  &  &  \\

$p_{N|e=20}(1)$
&  & -0.002 & & -0.079$^{*}$ & 0.102** &  \\
&  & (0.034) &  & (0.043) & (0.041) &  \\

$p_{N|e=30}(1)$ - $p_{N|e=10}(1)$
& & & 0.037 &  &  & 0.034 \\
&  &  & (0.038) &  &  & (0.041) \\

Spell duration (months)
&  &  &  &  & -0.003* & -0.004** \\
&  &  &  &  & (0.002) & (0.002)  \\

\midrule

Region fixed effects & \checkmark & \checkmark & \checkmark & \checkmark &  &  \\
Target occupation fixed effects & \checkmark & \checkmark & \checkmark & \checkmark &  &  \\
Panel fixed effects & \checkmark & \checkmark & \checkmark & \checkmark &  &  \\
Psychological traits &  &  &  & \checkmark &  &  \\
Individual fixed effects &  &  &  &  & \checkmark & \checkmark \\

\midrule

Observations & 6,034 & 1,845 & 1,809 & 911 & 1,208 & 1,206 \\
Job seekers  & 6,034 & 1,845 & 1,809 & 911 & 487 & 487 \\

\hline
\hline
\end{tabular}
\begin{tablenotes}[flushleft]
    \small
    \item \textit{Notes:}  This table presents estimate of the influence of subjective beliefs on reservation wage. The first four columns are cross-sectional regressions. Beliefs about the wage distribution are measured by the bias at the median ($B_{.5}$) and the bias at the interquartile range of the distribution ($B_{.75} - B_{.25}$). Measures of perceived offer arrival rates include the bias in the probability of receiving at least three job offers within three months, the probability of receiving at least one job offer conditional on 20 hours weekly effort on average and the increase of the arrival rate of offers with additional search effort. Column (4) additionally controls for psychological traits. Columns (5)--(6) use panel data with individual fixed effects. Weights are used. Standard errors in parentheses are either heteroskedastic robust (columns (1)--(4)) or clustered at the individual level (columns (5)--(6)). $^{*}p<0.1$, $^{**}p<0.05$, $^{***}p<0.01$.
\end{tablenotes}
\end{threeparttable}
\end{adjustbox}
\endgroup
\end{table}

\medskip

In columns (2) and (3), we use alternative measures of beliefs about the arrival rate of offers. The first is the perceived arrival rate of offers, conditional on an average of 20 hours of job search per week\footnote{This corresponds to the subjective probability of receiving at least one job offer within three months, denoted by $p_{N|e=20}(1)$}. This measure does not depend on the jobseeker's choice parameters, and is strongly correlated with the perceived offer arrival rate, making it a useful proxy for beliefs about the offer arrival rate. Our second measure is the perceived increase in the job offer arrival rate associated with higher weekly search effort.\footnote{This measure is computed as the difference between the subjective probability of receiving at least one job offer within three months when searching 30 hours per week and the corresponding probability when searching 10 hours per week.} Again, this belief does not depend on the actual level of effort, thus avoiding any simultaneity issues. This second measure complements the first one as it rather focuses on the shape of the arrival rate of job offers with respect to search effort rather than its level. Note that these two measures are only available from wave 6 of our panel.
The signs and magnitudes of the coefficients associated with beliefs about the wage distribution remain consistent with our first specification. In column 4, we include measures of psychological traits (Big Five and locus of control) to examine whether they affect our estimates. Overall, only the influence of the spread of the wage offer distribution seems to decrease when controlling for psychological traits.

\medskip

In columns (5) and (6), we take advantage of the panel structure of our data to further control for individual unobserved heterogeneity including individual fixed effects as well as spell duration.\footnote{However, restricting the sample to individuals with at least two survey responses raises issues of selection and statistical power.} The estimated impact of wage perceptions is smaller, but still significant: a 10 pp increase in the perceived probability that a job offer will propose a wage above median is associated with a 1\% increase in the reservation wage. The difference compared with cross-sectional estimation is likely to come from both unobserved heterogeneity affecting both labor market perceptions and reservation wage.  Consistently with theoretical predictions, the perceived level of the job offer arrival rate has a positive effect on the reservation wage: a 10pp increase in the subjective probability of receiving at least one job offer conditional on 20 hours of weekly search is associated with a 1\% increase in the reservation wage. In contrast, we find no significant effect of the perceived return to search effort on the level of the reservation wage. Finally, we also find a small but statistically significant 0.3\% decrease in the reservation wage per month over the course of the unemployment spell, consistent with theory. 


%

\paragraph{Information biases and search effort.}

We estimate similar specifications to assess the influence of subjective beliefs about labor market parameters on the level of search effort. Columns (1) to (4) of table \ref{tab:search_behaviors_effort} show estimates from cross-sectional regressions. Job seekers with optimistic beliefs about the median and dispersion of the wage distribution also exert more search effort: a 10 pp increase in the bias at the median is associated with an increase of 0.28 hours of search effort per week. Beliefs about the arrival rate of offers also have a positive effect on search effort. A 10pp increase in the bias about the arrival rate of offers ($\widetilde{p}_N(3) - p_N(3)$) is associated with an additional 0.2 hours of weekly search effort.  Columns (2)--(6) include alternative measures of perceived arrival rate of job offers that do not depend on job seekers' choice parameters: the perceived level of arrival rate of offers conditional on 20 hours of search per week ($p_{N|e=20}$) and the perceived return to search effort on the arrival rate of offers ($p_{N|e=30} - p_{N|e=10}$). These show that the level of search effort is an increasing function of perceived return to search on the arrival rate of offers: a +10pp increase in the gap between the arrival rate of offers conditional on 10 hours of search and conditional on 30 hours is associated with an additional +0.58 hours of search per week. In contrast, the perceived level of the arrival rate conditional on a fixed level of effort has little effect. Panel specifications confirm these findings: both the bias in median wage beliefs and the perceived return to search effort remain positively associated with search intensity. We also document a significant increase in search effort as the unemployment spell lengthens. These findings are consistent with potentially motivated beliefs: job seekers with higher perceived return to search effort actually search more. 

\subsection{Locus of control determines perceived arrival rate of offers}\label{ssec:psychic_beliefs}

In addition to beliefs about the fundamental parameters of the labor market, psychological traits have also been shown to predict search behavior. One of the most widely used models to describe individuals' personalities is the Big Five. Among the five major psychological traits, Conscientiousness and Neuroticism have been shown to be the most predictive of labor market outcomes, particularly in terms of job performance and wages (see \cite{almlund2011personality} for a review of studies linking economics and psychology). Alternatively, several papers have focused on the notion of locus of control to better understand the search behaviors. \cite{caliendo2015locus} and \cite{mcgee2016search} have documented the influence of internal locus of control on perceptions of returns to search effort and the observed level of exerted effort. Finally, self-esteem has also been shown to be a relevant psychological trait in predicting unemployment status, even after controlling for locus of control \citep[see][]{huysse2015low,mendolia2015youth}. Although these three types of personality measures target different aspects, \cite{almlund2011personality} suggests that self-esteem and locus of control are strongly related to neuroticism and emotional stability, implying that they may share common determinants.

\medskip

In this section, we investigate the psychological determinants of job seekers' labor market beliefs. We focus specifically on two components of the perceived job offer arrival rate. The first is the \textit{level} of the arrival rate conditional on a given search effort, here 20 hours per week, which captures job seekers' self-perceived employability independently of their search choices. We refer to these as \textit{baseline beliefs}, as they primarily reflect job seekers' assessment of their own value on the labor market, closely related to self-esteem. The second component is the return to search effort, measured as the difference in the perceived probability of receiving at least three offers between 30 and 10 hours of weekly search. These \textit{control beliefs} capture job seekers' perception of their ability to influence their environment through effort, a construct closely related to internal locus of control \citep[see][for a theoretical discussion]{spinnewijn2015unemployed}.
\medskip

We build measures of internal and external locus of control and the Big Five personality traits (see Appendix~\ref{ssec:app_psycho_items}), all normalized and standardized. Table~\ref{tab:psycho_perceptions} reports regressions of these two perceived parameters on standard demographic controls, region, panel, and target occupation fixed effects, with and without psychological traits. The sample is restricted to job seekers in panels 6 through 8 who hold consistent beliefs about 
the offer arrival rate.\footnote{Beliefs such that greater search effort leads to a higher offer arrival rate, and the probability of receiving at least one offer in the next month is strictly between 0 and 1.}

\medskip

The results point to locus of control as the dominant psychological correlate of both types of beliefs, with limited additional explanatory power from the Big Five traits. A one-standard-deviation increase in internal locus of control is associated with a 3.7~pp increase in the perceived probability of receiving at least one offer conditional on 20 hours of weekly search, and a 1.5~pp increase in the perceived return to effort. This pattern is consistent with the conceptual distinction above: job seekers who believe they have greater control over their environment not only perceive higher returns to search effort, but also report higher baseline employability, suggesting that internal locus of control shapes labor market beliefs more broadly than previously documented.



%
%

\section{The impact of informational biases}\label{sec:pred_outcome}

Stylized facts about job seekers' beliefs reveal important heterogeneity in informational biases that can strongly influence search behavior.  Using a job search model with subjective expectations, we quantify the causal effect of informational biases. This allows us to demonstrate how heterogeneity in reemployment biases can inform subsequent interventions. 

\subsection{Job search model with multiple sources of biases}

\paragraph{Model setup.}
We consider an extension from the standard McCall search model \citep[see, \emph{e.g.},][]{mccall1970economics,van1994effects}, in which job seekers face a flow of job offers that they decide to accept or reject. In this model, job offers arrive at job seeker $i$ according to a Poisson process with parameter $\lambda_{d,i}(\cdot)$.\footnote{We take the Poisson process as a parsimonious benchmark; Appendix~\ref{app:perceived_arrival_rate} discusses its empirical fit to the elicited offer probabilities and how we recover perceived arrival rates from the survey items.} Job seekers can influence their offer arrival rate by increasing their effort. However, the rate at which offers arrive may also depend on the duration of unemployment, e.g., due to human capital depreciation.  The wages proposed in each offer are randomly and independently drawn from a distribution $F_W$. 

\medskip

For tractability, we assume that $\lambda_{d,i}(e) = \lambda_{i}^0 e \theta^d$,  where $\lambda_{i}^0$ captures job seekers' employability and determines the  return to search effort. The arrival rate of job offers decreases over time following a geometric depreciation indexed by parameter $\theta$. While greater  search effort increases the arrival rate of job offers, it also entails a direct  cost of search  $c_i(\cdot)$, incurred in the current period. We consider this cost to be increasing and  convex in search intensity, i.e., $c' > 0$ and $c'' > 0$. During each period, job seekers jointly choose their reservation wage ($w^{*}$) and effort level ($e^{*}$), balancing two trade-offs: a higher reservation wage improves match quality but extends the unemployment spell, while greater search effort raises reemployment prospects at the expense of immediate search costs.

\paragraph{Beliefs about the arrival rate of job offers.} We introduce subjective beliefs into this model in two ways. First, as documented in Section~\ref{sec:information}, job seekers may have informational biases regarding the arrival rate of offers $\lambda_{d,i}(\cdot)$. We consider subjective beliefs about the individual employability parameter $\lambda_{i}^0$ and the depreciation rate $\theta$, respectively denoted by $\widetilde{\lambda}^0_{i}$ and $\widetilde{\theta}$. Thus, the perceived job offer arrival rate is $\widetilde{\lambda}_{d,i}(e) = \widetilde{\lambda}_{i}^0 e \widetilde{\theta}^d$.\footnote{This informational bias may also arise from misconceptions about the shape of the function $\lambda_{d,i}(\cdot)$. Among others, \cite{adams2023perceived} show that the perceived return to effort is close to constant, so the linear form seems a relevant approximation.} 

\medskip

This functional form aligns with the empirical findings of Section~\ref{ssec:psychic_beliefs}, in which the perceived conditional arrival rate and the returns to search effort are both determined by the same psychological dimension. The parameter $\widetilde{\lambda}_{i}^0$ can be interpreted as indexing the job seeker's self-esteem: it directly determines the perceived level of offers received and hence the job seeker's perceived attractiveness on the labor market.

\paragraph{Beliefs about wage distribution.} Second, we relax the assumption that job seekers have accurate perceptions of the wage distribution $F_{W}$. We assume that the true wage distribution is log-normal with parameters $\mu, \sigma$, denoted $F_{W} = \mathcal{LN}(\mu, \sigma)$. We allow job seeker $i$'s beliefs to deviate from the true distribution in both the median and the standard deviation, and denote by $\widetilde{\mu}_i$ and $\widetilde{\sigma}_i$ the parameters of the perceived wage distribution:
\begin{equation*}
    \widetilde{F}_{W_i} = \mathcal{LN}(\widetilde{\mu}_i, \widetilde{\sigma}_i).
\end{equation*}
Rational job seekers satisfy $\widetilde{\mu}_i = \mu$ and $\widetilde{\sigma}_i = \sigma$.

\paragraph{Job seeker's strategy.} The job seeker's strategy boils down to determining the reservation wage ($w^*$) in order to maximize her or his current unemployment value $U_{d,i}$: 
\begin{equation*}
    U_{d,i} = u(b_{d}) -  c_i(e) + \beta \left(U_{d+1,i} + \widetilde{\lambda}_{d,i}(e) \int_{w \geq w^*} [V_i(w) - U_{d+1,i} ] d\widetilde{F}_{W}(w)\right),
\end{equation*}
where $u(\cdot)$ denotes the utility function, taken as CRRA with parameter $\gamma$, $b_{d}$ is the unemployment benefits perceived in period $d$, and $V_i(w)$ is the value of being employed at wage $w$ which equals the actualized sum of future wages. Unemployment benefits decrease after 24 periods, similarly to the French unemployment insurance system.

\paragraph{Costs of low self-esteem.} We document a positive correlation between beliefs about the return to search effort and the actual level of search effort. The magnitude of this correlation suggests that these beliefs may have a motivational dimension. To allow for this behavioral effect in the model, we allow the cost of effort to depend on self-esteem: $c_i(e) = \left(c_1 + c_\lambda/\widetilde{\lambda}_{i}^0\right) e^2$.\footnote{Using this parametrization, we allow for more flexible $de^*/d\lambda^0$. Note that since we are using a linear functional form $\lambda_{d,i}(e) = \lambda^0_{i} e \theta^d$, we are already imposing $de^*/d \lambda^0 > 0$ \citep[see][]{cahuc2014labor}.} More broadly, $c_\lambda$ captures the extent to which anxiety and uncertainty associated with job search raise the perceived cost of effort. When $c_\lambda > 0$, job seekers with higher self-esteem perceive a lower marginal cost of effort.

\paragraph{Unobserved heterogeneity.}
We allow for heterogeneous beliefs among job seekers, as well as for heterogeneous true employability levels $\lambda_{i}^0$. Similarly to \cite{mueller2021job}, we categorize job seekers into two types based on their underlying employability. Low-type job seekers, with employability denoted by $\lambda^{0,l}$, account for a share $\phi$ of the population and have lower employability than high-type job seekers, i.e., $\lambda^{0,l} < \lambda^{0,h}$.

\subsection{Estimation procedure and results} \label{ssec:estimation}

\paragraph{Sample.} For the estimation, we further restrict the initial sample such that we can compute the initial beliefs distribution for each job seeker. More precisely, we consider (i) only job seekers of panels 6, 7, and 8, (ii) without missing variables among the search parameters, elicited bias about the wage distribution, marginal returns to effort, (iii) coherent beliefs, (iv) with less than 1 year unemployment experience. 

\paragraph{Perceived employability.}  

For every individuals in the restricted sample, we use the elicited perceived arrival rate of job offers under several effort scenarios to compute individual subjective baseline employability $\widetilde{\lambda}^0_i$. The procedure proceeds in three steps.

\medskip

\noindent\textit{Step 1: recovering individual Poisson parameters.} Under the  assumption that job offers arrive according to a Poisson process, each survey  question allows us to recover the underlying monthly Poisson parameter for a  given level of weekly search effort.\footnote{Specifically, if the perceived  probability of receiving at least three offers over three months under effort 
$s$ is $\tilde{p}_i(s)$, we recover the three-month Poisson parameter  $\widetilde{\Lambda}_{d,i}(s)$ by numerically solving 
$1 - \exp\!\left(-\widetilde{\Lambda}\right)\left(1 + \widetilde{\Lambda} + \widetilde{\Lambda}^2/2\right) = \tilde{p}_i(s)$, and set $\widetilde{\lambda}_{d,i}(s) = \widetilde{\Lambda}_{d,i}(s)/3$.} This yields three individual Poisson parameters $\widetilde{\lambda}_{d,i}(10)$,  $\widetilde{\lambda}_{d,i}(20)$, and $\widetilde{\lambda}_{d,i}(30)$.\footnote{As discussed in Appendix~\ref{app:perceived_arrival_rate}, the underlying Poisson 
parameters are estimated as if the survey questions referred to the event of  receiving at least three job offers within the next three months.}

\medskip

\noindent\textit{Step 2: identifying the perceived depreciation rate.} The panel structure of the survey allows us to identify the perceived depreciation of the arrival rate. For each individual, we compute the ratio of the average Poisson parameter across the three effort scenarios at the second interview (three months later) to its counterpart at the first interview, and take the cube root to obtain a monthly depreciation rate.\footnote{Formally, $\widetilde{\theta}_i = \left(\bar{\widetilde{\lambda}}_{d+3,i} \,/\, \bar{\widetilde{\lambda}}_{d,i}\right)^{1/3}$, where $\bar{\widetilde{\lambda}}_{d,i} = \sum_{s \in \{10,20,30\}} 
\widetilde{\lambda}_{d,i}(s)/3$.} We take the median value across individuals as our estimate of the common perceived depreciation parameter $\widetilde{\theta}$.

\medskip

\noindent\textit{Step 3: recovering individual baseline employability.} To recover individual perceived baseline employability $\widetilde{\lambda}^0_i$,  we exploit the variation in Poisson parameters across effort levels. Since $\widetilde{\lambda}_{d,i}(e) = e\,\widetilde{\lambda}^0_i\,\widetilde{\theta}^d$ is linear in $e$ with slope $\widetilde{\lambda}^0_i  \widetilde{\theta}^{d_i}$ and where $d_i$ denotes the number of months elapsed since the start of individual $i$'s unemployment spell at the time of the first interview. This slope identifies the product $\widetilde{\lambda}^0_i \widetilde{\theta}^{d_i}$. We estimate this slope for each individual by minimizing the sum of squared residuals across the three effort scenarios.\footnote{Formally, we estimate $(a_i, b_i) = \arg\min_{a,b} \sum_{e \in \{10,20,30\}} \left(a + be - \widetilde{\lambda}_{d,i}(e)\right)^2$, and retain $b_i$ as the estimated slope. The intercept $a_i$ is included to avoid imposing a zero-origin constraint but is not used further.} We then recover $\widetilde{\lambda}^0_i$ by dividing the estimated slope $b_i$ by $\widetilde{\theta}^{d_i}$.

\paragraph{Perceived wage distribution}

Our survey also allows us to identify individuals’ wage perception biases, as described in Section~\ref{ssec:wage_bias}. For each job seeker, we recover the perceived location and dispersion parameters of the log-normal wage distribution, denoted by $\widetilde{\mu}_i$ and $\widetilde{\sigma}_i$, as well as their true counterparts, $\mu_i$ and $\sigma_i$, estimated from administrative data on job vacancies posted on the PES platform. To abstract from heterogeneity in true wage distributions across occupations and locations, we express all wages relative to the minimum wage.

\paragraph{Generated Moments.}
We set the discount rate $\beta$ to 0.996 and the coefficient of relative risk aversion $\gamma$ to 2, following standard values in the job search literature \citep[see, e.g.,][]{mueller2021job}. Given these fixed parameters, the vector of parameters to be estimated, and the calibrated distributions of perceived employability and wage perception biases, we simulate moments as follows.

\medskip

For each job seeker in our estimation sample, we directly use their individually recovered belief parameters (perceived employability $\widetilde{\lambda}^0_i$ and wage perception parameters 
$\widetilde{\mu}_i$ and $\widetilde{\sigma}_i$) to solve for their optimal search parameters.\footnote{The three beliefs parameters are winsorized at the 5th and 95th percentile.} This yields what we refer to as our \textit{generated sample}: a dataset in which informational biases are taken directly from the data, while search behaviors and reemployment probabilities are generated by the model. By construction, this sample perfectly reproduces the empirical distribution of subjective beliefs about labor market primitives.

\medskip

We then assign true employability types to simulated job seekers. Lacking direct information on the correlation between perceived and true employability, we rely on a \textit{correct beliefs ranking} assumption: among all simulated job seekers, the share $\phi$ of job seekers with the lowest perceived employability is assigned to the low type, and the remainder to the high type. This assumption implies that perceived employability correctly ranks job seekers in terms of true employability, even if it does not perfectly predict its level.\footnote{We cannot test this assumption directly, as true employability types are not observed and are themselves estimated by the model. However, it is supported by the predictive power of perceived reemployment rates for the true reemployment probability and of the perceived arrival rate of job offers for its true counterpart (see Tables~\ref{tab:determinants_arrival} and~\ref{tab:predictive}), which show that higher subjective employability correlates with higher realized arrival rates of offers and reemployment.}

\paragraph{Selection of empirical moments and parameters estimation.}

We estimate the model parameters using the Generalized Method of Moments (GMM). Table~\ref{tab:moments_matching} summarizes the empirical target moments and their model-generated counterparts. The four moments based on realized reemployment rates identify the true employability parameters, the share of low-type workers, $\phi$, and the true depreciation rate. The cost-of-effort parameters are identified from the level and dispersion of search effort (measured in hours per week), conditional on perceived employability. The unemployment benefit schedule is identified through the reemployment bias and the wage acceptance rate. A higher benefit level increases the value of unemployment, thereby reducing the wage acceptance rate and affecting the reemployment bias. Overall, we estimate seven parameters using eight moments, implying that the model is overidentified.

\medskip

Imposing the joint distribution of job seekers' beliefs as an input to the model reduces the number of free parameters but also tightens the constraints on the fit of certain moments. Specifically, once perceived employability, wage perception biases, search effort, and the wage acceptance rate are conditioned, the perceived reemployment probabilities are fully determined by the model structure. Some of these objects are directly targeted as moments, while others are imposed as inputs. No free parameter thus provides sufficient flexibility to match average reemployment bias and wage acceptance rate simultaneously.

\medskip

To prevent the estimation from being distorted by moments that the model cannot freely target, and whose misfit reflects a structural limitation rather than parameter misspecification, we assign differential weights in the GMM objective function.\footnote{In practice, we assign a weight of 10 to the moments on search effort and reemployment rates, and a weight of 1 to the two remaining moments. This weighting scheme places substantially greater emphasis on the former moments while still retaining the latter in the estimation, thereby providing a weak form of regularization that prevents the model from drifting too far along these dimensions.} Table~\ref{tab:moments_matching} reports the fit of all eight moments. The results show that the model closely matches the moments related to conditional three-month reemployment rates, conditional levels of search effort, and average reemployment biases, whereas the moment on the average acceptance rate is poorly fitted.

\medskip

Table~\ref{tab:model_parameters} reports point estimates and confidence intervals for all model parameters. A notable result is that the estimated true depreciation rate reaches its upper bound of one, indicating that, within the set of economically admissible parameter values, the model relies entirely on dynamic selection, rather than genuine employability depreciation, to replicate the observed decline in reemployment rates over the unemployment spell. While the literature generally finds evidence for both mechanisms \citep[see e.g.][]{mueller2021job}, our model does not require true depreciation to fit the data: the quality of the moment fit reported in Table~\ref{tab:moments_matching} suggests that dynamic selection alone is sufficient to account for duration dependence in our sample. We interpret this result cautiously, as the estimator is constrained at the boundary of the parameter space, and do not rule out a role for true depreciation that our identification strategy may not separately recover.

\medskip

Two distinct selection mechanisms operate across the two perceived employability groups. The estimated share of low-type workers is 17.5\%, implying that both types coexist within the below-median perceived employability group. Because high-type workers exit unemployment more quickly, the composition of this group gradually shifts toward low-type workers over time, mechanically generating declining reemployment rates as unemployment duration increases. In the high perceived employability group, by contrast, all job seekers are assigned to the high true type, so selection on true employability is absent. Instead, selection arises from heterogeneity in perception biases. More optimistic job seekers set higher reservation wages, which lowers their reemployment probability and causes them to become increasingly overrepresented among the long-term unemployed. This mechanism also generates declining reemployment rates as unemployment duration increases.

\medskip

\paragraph{Depreciation versus selection: identification.} Because genuine depreciation and the two selection channels just described all produce reemployment rates that decline with unemployment duration, they are only weakly separated by our moments, which is what drives the estimate of $\theta$ to its boundary. The forces nonetheless leave distinct signatures. True depreciation scales down the arrival rate of \emph{every} surviving job seeker by the same monthly factor $\theta$, independently of type or beliefs, whereas selection operates through the changing \emph{composition} of the surviving pool, its mix of true employability types in the low perceived-employability group, and its distribution of perception biases in the high perceived-employability group. These composition shifts are disciplined not only by the reemployment-rate moments but also by the conditional search-effort and average-bias moments, which the model must reproduce jointly. In principle, $\theta$ is therefore identified by the portion of the duration decline that the estimated type gap ($\lambda^{0,l}$ versus $\lambda^{0,h}$, with population share $\phi$) and the perception-bias distribution cannot absorb. In our sample this residual is small: the flexibility of the two selection channels is enough to match the observed decline, so the moments are only weakly informative about $\theta$ and the estimator settles at the no-depreciation boundary. We view the boundary estimate as evidence that duration dependence in our data can be accounted for by selection, not as evidence against true depreciation, which our design is not well powered to detect separately.

\medskip

The estimated model illustrates how reemployment biases serve as concise and informative signals of underlying informational biases and search behavior. Our model provides valuable insights into the information content of reemployment biases, which are particularly convenient to measure as they require only a simple elicitation of job-finding expectations.

\medskip

Figure~\ref{fig:pepsi} plots, for each individual in our estimation sample, the bias in the location parameter of the wage distribution ($\widetilde{\mu}_i - \mu_i$) against the bias in perceived employability ($\widetilde{\lambda}^0_i - \lambda^0_i$), with point colors indicating the magnitude of the simulated reemployment bias. The figure reproduces the stylized facts documented in our descriptive analysis (Figure~\ref{fig:correlation_wage_arrival_biases}): reemployment biases are strongly correlated with both informational biases, confirming that they can be used to infer the underlying informational distortions.
\medskip

Splitting the analysis by true employability type reveals that the two groups are governed by distinct mechanisms. For low-type job seekers, reemployment biases are driven almost entirely by perceived employability bias. Because perceived employability is very low for this group, the optimal reservation wage lies below the minimum wage, implying a 100\% acceptance rate regardless of the wage perception bias. As a result, wage perception biases play virtually no role in determining reemployment outcomes.

\medskip

For high-type job seekers, by contrast, both informational biases matter. Wage perception biases become particularly influential when perceived employability is sufficiently high for the reservation wage to exceed the minimum wage. In that case, beliefs about the wage offer distribution directly affect wage acceptance decisions and, consequently, reemployment outcomes.

\medskip

We further examine whether reemployment biases are informative about search behavior. Table~\ref{tab:simulate_reemployment_bias_group} reports average informational biases and search outcomes across quartiles of the simulated reemployment bias distribution. Job seekers in higher reemployment bias quartiles tend to overestimate both their own employability and the median of the wage offer distribution. Search effort and reservation wages also increase monotonically across these quartiles. This pattern is consistent with the empirical relationships between informational biases and search behavior documented in Section~\ref{sec:information}: both search intensity and reservation wages are positively associated with more optimistic beliefs about employability and the level of the wage offer distribution.

\begin{table}[!htbp]
\centering
\caption{Average simulated outcomes by quartile of reemployment bias}
\label{tab:simulate_reemployment_bias_group}

\begin{adjustbox}{width=1\textwidth}
\begin{threeparttable}

\begin{tabular}{lcccccccc}
\hline
\textbf{Bias quartile}
& $\Pr(\text{high type})=1-\phi$
& 3-month bias
& $\mathbb{P}(T<3)$
& $\psi_3$
& $\widetilde{\mu}_i-\mu_i$
& $\widetilde{\lambda}_{0,i}-\lambda_{0,i}$
& $w^*$
& $e^*$ \\
\hline

Quartile 1 & 0.762 & -0.090 & 0.399 & 0.309 & -0.064 & -0.002 & 1.002 & 12.948 \\
Quartile 2 & 0.377 & 0.044 & 0.234 & 0.279 & -0.103 & 0.002 & 1.009 & 13.514 \\
Quartile 3 & 0.655 & 0.200 & 0.353 & 0.553 & -0.095 & 0.010 & 1.052 & 15.693 \\
Quartile 4 & 1.000 & 0.521 & 0.348 & 0.869 & 0.075 & 0.034 & 1.306 & 16.567 \\

\hline
\end{tabular}

\begin{tablenotes}[flushleft]
\small
\item \textit{Notes:} The simulated sample of job seekers is partitioned into quartiles according to the individual 3-month reemployment bias, defined as $\psi_3-\indic\{T<3\}$. For each quartile, the table reports average realized reemployment outcomes, beliefs, and search behavior. The variable $\psi_3$ denotes the perceived probability of finding a job within three months, while $\mathbb{P}(T<3)$ denotes the realized probability of reemployment within three months. The terms $\widetilde{\mu}_i-\mu_i$ and $\widetilde{\lambda}_{0,i}-\lambda_{0,i}$ measure belief errors regarding wage offers and job-offer arrival rates, respectively. The quantity $1-\phi$ denotes the share of high types in the quartile. Finally, $w^*$ denotes the average reservation wage relative to the minimum wage and $e^*$ denotes average search effort.
\end{tablenotes}

\end{threeparttable}
\end{adjustbox}
\end{table}

\FloatBarrier

\subsection{Counterfactual analysis}\label{ssec:counterfactual}

\paragraph{Average causal effects.} Table~\ref{tab:avg_counterfactual} presents the results of our counterfactual analysis, where we examine the average causal effects of each type of informational bias. We simulate four scenarios that differ in which informational biases are  corrected. The first scenario (A) corresponds to the calibrated model with both types of biases. We then consider two scenarios (B) and (C) in which we reduce only one informational bias by 50\%, either with respect to perceived employability ($\widetilde{\lambda}^0_i - 
\lambda^0_i$) or the median of the wage distribution ($\widetilde{\mu}_i - \mu_i$).\footnote{The latter only considers a reduction of the bias regarding the location parameter of the wage distribution, not its dispersion.} Finally, in scenario (D) we reduce 
both types of informational biases simultaneously. For each scenario, we report simulated search behaviors and 3-month reemployment probabilities for our estimation sample, evaluated at the start of the unemployment spell.

\medskip

The two types of informational bias have distinct effects on search behavior and reemployment. On the one hand, job seekers on average overestimate their employability. Correcting this bias leads them to lower their reservation wage and accept a larger share of the wage distribution, which increases their reemployment probability (+2.1~pp). On the other hand, the bias on the wage distribution is small on average. Correcting it leads to a small decrease in the  average reservation wage, slightly increasing the average reemployment rate by 0.7~pp. Although both average effects are modest in magnitude, they conceal substantial heterogeneity across true unobserved employability types and initial bias levels, which we examine in the next section.

\paragraph{Conditional causal effects.}

Tables~\ref{tab:conditional_counterfactual}  and \ref{tab:conditional_counterfactual_100}  present counterfactual outcomes across the four scenarios (A, B, C, D) for groups of job seekers defined by their baseline reemployment bias quartile and their true employability type, respectively reducing informational bias by 50\% or completely. First, we focus on Table~\ref{tab:conditional_counterfactual}. Reemployment bias quartiles are computed from the simulated baseline scenario~(A). The results reveal substantial heterogeneity in the effects of informational bias corrections across these groups. 

\medskip

Among low-type job seekers in the bottom quartile of the reemployment bias distribution, the dominant mechanism operates through the employability bias. In the baseline, these individuals have very low perceived employability, which leads them to exert little search effort. Partially correcting their employability bias raises their perceived employability, making search effort more profitable and less costly, thereby substantially increasing their effort level. Since their perceived employability remains low even after partial correction, their wage acceptance rate stays at 100\%, so the full benefit of the correction translates into higher reemployment probability (+1.3~pp).

\medskip

For high-type job seekers in this quartile, the reemployment bias mostly reflects pessimism regarding perceived employability. However, partially correcting this bias leads to only small adjustments in search effort, as the influence of the intervention on their perceptions is relatively limited. Wage biases also appear to have limited influence on these job seekers, as their perceived employability is too low for their reservation wages to bind above the minimum wage. Overall, the effect on reemployment outcomes for these individuals is small but positive (+ 0.2pp).

\medskip

Among low-type job seekers in the second and third quartiles, the main source of bias is optimism about their own employability. Partially correcting this bias lowers their perceived employability, which reduces the perceived return to search effort, leading to a 
decline in effort of up to $-$0.7 hours per week in quartile 2. This decline may be interpreted as a demotivation effect. Consistent with this mechanism, \cite{harmon2026job} find that informing optimistic job seekers of their high risk of long-term unemployment discourages active job search for a subset of them, those with the most over-optimistic beliefs, leading to transitions into passive support rather than employment. Such an intervention has no effect on reservation wages since, once again, these individuals have too low a perceived employability for their reservation wages to bind above the minimum wage. The overall impact of such a bias correction on these individuals is thus counterproductive, as the demotivation effect is not offset by any decrease in selectivity. The reemployment probability decreases by 0.04pp in the 2nd quartile and by 0.02pp in the 3rd quartile. 

\medskip

Among high-type job seekers in the highest bias quartile, both biases operate in the same direction. Correcting them leads to a substantial increase in reemployment probability of up to 
+8.9~pp, driven by a downward revision of reservation wages. In this group, the demotivation effect induced by a downward revision of perceived employability ($-$0.4 hours per week in quartile 4) is more than offset by a large decrease in selectivity (+19~pp in the acceptance rate), leading to a substantial improvement in reemployment probability, albeit at the potential cost of a lower expected reemployment wage. Our model predictions are consistent with \cite{altmann2025}, who show empirically that correcting optimistic wage beliefs increases reemployment probabilities, partly through a decline in reservation wages.

\medskip

Table~\ref{tab:conditional_counterfactual_100} presents counterfactual scenarios in which informational biases are fully corrected. The effects on search behaviors and reemployment outcomes are amplified relative to the partial correction scenarios. For low-type pessimists, correcting their underestimation of employability raises their 3-month reemployment probability by 2.1~pp, driven by a 55\% increase in search effort (from 7 to 10.9 hours per week). The demotivation effect for low-type job seekers in bias quartiles 2 and 3 is also of larger magnitude: their search effort decreases by 2.8 and 3.8 hours per week, respectively, resulting in a substantial decline in their 3-month reemployment probability ($-$1.5~pp and $-$2.2~pp). Finally, for high-type job seekers in quartiles 3 and 4, fully correcting their optimistic beliefs improves the probability of reemployment at 3-months by 1.6~pp and 12.5~pp respectively, driven by a large decline in wage selectivity (+8.4~pp and +32.7~pp in the acceptance rate) that more than offsets the demotivation effect ($-$0.8 and $-$1.3 hours of search effort respectively).

\medskip

The counterfactual analysis points to a heterogeneous influence of bias correction, depending on job seekers' initial beliefs and employability. This result advocates for targeted informational support, since providing the same objective information to all job seekers may not be an optimal policy: while pessimistic job seekers stand to gain from being reassured about their labor market prospects, optimistic job seekers, in contrast, may see negative feedback about their own employability undermine their motivation and search effort. Our model also confirms that heterogeneity in reemployment biases can serve as an efficient signal to predict heterogeneity in informational biases and, consequently, search behavior. In the following section, we provide new econometric tools that allow us to empirically and nonparametrically characterize the heterogeneity in reemployment biases and to form groups of job seekers that differ with respect to their initial biases and accordingly have distinct intervention needs. This procedure also allows us to observe additional differences within groups that do not naturally appear in the model.

\section{Stratifying support using reemployment-belief biases}\label{sec:heterogeneity}

We develop a data-driven procedure to predict reemployment-belief biases as flexibly as possible using machine learning, and then to summarize the resulting heterogeneity through an efficient stratification rule. Our analysis suggests that this stratification is informative about the type of support job seekers may need (for example, targeted information provision).

\subsection{Methodology}

Our goal is to identify groups of job seekers with different reemployment-belief biases without directly using measures of biases in labor-market perceptions or search behavior.

\subsubsection{Characterizing heterogeneity in biases using best linear predictors}

A key difficulty is that, for each job seeker \(i\), we observe only one realization of unemployment duration, \(T_i\), rather than the individual-specific job-finding distribution $ P(T_i \le t \mid \mathcal I_i)$. 
Hence, we cannot directly compare the elicited subjective probability $\widetilde P(T_i \le t \mid \mathcal I_i)$ to the corresponding objective probability for the same individual. The ideal object would be the individual bias
\[
\varepsilon_{t,i}^* := \widetilde P(T_i \le t \mid \mathcal I_i) - P(T_i \le t \mid \mathcal I_i),
\]
but instead we observe only
\[
\varepsilon_{t,i} := \widetilde P(T_i \le t \mid \mathcal I_i) - \mathbf 1\{T_i \le t\}.
\]

Still, \(\varepsilon_{t,i}\) is an unbiased signal of \(\varepsilon_{t,i}^*\). Indeed, for every vector of covariates \(X_i\) measurable with respect to the information set \(\mathcal I_i\),\footnote{More precisely, \(\sigma(X_i)\subset \mathcal I_i\), where \(\sigma(X_i)\) denotes the \(\sigma\)-algebra generated by \(X_i\).} $\mathbb E[\varepsilon_{t,i}\mid X_i] = \mathbb E[\varepsilon_{t,i}^*\mid X_i]$.
Therefore, rational expectations can be tested either unconditionally, using $\mathbb E[\varepsilon_{t,i}^*]=0$, using the observable counterpart \(\mathbb E[\varepsilon_{t,i}]=0\), or conditionally on selected characteristics, as in Section \ref{ssec:mbias}; see also \citet{spinnewijn2015unemployed,mueller2021job}. However, such conditional analyses require choosing groups ex ante, which may be restrictive and may leave substantial heterogeneity unexplored.

\medskip

To use the full set of covariates \(X_i\) as flexibly as possible, we rely on the Generic ML approach of \citet{chernozhukov2018generic}. For a given horizon \(t\), our target is the conditional mean bias
\begin{equation}\label{eq:error}
\Delta_t(x) := \mathbb E[\varepsilon_t^* \mid X=x].
\end{equation}
A positive value of \(\Delta_t(x)\) corresponds to over-optimism.

\medskip

Because \(X\) may be high-dimensional, we first construct an ML predictor \(\delta_t(X)\) of the bias signal. We then summarize the information contained in this predictor through the best linear predictor of \(\Delta_t(X)\) onto the span of \(1\) and \(\delta_t(X)\):
\[
BLP(\Delta_t(X)\mid \delta_t(X))
:=
\arg\min_{f\in \mathrm{span}(1,\delta_t(X))}
\mathbb E\!\left[\big(\Delta_t(X)-f(X)\big)^2\right].
\]
Letting
\[
\Delta_t(X)=\beta_1+\beta_2 \delta_t(X)+u_t,
\qquad
\mathbb E[u_t]=\mathbb E[u_t\delta_t(X)]=0,
\]
the coefficients are given by
\[
\beta_2=\frac{\mathrm{Cov}(\Delta_t(X),\delta_t(X))}{\mathrm{Var}(\delta_t(X))},
\qquad
\beta_1=\mathbb E[\Delta_t(X)]-\beta_2\mathbb E[\delta_t(X)].
\]
In particular, \(\beta_1\) captures the average reemployment bias in the population, so testing \(\beta_1=0\) amounts to testing average rationality. The coefficient \(\beta_2\) measures how informative the ML score \(\delta_t(X)\) is about the true conditional bias: values close to one indicate that the shape of \(\delta_t(X)\) closely tracks the shape of \(\Delta_t(X)\).

\medskip

Estimation proceeds by sample splitting. In one subsample, we train the ML predictor \(\delta_t\); in the other, we regress the observable signal \(\varepsilon_t\) on \(\delta_t(X)\). As emphasized by \citet{chernozhukov2018generic}, inference must account for the additional randomness generated by sample splitting. We therefore repeat the splitting procedure many times and report median-aggregated \(p\)-values and quantile-aggregated confidence intervals across splits.

\medskip

\subsubsection{Optimal stratification to maximize between-group heterogeneity}\label{ssec:select}

To simplify notation, we now suppress the horizon index \(t\), and write \(\delta(X)\) for the ML score and \(\varepsilon^*\) for the latent bias. Our goal is to summarize the function $B(d):=\mathbb E[\varepsilon^*\mid \delta(X)=d]$ through a partition of the support of \(\delta(X)\) into \(K\) groups.

\medskip

Let \(\mathcal G_K=\{G_1,\dots,G_K\}\) denote a partition of the support of \(\delta(X)\). We define the average bias in group \(G_k\) as $
\gamma(G_k):=\mathbb E[\varepsilon^* \mid \delta(X)\in G_k]$. We seek the partition that maximizes between-group heterogeneity:
\begin{align}
\max_{\{G_k\}\in \mathcal G_K}\;\overline{\Lambda}(\{G_k\})
&:= \sum_{k=1}^K P(\delta(X)\in G_k)\,\gamma(G_k)^2.
\label{eq:maxhet}
\end{align}
Because \(\varepsilon^*\) is not observed, we solve this problem in two steps.

\paragraph{Step 1: discretization.}
We first discretize the support of \(\delta(X)\) into a fine partition $
\mathcal I^M=\{I_{1,M},\dots,I_{M,M}\}$,
where each \(I_{j,M}=[\ell_{j-1},\ell_j]\) is a quantile bin of \(\delta(X)\), so that the bins have equal probability mass. Let \(\mathcal G_{K,M}\) denote the set of all partitions into \(K\) groups formed by unions of adjacent bins from \(\mathcal I^M\). The discretized version of \eqref{eq:maxhet} is
\begin{align}
\max_{\{G_k\}\in \mathcal G_{K,M}}\;\overline{\Lambda}(\{G_k\})
&:= \sum_{k=1}^K \gamma(G_k)^2\,P(\delta(X)\in G_k).
\label{eq:maxhet_M}
\end{align}

\paragraph{Step 2: estimation and implementation.}
For each fine bin \(I_{j,M}\), we estimate the group average $
\gamma(I_{j,M})=\mathbb E[\varepsilon^*\mid \delta(X)\in I_{j,M}]$
using the observable signal
\[
\widetilde{\gamma}(I_{j,M})
:=
\frac{1}{|\{i:\delta(X_i)\in I_{j,M}\}|}
\sum_{i:\delta(X_i)\in I_{j,M}} \varepsilon_i,
\]
where $|\{i:\delta(X_i)\in I_{j,M}\}|$ denotes the number of individuals with $\delta(X_i)$ falling into $I_{j,M}$. Under our monotonicity assumption on \(B\), the sequence \(j\mapsto \gamma(I_{j,M})\) is nondecreasing. We therefore use the increasing rearrangement of \(\{\widetilde{\gamma}(I_{j,M})\}_{j=1}^M\), denoted \(\{\widehat{\gamma}(I_{j,M})\}_{j=1}^M\), as in \citet{chernozhukov2009improving}. Each coarse group \(G_k\in\mathcal G_{K,M}\) is a union of consecutive fine bins. Since the quantile bins have equal probability mass,
\[
\gamma(G_k)
=
\frac{1}{|\{j:I_{j,M}\subset G_k\}|}
\sum_{j:I_{j,M}\subset G_k}\gamma(I_{j,M}),
\]
and we estimate this quantity using the plug-in estimator:
$\widehat{\overline{\gamma}}(G_k)
= \sum_{j:I_{j,M}\subset G_k}\widehat{\gamma}(I_{j,M})/|\{j:I_{j,M}\subset G_k\}|$. The empirical analogue of \eqref{eq:maxhet_M} is then
\begin{align}
\max_{\{G_k\}\in \mathcal G_{K,M}}\;
\widehat{\overline{\Lambda}}(\{G_k\})
&:=
\sum_{k=1}^K
\widehat{\overline{\gamma}}(G_k)^2\,
|\{j:I_{j,M}\subset G_k\}|.
\label{eq:gKM}
\end{align}

This maximization problem is equivalent to minimizing within-group heterogeneity:
\begin{equation}\label{eq:kmeans}
\min_{\{G_k\}\in \mathcal G_{K,M}}
\sum_{k=1}^K\sum_{j:I_{j,M}\subset G_k}
\Big(\widehat{\gamma}(I_{j,M})-\widehat{\overline{\gamma}}(G_k)\Big)^2.
\end{equation}
Thus, the problem reduces to one-dimensional \(k\)-means clustering with an order constraint. In one dimension, this problem admits fast polynomial-time solutions; see, for example, \citet{wu1991optimal}. Let \(\{G_{k,M}^*\}_{k=1}^K\) denote the solution to \eqref{eq:gKM}, and define the resulting step-function estimator
\[
\widehat B_{n,M}(d)
:=
\sum_{k=1}^K
\widehat{\overline{\gamma}}(G_{k,M}^*)\,
\mathbf 1\{d\in G_{k,M}^*\}.
\]
Proposition \ref{prop:k_means_lit} shows that \(\widehat B_{n,M}\) converges to an optimal \(K\)-step approximation of \(B\).

\begin{prop}\label{prop:k_means_lit}
Let \(D=\delta(X)\), and define \(B(d)=\E[\varepsilon^*\mid D=d]\).
Assume: (i) \(B\) is càdlàg and nondecreasing on the support of \(D\); (ii) \(\E[\varepsilon^2]<\infty\); (iii) the support of \(D\) is a compact interval and admits a density bounded away from zero and infinity; (iv) the partition \(\mathcal I^M\) consists of equal-probability (quantile) bins; (v) \(M\to\infty\) and \(M^2/n\to 0\); (vi) the best \(K\)-step \(L^2(P_D)\)-approximation of \(B\) is unique.\footnote{This is a generic condition rather than a substantive restriction. Primitive sufficient conditions can be stated on the law of $Y=B(D)$. \citet{mease2006unique} give conditions
for uniqueness of the optimal $K$-partition of a distribution.}
\medskip

Let \(\widehat B_{n,M}\) be defined by \eqref{eq:kmeans}. Then
\[
\widehat B_{n,M} \to B_K^*
\quad \text{in } L^2(P_D),
\]
where \(B_K^*\) is a solution to \eqref{eq:maxhet}, i.e. a best \(K\)-step approximation of \(B\).
\end{prop}

The number of groups \(K\) can be selected using a criterion analogous to the standard \(k\)-means choice of \(K\). The different steps of the procedure are illustrated in Figure \ref{fig:steps_xgb}. Figure \ref{fig:steps_a_xgb} displays the distribution of the predicted score \(\delta(X)\) in one sample split. Figure \ref{fig:steps_b_xgb} shows the estimated average biases \(\widetilde{\gamma}(I_{j,M})\) on the quantile grid. Figure \ref{fig:steps_c_xgb} displays the resulting \(K=4\) groups after applying ordered \(k\)-means to the rearranged values \(\widehat{\gamma}(I_{j,M})\). Finally, Figure \ref{fig:steps_d_xgb} reports the estimated average bias \(\widehat{\overline{\gamma}}(G_{k,M}^*)\) in each group.

\medskip

As in \citet{chernozhukov2018generic}, the quantities of interest are group averages indexed by the ML score \(\delta(X)\). We compare these group-level characteristics using median-aggregated \(p\)-values (Definition 4.2 in \citet{chernozhukov2018generic}) and quantile-aggregated confidence intervals (Definition 4.3 therein). In practice, we use 100 sample splits.

\subsection{Results of the classification}\label{sec:results_classif}

\paragraph{Selection of the ML method.} Table \ref{tab:ML_comparison} presents two performance measures used to select the ML predictor of the 3 months reemployment bias: one based on the correlation between the predictor $\delta(X)$ and the true conditional bias $\mathbb{E}(\varepsilon^*|X)$, namely $|\widehat{\beta}_2|^2\text{Var}(\delta(X))$, and the other based on the part of the variation in $\mathbb{E}(\varepsilon^*|X)$  explained by the $K$ groups \eqref{eq:maxhet}. To ensure comparability between methods for this decision, we use quartiles of $\delta(X)$ for the partition $\mathcal{G}_K$ instead of the optimization method described in section \ref{ssec:select}.  

\medskip

Table \ref{tab:ML_comparison} shows that out of the four predictors, three perform similarly in terms of maximizing the heterogeneity between groups and in terms of the measure associated with the BLP: glm net, Random Forest and XGBoost. Therefore, we present the classification analysis for both the XGBoost and the Random Forest predictors in order to test how sensitive our results are to the ML predictor used. We find the main results of the analysis to be very comparable between the two, reinforcing their robustness. We note that the estimated coefficient $\widehat{\beta}_2$, which is the heterogeneity loading in the best linear predictor (BLP) of the true conditional bias on our proxy,  takes values close to 1 across all methods, indicating that the proxy is well calibrated and that it neither materially attenuates nor inflates the underlying heterogeneity in the biases.

\paragraph{Heterogeneity across groups with respect to actual job-finding rates and expectations.}

\begin{figure}[!htbp]
    \centering
    \caption{Perceived and realized reemployment probabilities by reemployment-bias group (XGBoost)}
    \label{fig:GLM_bias_grps_horizon_xgb}

    \includegraphics[width=\textwidth]{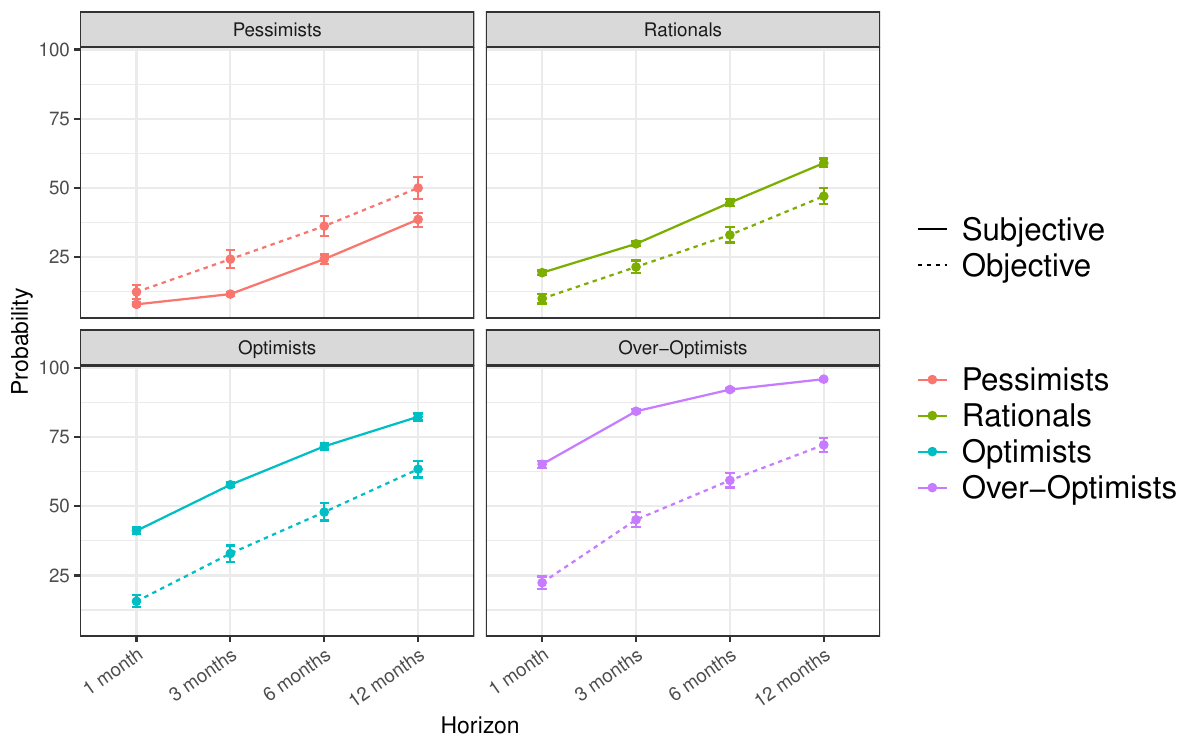}

    \caption*{\footnotesize
    \textit{Notes:} This figure compares average perceived and realized reemployment probabilities across 3-month reemployment-bias groups at different time horizons. Realized reemployment probabilities are computed as the share of individuals who are reemployed by each horizon following the survey interview. Reemployment-bias group membership is estimated using the XGBoost classification procedure and is recomputed across 100 sample splits. Group-specific means and confidence intervals are estimated separately for each split, and the reported values correspond to the median estimates across splits. The sample is restricted to individuals whose reported reemployment beliefs satisfy the consistency restrictions. Sampling weights are used throughout.}
\end{figure}

Using the elbow method, we select $K=4$ groups. We label each group based on the average 3-month reemployment (see  Figure \ref{fig:GLM_bias_grps_horizon_xgb}). The first group is the only one with a negative bias (-13pp for the XGBoost predictor and -10pp with the Random Forest), meaning that individuals in this group underestimate their reemployment probability on average. We call this group pessimists. The second group has the smallest average bias in absolute value (+8pp for the XGBoost predictor and +9.8pp with the Random Forest), so we label this group as rational. Groups 3 and 4 have both very high and positive average 3-month reemployment biases (+25pp and +39pp with the XGBoost). We label these two groups as optimists and over-optimists, respectively. In the following, we refer to these groups by their labels. The estimated shares are respectively 16.3\%, 27.4\%, 23.8\% and 32.7\% when using the XGBoost algorithm and 20.3\%, 22.7\%, 29.8\%, and 27.3\% when using the Random Forest. Our classification shows how heterogeneous the reemployment biases are. Although the average bias for the entire sample is strongly positive for all horizons (see Figure \ref{fig:avg_bias_by_panel}), our procedure allows us to identify a significantly large group for which the reemployment bias is negative.

\medskip

 Figure \ref{fig:GLM_bias_grps_horizon_xgb} shows the expected and actual job-finding rates for each of the four bias groups formed and for four different time horizons (1, 3, 6, and 12 months). These plots confirm the strong differences in job search biases between the four groups. In addition to the difference between the actual and expected job-finding rate, the four groups also differ in their actual job-finding rate. At the 6-month horizon following the survey, 36.2\% of the job seekers in the pessimistic group have found a job, compared to 33.0\% in the rational group, 47.9\% in the optimistic group, and 59.3\% in the over-optimistic group.\footnote{Using the Random Forest predictor, we find similar 6-month reemployment rates in each group: 34\%, 35\%, 50\% and 58\% respectively, see Figure \ref{fig:GLM_bias_grps_horizon}.}

 \medskip
 
 This evidence may indicate that job seekers' expectations are well ranked among individuals, but that they systematically overestimate the magnitude of these differences.

\subsection{Heterogeneity in groups' characteristics and behavior}\label{ssec:res1}

Job-finding biases are strongly heterogeneous among job seekers to the extent that we can build groups of individuals with different average biases. We now investigate the composition of these groups in order to characterize this heterogeneity. We leverage our rich set of data to compare groups along demographics, past unemployment history, and psychology. Importantly, we also compare them with respect to beliefs regarding the labor market and search behaviors, which have not been used to create the group.\footnote{This is referred to as CLAN in \cite{chernozhukov2018generic}.}

\subsubsection{On the variables used for the classification}

\paragraph{Demographic composition by bias groups.}
Table \ref{tab:bias_group_stat_desc_xgb} shows that our four bias groups have different demographics. One of the main differences regards job seekers' age. Pessimistic job seekers are 41.6 years old on average, which is 2.8 years older than the optimists and 3.8 years older than the over-optimists. They also tend to be less qualified, as they are slightly less educated (34.8\% with college education against 39.5\% for the rationals, 40.8\% for the optimists and 46.3\% for the over-optimists). Finally, more optimistic job seekers are less likely to be women since the share of women is 54.6\% in the pessimistic group against 47.6\% in the over-optimistic group. We find no significant differences in group composition when it comes to family-related variables. These demographic differences remain similar when using the Random Forest predictor to classify job seekers into bias groups (see Table \ref{tab:bias_group_stat_desc}).

\paragraph{Unemployment status by bias groups.}
Table \ref{tab:bias_group_stat_desc_xgb} also depicts strong heterogeneity in unemployment status across the four bias groups. Optimism is strongly correlated with shorter current unemployment spells and higher remaining benefits duration: 30\% of job seekers are no longer eligible to receive unemployment benefits in the optimistic group whereas this proportion is 43.5\% in the pessimistic group. Pessimistic job seekers also have a significantly higher cumulative unemployment duration, accumulating around 24 more months of unemployment over their lifetime compared with individuals in the rational group. These results remain similar when considering the Random Forest predictor as shown in Table \ref{tab:bias_group_stat_desc}.

\subsubsection{On the variables not used for the classification}

\paragraph{Psychological traits by bias groups.}

\begin{figure}[htbp]
\caption{Average psychological traits among bias groups}
\label{fig:gml_psycho}
     \centering
    \begin{subfigure}{0.48\linewidth}
        \centering
        \includegraphics[width=\linewidth]{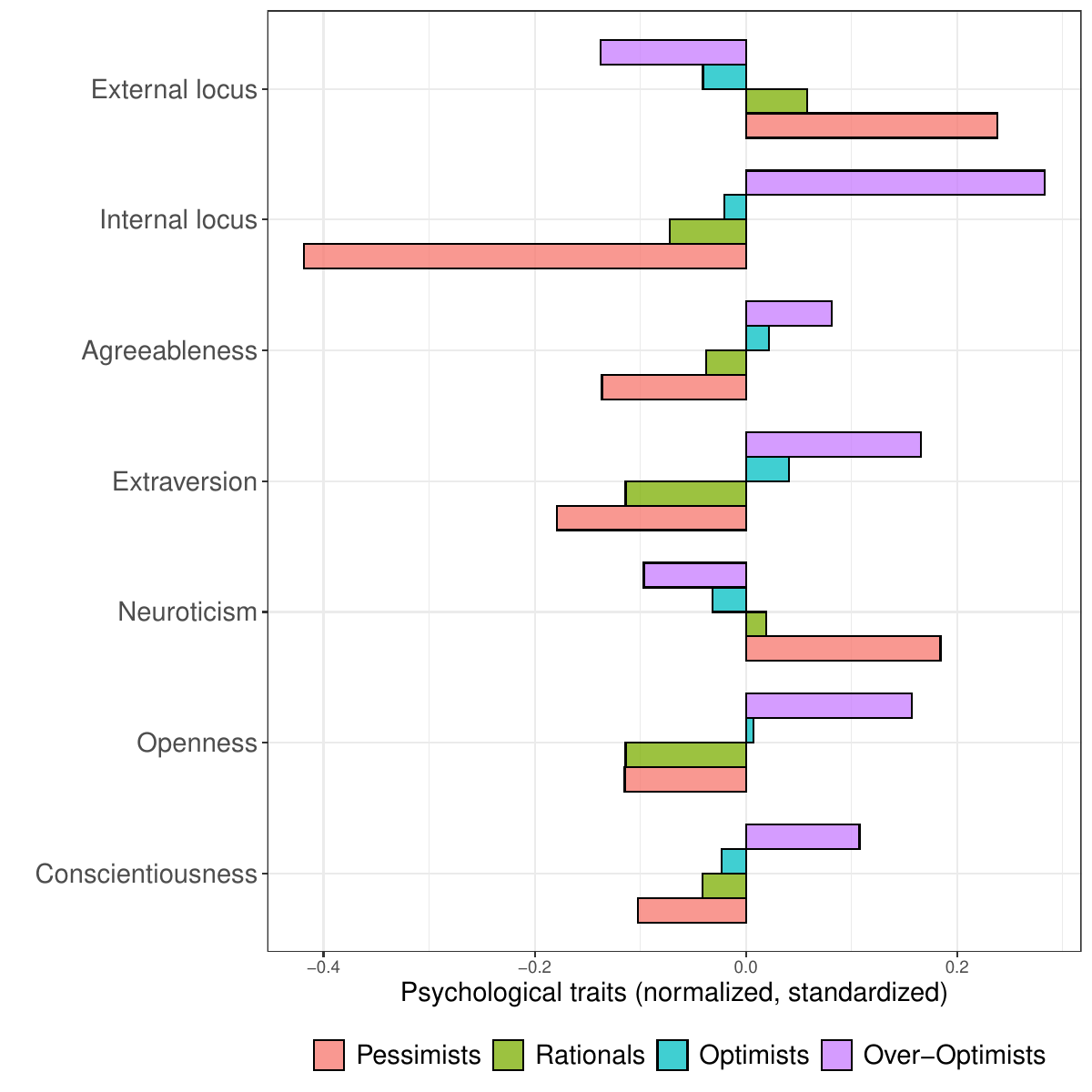}
        \caption{XGBoost}
        \label{fig:gml_psycho_xgb}
    \end{subfigure}
\hfill  
    \begin{subfigure}{0.48\linewidth}
        \centering
        \includegraphics[width=\linewidth]{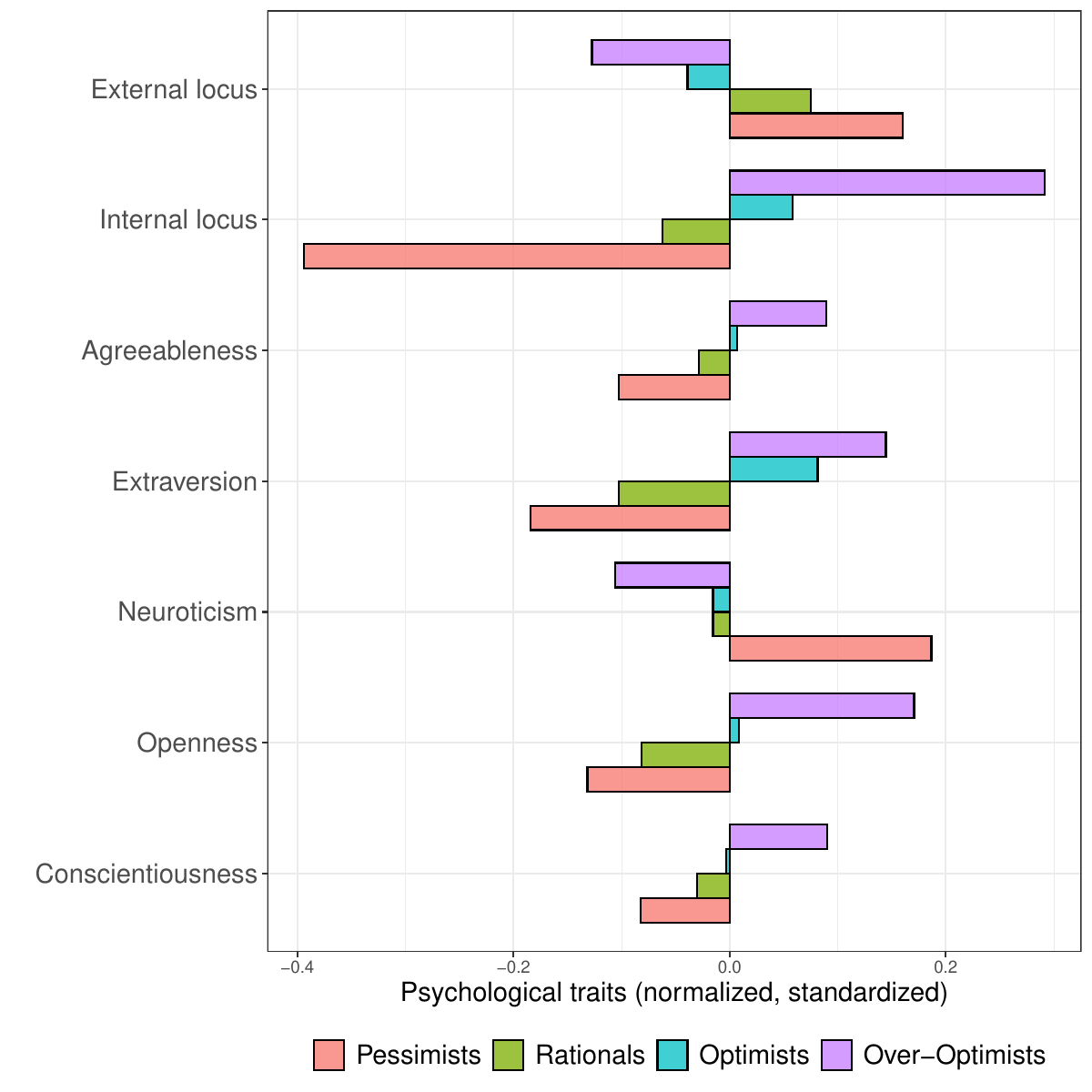}
        \caption{Random Forest}
        \label{fig:gml_psycho_rf}
    \end{subfigure}
\caption*{\footnotesize \textit{Notes: }Psychological traits are normalized and standardized over the whole sample. Reemployment bias group memberships are predicted on 100 different sample splits. In each split, we compute average psychological index for each traits in each group. The median of means over the 100 splits are then reported. XGBoost algorithm is used to predict bias groups in panel (a) and Random Forest is used in panel (b).}
\end{figure}

In addition to standard demographics and unemployment status, we investigate whether job-finding biases are correlated with common psychological traits. Figure \ref{fig:gml_psycho} compares average measures of the Big Five personality traits and locus of control in the four bias groups. Among the Big Five, neuroticism and extraversion are the two main psychological traits that differ between groups. There is a 0.28 standard deviation difference between the average neuroticism measure in the pessimist and the over-optimist group. Neuroticism captures the tendency to experience strong negative emotions such as anger or anxiety. Our measure is based on how individuals see themselves as \textit{anxious}, \textit{irritable}, \textit{composed} and \textit{emotionally stable}. Second, we find extraversion to be correlated with optimism regarding reemployment probabilities as the mean difference between pessimistic and over optimistic groups is 0.35 standard deviations. Finally, locus of control is the most discriminating psychological trait: average internal locus of control index difference rises to 0.7 standard deviations. Job seekers with more pessimistic beliefs tend to feel that they have less influence on their future trajectory. Importantly, locus of control correlates with perceived return to search effort as shown in Table \ref{tab:psycho_perceptions} and with actual search effort intensity \citep{caliendo2015locus}. Overall, these results show that job seekers in different bias groups differ significantly across several psychological dimensions, several of which have themselves been linked to search behavior. The difference in psychological traits across bias groups is preserved when using the Random Forest predictor to classify job seekers (see Figure \ref{fig:gml_psycho_rf}).

\paragraph{Beliefs and informational biases.}

We now examine whether our reemployment bias groups also correlate with beliefs regarding the labor market. Figure \ref{fig:individual_info_biases_and_group} displays the individuals in our sample on the plane generated by the two informational biases that we consider, with point color referring to their bias group, using either XGBoost (a) or Random Forest predictor (b). On the one hand, pessimistic job seekers are more often located at the bottom left corner, meaning that they have, on average, pessimistic beliefs regarding the two main parameters of the labor market. On the other hand, optimist and over-optimist job seekers are located more towards the upper right corner of the graph, meaning that they also hold optimistic beliefs regarding the parameters of the labor market. Thus, the main takeaway is that our stratification replicated the stylized facts described in our descriptive evidence from Figure \ref{fig:correlation_wage_arrival_biases} and our model in Section \ref{sec:pred_outcome}.

\medskip

Focusing now more closely on wage beliefs, Table \ref{tab:informational_biases_by_group_xgb} (Panel A) displays average wage biases by group at three different points of the wage distribution. Importantly, even though wage biases were not used to classify job seekers, they are strongly correlated with the groups formed. Whereas optimistic and over-optimistic individuals have low wage biases at the median of the distribution, pessimistic individuals underestimate the probability that a job offer proposes a wage higher than the median by 21.4pp on average. Figure \ref{fig:true_perceived_cdf_by_group_xgb} illustrates different wage biases by group, when approximating the perceived distribution with a log-normal. Compared to Figure~\ref{fig:fitted_wage_distrib}, which presents the population-average distribution, it reveals substantial heterogeneity across groups. The bias of pessimists lies mainly in a systematic underestimation, particularly at the lower quantiles of the distribution whereas the bias of the optimists mainly concerns the overestimation of the spread of the distribution. 

\medskip

The same holds for biases regarding the arrival rate of job offers. Pessimistic job seekers underestimate the probability of receiving at least three job offers within three months, both when we predict the true probability (Panel B) and when we restrict the sample to individuals who took the survey twice, allowing us to observe the actual number of job offers received (Panel C). Once again, the bias with respect to the arrival rate of job offers monotonically increases across reemployment bias groups, ranging from -3.6pp for the pessimists to 24.7pp for the over-optimists. However, there is a significant difference in the \textit{true} probability of receiving 3 job offers or more in a three-month span (10 pp difference between pessimists and over-optimists) which shows that job seekers' biases may be cardinal rather than ordinal
Table \ref{tab:informational_biases_by_group} mirrors the results of Table \ref{tab:informational_biases_by_group_xgb} when using the Random Forest algorithm to predict bias group memberships.

\medskip

\paragraph{Perceived return to search effort.} Thanks to more specific items in the survey after wave 6, we are able to elicit beliefs about the arrival rate of offers more precisely, conditioning on different levels of search effort. Figure \ref{fig:perceived_return} displays the average perceived arrival rate of offers conditional on 10, 20, and 30 hours of search effort per week. Job seekers in the four bias groups differ substantially in their perception of arrival rate of offers, regardless of the effort level. Interestingly, this graph shows that individuals across reemployment bias groups perceive different returns to search effort: pessimists perceive on average 0.023 additional job offers within 3 months per additional hour of weekly search effort, compared with 0.027 for the rationals and 0.042 for the over-optimists.\footnote{This is using the XGBoost algorithm. These arrival rates are recovered from the elicited effort-conditional probabilities under the Poisson benchmark discussed in Appendix~\ref{app:perceived_arrival_rate}; the same monotone ordering across groups is apparent directly in the raw elicited probabilities and does not rely on this benchmark.} This is consistent with the relationship between the locus of control and the perceived returns to search effort discussed in Section \ref{ssec:psychic_beliefs}: People who are more optimistic with respect to their probability of reemployment have a higher internal locus, which is correlated with a higher perception of the returns to effort.

\begin{table}[!htbp]
\centering
\caption{Reemployment-bias groups and search behaviors (XGBoost)}
\label{tab:behaviors_gml_xgb}

\begin{adjustbox}{width=\textwidth}
\begin{threeparttable}

\begin{tabular}{lcccccc}
\hline
\hline
& Reservation wage & Reservation wage & Search effort & Reservation & Number of applications & Number of applications \\
& (log, administrative) & (log, self-reported) & (hours/week) & mobility & (PES website) & (other occupations) \\
& (1) & (2) & (3) & (4) & (5) & (6) \\
\midrule

Pessimists
& -0.029*** & -0.038*** & -0.084 & 0.711 & 0.788*** & 0.626*** \\
& (0.010) & (0.014) & (0.466) & (0.658) & (0.226) & (0.188) \\

Rationals
& -- & -- & -- & -- & -- & -- \\
& & & & & & \\

Optimists
& -0.001 & -0.020 & 1.797*** & 1.148** & 0.202 & 0.136 \\
& (0.010) & (0.012) & (0.399) & (0.557) & (0.169) & (0.140) \\

Over-optimists
& 0.010 & 0.024* & 1.755*** & 0.838 & -0.185 & -0.178 \\
& (0.010) & (0.013) & (0.386) & (0.537) & (0.148) & (0.121) \\

\midrule

Demographics & \checkmark & \checkmark & \checkmark & \checkmark & \checkmark & \checkmark \\
Target occupation & \checkmark & \checkmark & \checkmark & \checkmark & \checkmark & \checkmark \\
Region fixed effects & \checkmark & \checkmark & \checkmark & \checkmark & \checkmark & \checkmark \\

\midrule

Mean (Rationals) & \euro 2,413 & \euro 2,219 & 12.1 & 23.5 km & 0.95 & 0.846 \\
$p$-value (joint test) & 0.000 & 0.000 & 0.000 & 0.112 & 0.000 & 0.000 \\
Observations & 9,571 & 6,748 & 9,350 & 10,930 & 10,938 & 10,938 \\

\hline
\hline
\end{tabular}

\begin{tablenotes}[flushleft]
\small
\item \textit{Notes:} This table reports the relationship between reemployment-bias groups and search behaviors. Each outcome is regressed on reemployment-bias group indicators, demographic characteristics, target-occupation fixed effects, and region fixed effects, with the Rational group serving as the reference category. Group membership is estimated using the XGBoost classification procedure and is recomputed across 100 sample splits. Regressions are estimated separately for each split, and the reported coefficients, standard errors, sample sizes, and test statistics correspond to the median across splits.

The administrative reservation wage is trimmed at \euro 10,000, while the self-reported reservation wage is winsorized at the 2.5th and 97.5th percentiles. The number of applications is winsorized at the 98th percentile among positive values. The reported $p$-values correspond to joint tests of whether all reemployment-bias group coefficients are equal to zero and are computed separately for each split before taking the median across splits.

The sample is restricted to individuals whose reported reemployment beliefs satisfy the consistency restrictions described in Section~\ref{ssec:panel_data}. Sampling weights are used throughout.

Significance levels: $^{*}p<0.10$, $^{**}p<0.05$, and $^{***}p<0.01$.
\end{tablenotes}

\end{threeparttable}
\end{adjustbox}
\end{table}

\paragraph{On search behaviors.} A consequential point of our stratification for policy recommendations is that, although we do not include search behavior measures to perform the classification, reemployment bias groups are still useful for predicting job seekers' behaviors. Tables \ref{tab:behaviors_gml_xgb} and \ref{tab:behaviors_gml} present estimates from regressions of several search measures on group dummies when using XGBoost and Random Forest predictors, respectively. These regressions test whether, even when controlling for demographic profiles, target occupations and location, the bias groups contain relevant information about job search behaviors. Consistent with beliefs regarding the wage distribution, we find that job seekers' reservation wage is correlated with bias group, the pessimists setting lower wages than the rationals (-3.8\% with the XGBoost predictor and -3.2\% with the Random Forest). Search effort also increases significantly with optimism, with a gap of close to 2 hours per week between pessimists and optimists (around 1.9 hours with XGBoost and 2.2 hours with the Random Forest predictor). Finally, we find pessimistic job seekers to send more online applications both in their target occupation and in other occupations. Pessimistic job seekers thus seem to be less selective in their search, setting lower wages and applying to more job vacancies. In contrast, over-optimists set higher wages, spend significantly more time searching and apply less, meaning that they are more selective in their choice of vacancies. This behavior is also consistent with the avoidance strategies described in \cite{kHoszegi2022fragile} among individuals with fragile beliefs about their own ability. 

\subsection{Takeaway: policy implications}

In light of our model results, the stratification analysis of Section~\ref{sec:heterogeneity} has important policy implications for the design of targeted informational interventions. In the spirit of the Bayesian persuasion and more generally information provision literature policymakers can use knowledge of job seekers' beliefs to tailor the content of informational treatments. Our stratification methodology provides a low-cost tool to distinguish between job seekers using only survey questions about reemployment expectations, without requiring the full elicitation of informational biases. The link between the bias groups and behavioral needs identified in Section~\ref{sec:heterogeneity} suggests using this 
information to design targeted interventions.

\medskip

First, for the pessimists, biases on both wages and employability are pessimistic, and search effort is suboptimal. Providing correct information about employability and the wage distribution would increase reservation wages and search effort, with ambiguous net effects on reemployment duration but likely positive effects on match quality. These individuals would benefit from receiving correct information in cases where self-confidence has instrumental value for motivation \citep[see, \emph{e.g.}][]{benabou2002self}. The strong correlations we find with psychological components (locus of control, neuroticism) suggest that interventions aiming at remotivating these individuals, possibly associated with a psychological help, could restore self-esteem and magnify those effects. 

\medskip

Second, on the other types of job seekers, our results suggest that providing objective information about employability carries a risk of demotivation. As shown in Section~\ref{ssec:counterfactual}, correcting optimistic employability biases reduces reservation wages but also lowers the perceived return to search effort, partially offsetting the gain. This is consistent with frameworks in which individuals hold fragile beliefs that serve a protective function for motivation \citep[see, \emph{e.g.},][]{kHoszegi2022fragile}: negative feedback may undermine self-esteem and reduce search effort in ways that outweigh the benefits of more accurate beliefs.\footnote{\cite{harmon2026job} show that informing job seekers about their risk of long-term unemployment can discourage active job search for a subset of them. In a related but distinct context, \cite{hakimov2023confidence} find that information provision targeting more optimistic students has no positive influence on their educational outcomes.} For this population, interventions encouraging greater search diversity do not carry this risk and appear better suited to their needs. Our data show that rational, optimistic and especially over-optimistic job seekers tend to make fewer applications (Table~\ref{tab:behaviors_gml_xgb}) than the pessimists, suggesting that they have the most to gain from guidance aimed at broadening their search and increasing their number of applications.

\section{Conclusion}\label{sec:conclude}

We develop new measures of multiple dimensions of job seekers' subjective beliefs and expectations and combine them, through a machine-learning methodology, with detailed administrative records on search behavior and reemployment. This unique dataset allows us to establish new facts about the job search process and to turn easily elicited reemployment expectations into an operational tool for targeting support.

\medskip

We first show that reemployment expectations are informative. They predict realized reemployment, are revised differently at short and long horizons, and are strongly associated with biases in beliefs about the two labor market fundamentals, the arrival rate of offers and the wage distribution. These underlying informational biases, together with psychological traits such as the locus of control, are in turn important determinants of search behavior.

\medskip

We then develop a structural job search model that accounts \emph{a priori} for multiple sources of uncertainty about the labor market and one's own employability. Job seekers with pessimistic reemployment expectations tend to be both less employable and more pessimistic about the labor market, and to search less. Correcting their beliefs raises reservation wages and, for the most pessimistic, search effort and reemployment, with ambiguous effects on unemployment duration but positive effects on match quality. Importantly, the effect is not uniform, however: for optimistic job seekers accurate information can lower the perceived return to effort and reduce search, so that its value ultimately depends on underlying unobserved employability.

\medskip

Finally, we develop a machine-learning stratification that recovers these pessimistic, rational, optimistic, and over-optimistic  profiles  from reemployment expectations alone. The resulting groups differ sharply in their beliefs, search behavior, and informational needs, providing public employment services with a simple and inexpensive basis for tailoring interventions.

\medskip

These findings call for more personalized information provision. For pessimistic job seekers, accurate information about their prospects,  especially when combined with support targeting motivation and the locus of control, can raise search effort and improve match quality. For optimistic job seekers, by contrast, uniform information provision carries a risk of demotivation: it may discourage low-employability individuals while benefiting high-employability ones. This heterogeneity is precisely what our stratification  leveraging reemployment expectations is designed to capture, allowing services to target both the content and the recipients of information rather than delivering the same message to everyone.

\medskip

Several limitations point to avenues for future work. First, our structural model makes simplifying assumptions, notably on the functional form of the arrival rate and the independence of wage offers, that could be relaxed to refine the estimated costs of biased beliefs. Second, while we document strong associations between beliefs and behavior, establishing causality remains challenging; randomized information experiments could identify the causal effects of correcting specific biases. Third, our sample is drawn from French job seekers, and the external validity of our findings to labor markets with different unemployment insurance systems is an open question. Fourth, we do not observe how the same individuals respond to changing macroeconomic conditions; the stability of our classification over the business cycle, and transitions across groups, are important directions for future work. Finally, the optimal design of information interventions, namely what to provide, how to frame it, and when to deliver it, warrants further investigation, potentially drawing on the information-design literature.

\medskip

More broadly, our results illustrate how subjective expectations, tracked over time and combined with machine-learning tools, can become part of the standard toolkit of public employment services seeking to personalize their support.

\FloatBarrier
\bibliographystyle{myagsm}
\bibliography{ref.bib}

@article{altmann2025,
  title={Wage Expectations and Job Search},
  author={Altmann, Steffen and  Malte, Robert Mahlstedt and Rattenborg, Jacob and
Sebald, Alexander and  Settele, Sonja and Wohlfart, Johannes},
  journal={Working paper},
  year={2025}
}

@article{kHoszegi2022fragile,
  title={Fragile self-esteem},
  author={K{\H{o}}szegi, Botond and Loewenstein, George and Murooka, Takeshi},
  journal={The Review of Economic Studies},
  volume={89},
  number={4},
  pages={2026--2060},
  year={2022},
  publisher={Oxford University Press}
}

@article{bandiera2025search,
  title={The search for good jobs: evidence from a six-year field experiment in Uganda},
  author={Bandiera, Oriana and Bassi, Vittorio and Burgess, Robin and Rasul, Imran and Sulaiman, Munshi and Vitali, Anna},
  journal={Journal of Labor Economics},
  volume={43},
  number={3},
  pages={885--935},
  year={2025},
  publisher={The University of Chicago Press Chicago, IL}
}

@article{hakimov2023confidence,
  title={Confidence and college applications: Evidence from a randomized intervention},
  author={Hakimov, Rustamdjan and Schmacker, Renke and Terrier, Camille},
  year={2023},
  journal={Working Paper}
}

@article{benabou2002self,
  title={Self-confidence and personal motivation},
  author={B{\'e}nabou, Roland and Tirole, Jean},
  journal={The Quarterly Journal of Economics},
  volume={117},
  number={3},
  pages={871--915},
  year={2002},
  publisher={MIT Press}
}

@article{adams2023perceived,
  title={Perceived returns to job search},
  author={Adams-Prassl, Abi and Boneva, Teodora and Golin, Marta and Rauh, Christopher},
  journal={Labour Economics},
  volume={80},
  pages={102307},
  year={2023},
  publisher={Elsevier}
}

@article{von2021heterogeneity,
  title={Heterogeneity in households’ stock market beliefs: Levels, dynamics, and epistemic uncertainty},
  author={Von Gaudecker, Hans-Martin and Wogrolly, Axel},
  journal={Journal of Econometrics},
  year={2021},
  publisher={Elsevier}
}

@article{belot2019providing,
  title={Providing advice to jobseekers at low cost: An experimental study on online advice},
  author={Belot, Michele and Kircher, Philipp and Muller, Paul},
  journal={The Review of Economic Studies},
  volume={86},
  number={4},
  pages={1411--1447},
  year={2019},
  publisher={Oxford University Press}
}

@article{cooper2020behavioral,
  title={Behavioral job search},
  author={Cooper, Michael and Kuhn, Peter},
  journal={Handbook of Labor, Human Resources and Population Economics},
  pages={1--22},
  year={2020},
  publisher={Springer}
}

@article{mueller2023expectations,
  title={Expectations data, labor market, and job search},
  author={Mueller, Andreas I and Spinnewijn, Johannes},
  journal={Handbook of Economic Expectations},
  pages={677--713},
  year={2023},
  publisher={Elsevier}
}

@book{heckman2021some,
  title={Some contributions of economics to the study of personality.},
  author={Heckman, James J and Jagelka, Tom{\'a}{\v{s}} and Kautz, Tim},
  year={2021},
  publisher={The Guilford Press}
}

@article{marinescu2021unemployment,
  title={Unemployment insurance and job search behavior},
  author={Marinescu, Ioana and Skandalis, Daphn{\'e}},
  journal={The Quarterly Journal of Economics},
  volume={136},
  number={2},
  pages={887--931},
  year={2021},
  publisher={Oxford University Press}
}

@article{kamenica2011bayesian,
  title={Bayesian persuasion},
  author={Kamenica, Emir and Gentzkow, Matthew},
  journal={American Economic Review},
  volume={101},
  number={6},
  pages={2590--2615},
  year={2011},
  publisher={American Economic Association}
}

@article{manski1999worker,
  title={Worker perceptions of job insecurity in the mid-1990s: Evidence from the survey of economic expectations},
  author={Manski, Charles F and Straub, John D},
  year={2000},
  journal={The Journal of Human Resources},
  volume={35},
  number={3},
  pages={447--479}
}

@article{he2024understanding,
  title={Understanding Expectations in Job Search: Subjective Duration Dependence, Aggregate Labor Market Shocks and Perceived Aggregate Labor Market Shocks},
  author={He, Qiwei and Kircher, Philipp},
  year={2026},
  journal={Working paper}
}

@article{spinnewijn2015unemployed,
  title={Unemployed but optimistic: Optimal insurance design with biased beliefs},
  author={Spinnewijn, Johannes},
  journal={Journal of the European Economic Association},
  volume={13},
  number={1},
  pages={130--167},
  year={2015},
  publisher={Oxford University Press}
}

@article{van1994effects,
  title={The effects of changes of the job offer arrival rate on the duration of unemployment},
  author={{V}an den {B}erg, Gerard J},
  journal={Journal of Labor Economics},
  volume={12},
  number={3},
  pages={478--498},
  year={1994},
  publisher={University of Chicago Press}
}

@article{tekleselassie2025feedback,
  title={Feedback, Confidence and Job Search Behavior},
  author={Tekleselassie, Tsegay and Witte, Marc and Radbruch, Jonas and Hensel, Lukas and Isphording, Ingo E},
  year={2025},
  journal={IZA Discusison paper}
}

@article{roussille2021central,
  title={The central role of the ask gap in gender pay inequality},
  author={Roussille, Nina},
  journal={The Quarterly Journal of Economics},
  year={2024},
  volume={139},
  number={3},
  pages={1557--1610}
}

@article{wu1991optimal,
  title={Optimal quantization by matrix searching},
  author={Wu, Xiaolin},
  journal={Journal of algorithms},
  volume={12},
  number={4},
  pages={663--673},
  year={1991},
  publisher={Elsevier}
}

@article{chernozhukov2009improving,
  title={Improving point and interval estimators of monotone functions by rearrangement},
  author={Chernozhukov, Victor and Fernandez-Val, Ivan and Galichon, Alfred},
  journal={Biometrika},
  volume={96},
  number={3},
  pages={559--575},
  year={2009},
  publisher={Oxford University Press}
}

@article{mueller2021job,
  title={Job seekers' perceptions and employment prospects: Heterogeneity, duration dependence, and bias},
  author={Mueller, Andreas I and Spinnewijn, Johannes and Topa, Giorgio},
  journal={American Economic Review},
  volume={111},
  number={1},
  pages={324--63},
  year={2021}
}

@article{caliendo2023accuracy,
  title={The Accuracy of Job Seekers' Wage Expectations},
  author={Caliendo, Marco and Mahlstedt, Robert and Schmei{\ss}er, Aiko and Wagner, Sophie},
  journal={arXiv preprint arXiv:2309.14044},
  year={2023},
  note={Revised September 2024}  
}

@article{katz1990unemployment,
  title={Unemployment insurance, recall expectations, and unemployment outcomes},
  author={Katz, Lawrence F and Meyer, Bruce D},
  journal={The Quarterly Journal of Economics},
  volume={105},
  number={4},
  pages={973--1002},
  year={1990},
  publisher={MIT Press}
}

@article{caliendo2015locus,
  title={Locus of control and job search strategies},
  author={Caliendo, Marco and Cobb-Clark, Deborah A and Uhlendorff, Arne},
  journal={Review of Economics and Statistics},
  volume={97},
  number={1},
  pages={88--103},
  year={2015},
  publisher={The MIT Press}
}

@article{mccall1970economics,
  title={Economics of information and job search},
  author={McCall, John Joseph},
  journal={The Quarterly Journal of Economics},
  pages={113--126},
  year={1970},
  publisher={JSTOR}
}

@article{conlon2018labor,
  title={Labor market search with imperfect information and learning},
  author={Conlon, John J and Pilossoph, Laura and Wiswall, Matthew and Zafar, Basit},
  year={2018},
  journal={NBER Working Paper No. 24988.}
}

@article{le2017unemployment,
  title={Unemployment insurance and reservation wages: Evidence from administrative data},
  author={Le Barbanchon, Thomas and Rathelot, Roland and Roulet, Alexandra},
  journal={Journal of Public Economics},
  year={2017},
  publisher={Elsevier}
}

@article{chernozhukov2018generic,
  author = {Chernozhukov, Victor and Demirer, Mert and Duflo, Esther and Fernández-Val, Iván},
title = {Fisher–Schultz Lecture: Generic Machine Learning Inference on Heterogeneous Treatment Effects in Randomized Experiments, With an Application to Immunization in India},
journal = {Econometrica},
volume = {93},
number = {4},
pages = {1121-1164},
  year={2025}
}

@article{manski2004measuring,
  title={Measuring expectations},
  author={Manski, Charles F},
  journal={Econometrica},
  volume={72},
  number={5},
  pages={1329--1376},
  year={2004},
  publisher={Wiley Online Library}
}

@article{mcgee2016search,
  title={Search, effort, and locus of control},
  author={McGee, Andrew and McGee, Peter},
  journal={Journal of Economic Behavior \& Organization},
  volume={126},
  pages={89--101},
  year={2016},
  publisher={Elsevier}
}

@book{cahuc2014labor,
  title={Labor economics},
  author={Cahuc, Pierre and Carcillo, St{\'e}phane and Zylberberg, Andr{\'e}},
  year={2014},
  publisher={MIT press}
}

@article{jones2022can,
  title={Can information correct optimistic wage expectations? Evidence from Mozambican job-seekers},
  author={Jones, Sam and Santos, Ricardo},
  journal={Journal of Development Economics},
  volume={159},
  pages={102987},
  year={2022},
  publisher={Elsevier}
}

@article{giustinelli2022expectations,
  title={Expectations in Education},
  author={Giustinelli, P},
  journal={Handbook of Economic Expectations},
  pages={193--224},
  year={2023},
  publisher={Elsevier}
}

@article{pollard1982quantization,
  author  = {Pollard, David},
  title   = {Quantization and the Method of $k$-Means},
  journal = {IEEE Transactions on Information Theory},
  volume  = {28}, number = {2}, pages = {199--205}, year = {1982}
}

@book{graf2000foundations,
  author    = {Graf, Siegfried and Luschgy, Harald},
  title     = {Foundations of Quantization for Probability Distributions},
  series    = {Lecture Notes in Mathematics}, volume = {1730},
  publisher = {Springer}, address = {Berlin}, year = {2000},
  doi       = {10.1007/BFb0103945}
}

@book{vandervaart1998asymptotic,
  author    = {van der Vaart, A. W.},
  title     = {Asymptotic Statistics},
  series    = {Cambridge Series in Statistical and Probabilistic Mathematics},
  publisher = {Cambridge University Press}, address = {Cambridge}, year = {1998}
}

@article{mease2006unique,
  author  = {Mease, David and Nair, Vijayan N.},
  title   = {Unique Optimal Partitions of Distributions and Connections to
             Hazard Rates and Stochastic Ordering},
  journal = {Statistica Sinica},
  volume  = {16}, number = {4}, pages = {1299--1312}, year = {2006}
}

@article{hall2018wage,
  title={Wage dispersion and search behavior: The importance of nonwage job values},
  author={Hall, Robert E and Mueller, Andreas I},
  journal={Journal of Political Economy},
  volume={126},
  number={4},
  pages={1594--1637},
  year={2018},
  publisher={University of Chicago Press Chicago, IL}
}

@article{van2024predicting,
  title={Predicting re-employment: machine learning versus assessments by unemployed workers and by their caseworkers},
  author={Van Den Berg, Gerard J and Kunaschk, Max and Lang, Julia and Stephan, Gesine and Uhlendorff, Arne},
  journal={IAB-Discussion Paper},
  year={2024}
}

@article{altmann2018learning,
  title={Learning about job search: A field experiment with job seekers in Germany},
  author={Altmann, Steffen and Falk, Armin and J{\"a}ger, Simon and Zimmermann, Florian},
  journal={Journal of Public Economics},
  volume={164},
  pages={33--49},
  year={2018},
  publisher={Elsevier}
}

@article{gosling2003very,
  title={A very brief measure of the Big-Five personality domains},
  author={Gosling, Samuel D and Rentfrow, Peter J and Swann Jr, William B},
  journal={Journal of Research in Personality},
  volume={37},
  number={6},
  pages={504--528},
  year={2003},
  publisher={Elsevier}
}

@article{armantier2017overview,
  title={An overview of the survey of consumer expectations},
  author={Armantier, Olivier and Topa, Giorgio and Van der Klaauw, Wilbert and Zafar, Basit},
  journal={Economic Policy Review},
  number={23-2},
  pages={51--72},
  year={2017}
}

@article{almlund2011personality,
  title={Personality psychology and economics},
  author={Almlund, Mathilde and Duckworth, Angela Lee and Heckman, James and Kautz, Tim},
  journal={Handbook of the Economics of Education},
  pages={1--181},
  year={2011},
  publisher={Elsevier}
}

@article{huysse2015low,
  title={Low self-esteem predicts future unemployment},
  author={Huysse-Gaytandjieva, Anna and Groot, Wim and Pavlova, Milena and Joling, Catelijne},
  journal={Journal of Applied Economics},
  volume={18},
  number={2},
  pages={325--346},
  year={2015},
  publisher={Elsevier}
}

@article{mendolia2015youth,
  title={Youth unemployment and personality traits},
  author={Mendolia, Silvia and Walker, Ian},
  journal={IZA Journal of Labor Economics},
  volume={4},
  pages={1--26},
  year={2015},
  publisher={Springer}
}

@article{almaas2023economics,
  title={Economics and measurement: New measures to model decision making},
  author={Alm{\aa}s, Ingvild and Attanasio, Orazio and Jervis, Pamela},
  year={2024},
  journal={Econometrica},
  volume={92},
  number={4},
  pages={947--978}
}

@article{christensen2026perceived,
  title={Perceived Wage Differences and Early Career Job Search Choices},
  author={Christensen Nygaard, Asker and Harmon Arpe, Nikolaj and Settelt, Sonja and Skandalis, Daphné},
  journal={Working Paper},
  year={2026}
}

@article{harmon2026job,
  title={Job Search, Overoptimism and Statistical Profiling: Can Information Provision Improve Job Search Outcomes?},
  author={Harmon, Nikolaj and Mahlstedt, Robert and Rasmussen, Mette},
  year={2026},
  journal={Working Paper}
}

\appendix

\newpage 
\FloatBarrier

\section{Survey description}\label{app:survey_description}
\renewcommand{\thetable}{\thesection.\arabic{table}}
\setcounter{table}{0}

\renewcommand{\thefigure}{\thesection.\arabic{figure}}
\setcounter{figure}{0}

\begin{figure}[!htbp]
    \centering
    \caption{Contact email sent to the sampled job seekers registered at the PES.}
    \label{fig:mail_de_contact}
    \includegraphics[height=0.9\textheight]{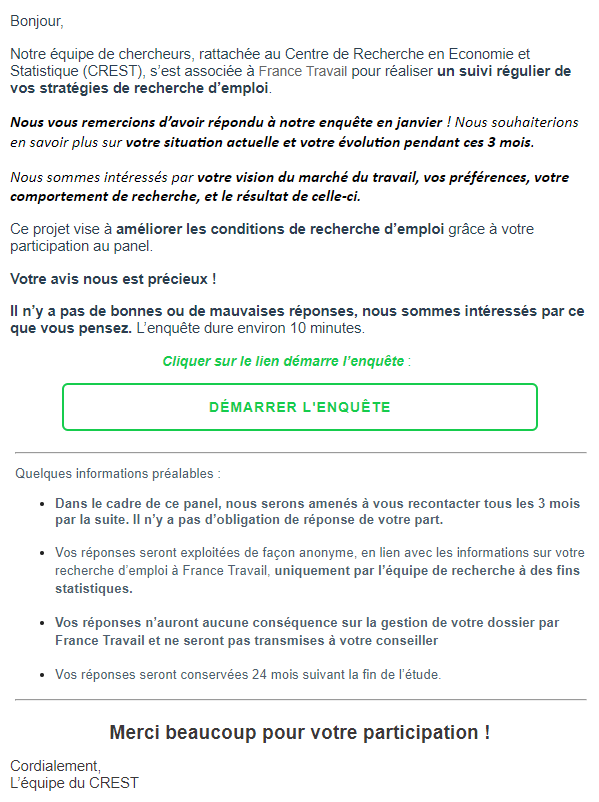}
\end{figure}

\begin{figure}[!htbp]
    \centering
    \caption{Survey page eliciting the subjective reemployment expectations at various horizons.}
    \label{fig:remployment_slider}
    \includegraphics[height=0.9\textheight]{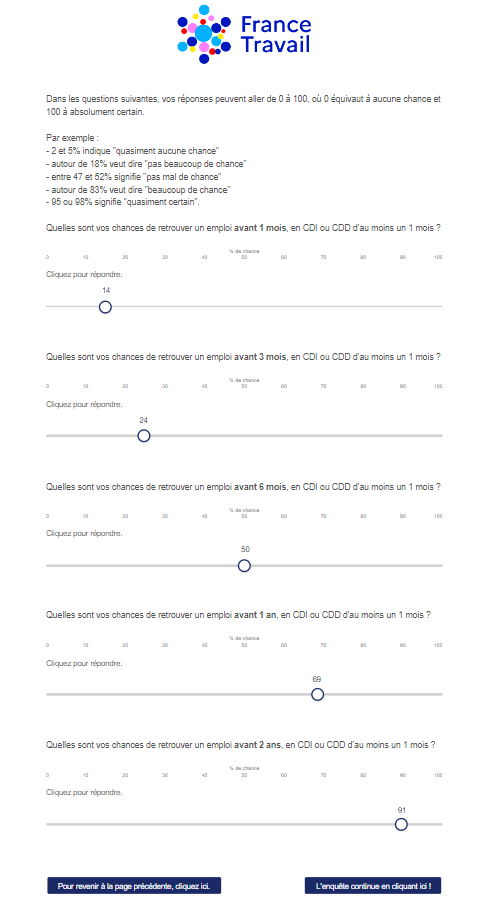}
\end{figure}

\begin{table}[!htbp]
\centering
\caption{Descriptive statistics: Survey variables}
\label{tab:subj_exp}

\begin{adjustbox}{width=\textwidth}
\begin{threeparttable}

\begin{tabular}{lcccccccc}
\hline
\hline
& Mean & SD & Min & P25 & Median & P75 & Max & Observations \\
\cmidrule{2-9}

\textbf{Subjective unemployment duration} & & & & & & & & \\
\cmidrule{2-9}
Probability of reemployment within 1 month & 39.0 & 28.8 & 0 & 16 & 34 & 52 & 100 & 25,551 \\
Probability of reemployment within 3 months & 50.5 & 30.1 & 0 & 25 & 50 & 75 & 100 & 25,846 \\
Probability of reemployment within 6 months & 60.6 & 30.9 & 0 & 40 & 61 & 90 & 100 & 25,846 \\
Probability of reemployment within 12 months & 69.3 & 31.3 & 0 & 50 & 80 & 100 & 100 & 25,594 \\
Probability of reemployment within 24 months & 74.1 & 32.2 & 0 & 50 & 91 & 100 & 100 & 25,449 \\

\cmidrule{2-9}
\textbf{Subjective job-offer arrival rates} & & & & & & & & \\
\cmidrule{2-9}
Probability of receiving at least 1 offer & 45.2 & 31.6 & 0 & 18 & 49 & 70 & 100 & 19,589 \\
\hspace{3pt}Conditional on 10 hours/week of search & 35.4 & 27.3 & 0 & 11 & 30 & 51 & 100 & 6,535 \\
\hspace{3pt}Conditional on 20 hours/week of search & 41.9 & 28.4 & 0 & 18 & 41 & 60 & 100 & 5,799 \\
\hspace{3pt}Conditional on 30 hours/week of search & 48.8 & 31.4 & 0 & 20 & 50 & 75 & 100 & 5,648 \\
Probability of receiving at least 3 offers & 41.3 & 30.8 & 0 & 13 & 40 & 61 & 100 & 19,268 \\

\cmidrule{2-9}
\textbf{Subjective wage-offer distribution} & & & & & & & & \\
\cmidrule{2-9}
Probability wage exceeds the 25th percentile & 46.5 & 30.9 & 0 & 19.8 & 46 & 71.1 & 100 & 16,261 \\
Probability wage exceeds the median & 41.4 & 31.2 & 0 & 13.5 & 38.4 & 64.3 & 100 & 16,259 \\
Probability wage exceeds the 75th percentile & 35.1 & 31.6 & 0 & 7.4 & 27.2 & 55.5 & 100 & 16,259 \\ 

\cmidrule{2-9}
\textbf{Search behavior} & & & & & & & & \\
\cmidrule{2-9}
Search effort (hours/week) & 13.1 & 9.2 & 0 & 5 & 10 & 20 & 65 & 22,054 \\
Reservation wage (relative to the minimum wage) & 1.3 & 0.5 & 1 & 1 & 1.1 & 1.4 & 6 & 15,587 \\

\hline
\hline
\end{tabular}

\begin{tablenotes}[flushleft]
\small
\item \textit{Notes:} This table reports descriptive statistics for the survey variables used in the analysis. The sample pools Waves 1--8 of the panel survey (October 2021--July 2023) and is restricted to the 25,846 individuals who reported their perceived 3-month reemployment probability and a main occupation.

The first panel reports beliefs about unemployment duration, measured as the subjective probability of finding a job within different time horizons. The second panel reports beliefs about the arrival rate of job offers, including beliefs conditional on alternative levels of weekly search effort, which are available only in Waves 6--8. The third panel reports beliefs about the wage-offer distribution. Specifically, ``Probability wage exceeds the $X$th percentile'' denotes the subjective probability that a received wage offer exceeds the $X$th percentile of the objective wage-offer distribution. The final panel reports selected measures of search behavior.

Means and standard deviations are computed using survey weights. The number of observations varies across variables because some questions were asked only in specific survey waves or were subject to item nonresponse.
\end{tablenotes}

\end{threeparttable}
\end{adjustbox}
\end{table}

\begin{table}[!htbp] 
\centering 
\caption{Descriptive Statistics: response rate by panel, wave and status} 
\label{tab:non_resp_waves}  
\begin{adjustbox}{width = \textwidth}
\begin{threeparttable}
\begin{tabular}{|c|c|cc|l|cc|cc|} 
\hline 
\multirow{3}{*}{\textbf{Panel}} & \multirow{3}{*}{\textbf{Date}} & \multicolumn{2}{c|}{\textbf{Wave 1}} & \multirow{3}{*}{Status} & \multicolumn{2}{c|}{\textbf{Wave 2}} & \multicolumn{2}{c|}{\textbf{Wave 3}}\\
& & Response & Nb. useable & & Response & Shares & Response & Shares \\
& & rate & responses & & rate & by status & rate & by status \\
\hline 
 \multirow{4}{*}{\textbf{Panel 1}}  & \multirow{4}{*}{\textbf{Oct. 2021}} & \multirow{4}{*}{10.1} & \multirow{4}{*}{2,731} &   Satisfactory return  & \multirow{4}{*}{44.2} & 13.9 & \multirow{4}{*}{39.7} & 14.5 \\ 
&  & & & Unsatisfactory return & & 23.1 & & 21.7 \\ 
&  & & & Stop searching & & 5.5 & & 4.5  \\ 
&  & & & Continue searching & & 57.4 &  & 59.3 \\ 
 \hline
\multirow{4}{*}{\textbf{Panel 2}} & \multirow{4}{*}{\textbf{Jan. 2022}} &  \multirow{4}{*}{21.3}  & \multirow{4}{*}{4,581} &  Satisfactory return  & \multirow{4}{*}{31.7} & 16.3 & \multirow{4}{*}{35.2} & 12.6  \\ 
 & & & & Unsatisfactory return & & 23.1  & & 27.0  \\ 
 & & & & Stop searching & & 7.1 & & 5.5  \\ 
 & & & & Continue searching  & & 53.5 & & 54.9  \\ 
 \hline
\multirow{4}{*}{\textbf{Panel 3}} & \multirow{4}{*}{\textbf{Apr. 2022}} & \multirow{4}{*}{17.6}  & \multirow{4}{*}{2,854} &  Satisfactory return  &\multirow{4}{*}{32.4} & 16.5 &  \multirow{4}{*}{36.1} & 14.1  \\ 
 & & & & Unsatisfactory return  & & 17.8 &  & 25.9  \\ 
 & & & & Stop searching & & 9.7 &  & 6.1  \\ 
 & & & & Continue searching & & 56.1 &  & 53.9  \\ 
  \hline
 \multirow{4}{*}{\textbf{Panel 4}} &\multirow{4}{*}{\textbf{Jul. 2022}} & \multirow{4}{*}{15.8}  & \multirow{4}{*}{3,393} &  Satisfactory return  &\multirow{4}{*}{32.9} & 13.8  &  \multirow{4}{*}{40.3} & 12.0  \\ 
 & & & & Unsatisfactory return  & & 20.7 &  & 18.2 \\ 
 & & & & Stop searching & & 6.1 &  & 6.7  \\ 
 & & & & Continue searching & & 59.5 &  & 63.2  \\ 
  \hline
 \multirow{4}{*}{\textbf{Panel 5}} &\multirow{4}{*}{\textbf{Oct. 2022}} & \multirow{4}{*}{16.1}  & \multirow{4}{*}{3,854} &  Satisfactory return  &\multirow{4}{*}{33.0} & 11.8 & \multirow{4}{*}{38.7} & 11.6 \\ 
& & & & Unsatisfactory return & & 14.3 &  & 17.5 \\ 
& & & & Stop searching & & 7.8 &  & 6.2  \\ 
& & & & Continue searching & & 66.1 &  & 64.7  \\ 
 \hline
 \multirow{4}{*}{\textbf{Panel 6}} &\multirow{4}{*}{\textbf{Jan. 2023}} & \multirow{4}{*}{14.3}  & \multirow{4}{*}{3,145} &  Satisfactory return  &\multirow{4}{*}{32.9} & 12.0 & \multirow{4}{*}{45.4} & 11.7 \\ 
& & & & Unsatisfactory return & & 14.8 &  & 20.7 \\ 
& & & & Stop searching & & 6.8 &  & 6.2  \\ 
& & & & Continue searching & & 66.5 &  & 61.4  \\ 
\hline
\multirow{4}{*}{\textbf{Panel 7}} &\multirow{4}{*}{\textbf{Apr. 2023}} & \multirow{4}{*}{12.1}  & \multirow{4}{*}{2,514} &  Satisfactory return  &\multirow{4}{*}{35.0} & 10.6 & \multirow{4}{*}{45.6} & 14.7 \\ 
& & & & Unsatisfactory return & & 17.0 &  & 20.0 \\ 
& & & & Stop searching & & 7.4 &  & 4.4  \\ 
& & & & Continue searching & & 65.0 &  &  61.0 \\ 
\hline
\multirow{4}{*}{\textbf{Panel 8}} &\multirow{4}{*}{\textbf{Jul. 2023}} & \multirow{4}{*}{12.9}  & \multirow{4}{*}{2,774} &  Satisfactory return  & \multirow{4}{*}{35.9} & 12.5 & \multirow{4}{*}{46.4} & 8.6 \\ 
& & & & Unsatisfactory return & & 15.7 &  & 17.9 \\ 
& & & & Stop searching & & 8.1 &  &  4.4 \\ 
& & & & Continue searching & & 63.7 &  & 69.0 \\ 
\hline 
\end{tabular}
\begin{tablenotes}[flushleft]
    \small
    \item \textit{Notes}: The sample is from waves 1-8 of the panel (Oct 2021- July 2023). Columns ``Wave X" gather the average response rate and number of usable responses. Usable responses stands for survey answers containing at least the perceived reemployment probability within 3 months. ``Shares by status" gives the decomposition by Status of the respondents in the second and third time survey. ``Satisfactory return" refers to a declared return to employment where the job seekers stops searching, and  ``Unsatisfactory return" when he or she continues to search. ``Stop searching" refers to the situation where he or she did not find a job and stop searching, and the contrary for ``Continue searching". The panel 1 are first interviewed (first wave) in October 2021 with a response rate of 10.2\% providing 2,737 usable responses for the analysis. Among these individuals, 44.3\% have answered the second wave of the survey (in January 2022). From the second interview and onwards, jobseekers are asked whether they have found a job since the previous interview and whether they are still actively looking for a job. Individuals from panel 1 who answered the second wave of the survey are for instance 13.6\% to have found a job and stopped looking for a job.  
\end{tablenotes}
\end{threeparttable}
\end{adjustbox}
\end{table} 

\newpage

\subsection{Eliciting Psychological Traits}\label{ssec:app_psycho_items}

In this section, we describe how we elicit psychological traits
corresponding to the \textit{Big Five} personality dimensions and the
locus of control. One important constraint for our online survey is to
limit the number of items, as keeping the survey short ensures both a
higher participation rate and a higher completion rate.

\paragraph{Big Five personality traits.}
In its shorter version, the Big Five can be measured by asking
participants to rate how accurately a set of characteristics describes
their personality. To reduce the number of items, we ask job seekers to
evaluate 10 pairs of characteristics, with two pairs assigned to each of
the five personality traits. Each pair is scored using a slider ranging
from 0 to 10. The items are translated from \cite{gosling2003very} and
are designed to capture the main dimensions of each Big Five trait.

\medskip

As shown in Table~\ref{tab:psycho_items}, for each trait, one pair of
characteristics is \emph{positively} associated with the trait and the
other is \emph{negatively} associated with it. For instance, being
\textit{dependable} and \textit{self-disciplined} is associated with
higher conscientiousness, whereas being \textit{disorganized} and
\textit{careless} is associated with lower conscientiousness. We
construct our final index for each trait as the score of the positively
associated pair minus the score of the negatively associated pair.

\paragraph{Locus of control.}
We measure the locus of control by asking job seekers to indicate, using
a slider from 0 to 10, how much they agree with six statements. Three
statements correspond to an internal locus of control and three to an
external locus of control. We build two separate measures: the
\emph{internal locus} score is the average rating given to the three
internal statements, and the \emph{external locus} score is the average
rating given to the three external statements.

\begin{table}[!htbp]
    \centering
    \caption{Big Five and locus of control survey items}
    \label{tab:psycho_items}
    \begin{tabularx}{\textwidth}{|l|X|}
        \hline
        \textbf{Big Five} & \textbf{\textit{For each of the following personality trait pairs, please
          assign a score (0--10) reflecting how well they apply to you:}} \\
        \hline
        Conscientiousness (+) & Dependable, self-disciplined \\
        Conscientiousness ($-$) & Disorganized, careless \\
        Openness (+)           & Open to new experiences \\
        Openness ($-$)         & Conventional, not very original \\
        Neuroticism (+)        & Anxious, irritable \\
        Neuroticism ($-$)      & Calm, emotionally stable \\
        Extraversion (+)       & Extraverted, enthusiastic \\
        Extraversion ($-$)     & Reserved, quiet \\
        Agreeableness (+)      & Sympathetic, warm \\
        Agreeableness ($-$)    & Critical, quarrelsome \\
        \hline
        \textbf{Locus of control} & \textbf{\textit{For each of the following statements, please indicate how
          much you agree by assigning a score (0--10):}} \\
        \hline
        Internal locus & What happens to me is my own doing. \\
        Internal locus & My achievements in life are due more to my own
                         abilities than to luck. \\
        Internal locus & When I make plans, I am confident in my ability
                         to carry them out. \\
        External locus & Many of the negative events in people's lives
                         are partly due to bad luck. \\
        External locus & To get a good job, you have to be in the right
                         place at the right time. \\
        External locus & I often feel like I have little control over the
                         things that happen to me. \\
        \hline
    \end{tabularx}
\end{table}

\FloatBarrier
\newpage

\section{Sample description}
\label{app:sample_description}
\renewcommand{\thetable}{\thesection.\arabic{table}}
\setcounter{table}{0}

\renewcommand{\thefigure}{\thesection.\arabic{figure}}
\setcounter{figure}{0}

\begin{table}[!htbp]
\centering

\begin{adjustbox}{width = \textwidth}

\begin{threeparttable}

\caption{Descriptive Statistics: survey non-response (Part 1)}
\label{tab:non_resp1}

\begin{tabular*}{\textwidth}{@{\extracolsep{\fill}} rcccc}
\toprule
Sample  & Respondents & Received the email & Respondents & Respondents \\
Take up weights & No & No & Yes & Yes \\
Sampling weights & No & No & No & Yes \\
\midrule

\textbf{Women} & 56.2 & 47.7 & 49.0 & 52.0 \\

\addlinespace
\textbf{Age} & 42.5 & 37.5 & 37.8 & 40.6 \\
$\leq$ 25 & 7.0 & 15.0 & 15.2 & 10.8 \\
25--35 & 24.8 & 34.1 & 34.0 & 29.3 \\
35--45 & 24.7 & 23.6 & 23.5 & 23.7 \\
45--55 & 24.5 & 16.5 & 16.5 & 19.2 \\
$>$ 55 & 19.0 & 10.8 & 10.8 & 17.0 \\

\addlinespace
\textbf{Married} & 48.2 & 38.6 & 42.6 & 44.4 \\

\addlinespace
\textbf{At least 1 child} & 47.7 & 42.3 & 42.3 & 43.8 \\

\addlinespace
\textbf{Education level} \\
High school & 21.4 & 24.4 & 24.0 & 23.4 \\
HS + 2y college & 17.3 & 13.7 & 15.8 & 15.8 \\
Bachelor degree & 12.9 & 9.7 & 11.4 & 11.3 \\
Master degree & 15.2 & 10.4 & 11.8 & 11.5 \\
Professional degree & 23.1 & 27.2 & 24.7 & 25.7 \\
No degree & 23.1 & 27.2 & 24.7 & 25.7 \\

\addlinespace
\textbf{Qualification} \\
High skilled worker & 19.8 & 12.8 & 12.8 & 12.1 \\
Qualified employee & 43.8 & 44.5 & 44.6 & 45.9 \\
Non qualified employee & 15.2 & 19.9 & 20.0 & 19.4 \\

\addlinespace
\textbf{Nb. year of experience} & 7.8 & 5.9 & 6.1 & 6.9 \\

\bottomrule
\end{tabular*}

\begin{tablenotes}[flushleft]
\small
\item \textit{Notes:} The sample combines waves 1--8 of the panel (October 2021--July 2023). 
Column~1 reports average characteristics of job seekers who responded to the survey (i.e., answered at least the first question). Column~2 reports average characteristics of the full sample invited to take the survey. Column~3 reweights respondents using participation weights, so as to match the characteristics of the group invited to take the survey. Column~4 additionally applies sampling weights to match the average characteristics of the overall population of job seekers registered with the PES.
\end{tablenotes}

\end{threeparttable}
\end{adjustbox}
\end{table}

\begin{table}[!htbp]
\centering
\begin{adjustbox}{width=\textwidth}
\begin{threeparttable}

\caption{Descriptive Statistics: survey non-response (Part 2)}
\label{tab:non_resp2}

\begin{tabular}{rcccc}
\toprule
Sample  & Respondents & Received the email & Respondents & Respondents \\
Take up weights & No & No & Yes & Yes \\
Sampling weights & No & No & No & Yes \\
\midrule

\textbf{Spell duration} \\
$\leq$ 1 month & 31.2 & 31.7 & 31.4 & 10.8 \\
1--3 months & 23.6 & 23.5 & 23.4 & 11.4 \\
3--6 months & 11.3 & 10.7 & 10.9 & 15.9 \\
6--12 months & 15.2 & 16.2 & 15.9 & 22.0 \\
12--24 months & 9.1 & 9.0 & 9.2 & 18.8 \\
$>$ 24 months & 9.7 & 8.9 & 9.2 & 21.1 \\

\addlinespace
\textbf{Remaining benefits} \\
Exhausted & 20.5 & 22.8 & 23.0 & 34.0 \\
$\leq$ 6 months & 9.9 & 13.3 & 12.9 & 12.9 \\
6--12 months & 13.2 & 16.5 & 15.8 & 13.4 \\
12--24 months & 45.5 & 41.2 & 42.0 & 32.1 \\
$>$ 24 months & 10.9 & 6.3 & 6.3 & 7.6 \\

\addlinespace
\textbf{Annual reservation wage (\euro)} \\
$\leq$ 20k & 22.3 & 28.9 & 27.0 & 24.8 \\
20k--25k & 31.0 & 34.8 & 34.8 & 37.1 \\
25k--35k & 23.5 & 21.3 & 21.9 & 21.5 \\
$>$ 35k & 23.1 & 14.9 & 16.3 & 16.6 \\

\addlinespace
At least one application & \multirow{2}{*}{63.0} & \multirow{2}{*}{62.8} & \multirow{2}{*}{62.6} & \multirow{2}{*}{66.2} \\
through the PES & & & & \\

\bottomrule
\end{tabular}

\begin{tablenotes}[flushleft]
\small
\item \textit{Notes:} The sample combines waves 1--8 of the panel (October 2021--July 2023). 
Column~1 reports average characteristics of job seekers who responded to the survey (i.e., answered at least the first question). Column~2 reports average characteristics of the full sample invited to take the survey. Column~3 reweights respondents using participation weights, so as to match the characteristics of the group invited to take the survey. Column~4 additionally applies sampling weights to match the average characteristics of the overall population of job seekers registered with the PES.
\end{tablenotes}
\end{threeparttable}
\end{adjustbox}
\end{table}

\section{Additional Tables and Figures of section \ref{sec:information}}\label{app:evidence}

\renewcommand{\thetable}{\thesection.\arabic{table}}
\setcounter{table}{0}

\renewcommand{\thefigure}{\thesection.\arabic{figure}}
\setcounter{figure}{0}

\begin{figure}[!htbp]
\caption{Informational biases empirical distributions}
\centering
\begin{subfigure}{0.5\linewidth}
    \centering
    \includegraphics[width = \textwidth]{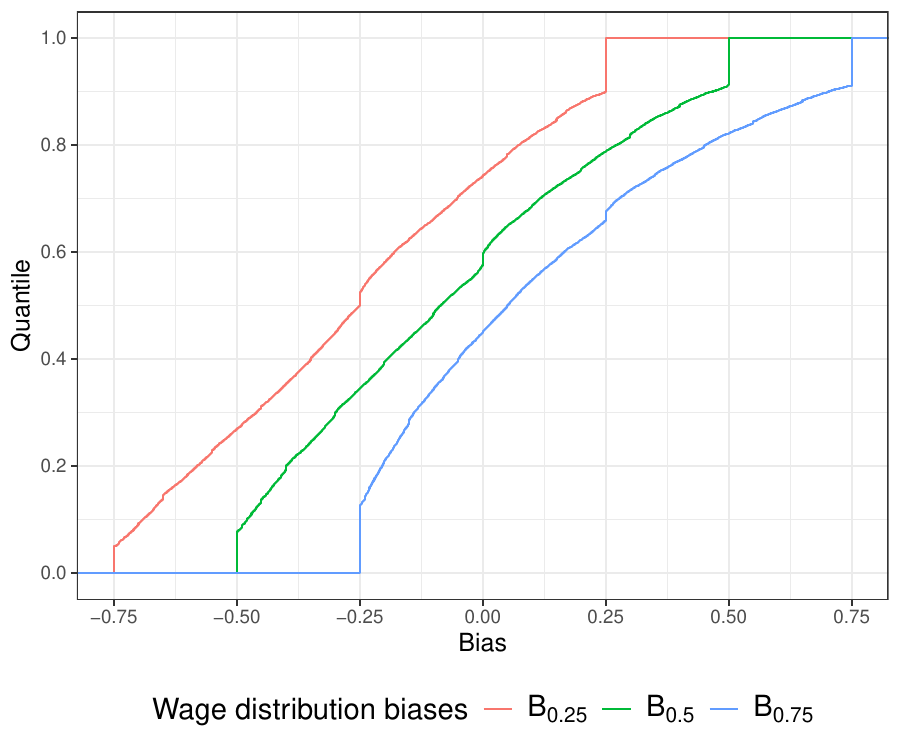}
    \caption{Wage biases} 
     \label{fig:wage_biases_ecdf}
\end{subfigure}

\begin{subfigure}{0.5\linewidth}
    \includegraphics[width = \textwidth]{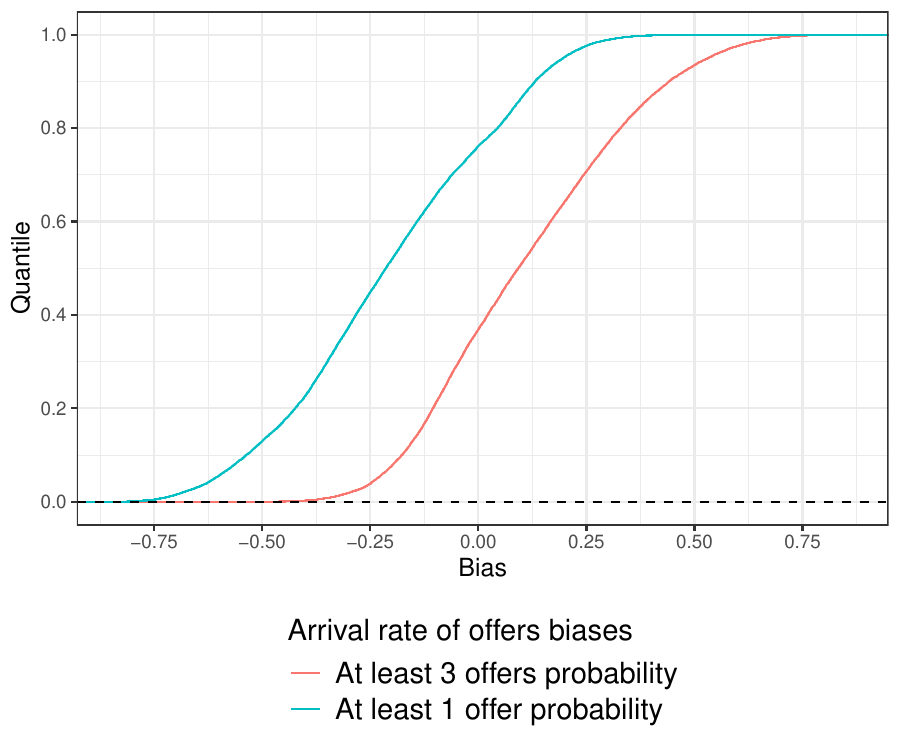}
    \caption{Arrival rate of offers biases}
    \label{fig:ecdf_biases_lambda}
\end{subfigure}
    
\caption*{\footnotesize \textit{Notes}: These figures plot the empirical distribution of both informational biases. Panel (a) plots individual biases $B_i(\alpha) = \widetilde{P}_i(W > F_i^{-1}(\alpha)) - \alpha $ at different quantiles of the wage distribution: relatively low wages $\alpha=0.25$ (red), median $\alpha=0.5$ (green),  and high wages $\alpha=0.75$ (blue). Panel (b) plots individual biases at several point of the arrival rate of offer distribution : $\widetilde{P}( N >x) - P(N > x)$ for $x = 1,3$. This is over the panels 1-8.}

\end{figure}

\begin{table}[!htpb]
\centering
\footnotesize
\setlength{\tabcolsep}{3pt}            

\scalebox{0.8}{
\begin{threeparttable}
\caption{Determinants of the biases on the perception of the offers' wage distribution}
\label{tab:correlations_X_wages_biases}

\begin{tabular}{lccccc}
\toprule
\toprule
& \multicolumn{5}{c}{Biases over the wage distribution} \\
\cmidrule(lr){2-6}
 & $B(0.25)$ & $B(0.5)$ & $|B(0.5)|$ & $B(0.75)$ & $B(0.75)-B(0.25)$ \\ 
\midrule

\textbf{Age bin (Ref : $<$ 25)} \\
Age 25--35 & -0.008         & -0.014         & -0.0006         & 0.0001  & 0.008 \\ 
Age 35--45 & 0.005          & 0.006          & 0.010           & 0.021                 & 0.016 \\  
Age 45--55 & -0.013         & -0.014         & 0.012           & -0.001                & 0.012 \\ 
Age $>$ 55 & -0.103$^{***}$ & -0.102$^{***}$ & 0.036$^{***}$   & -0.086$^{***}$        & 0.017 \\

\addlinespace
\textbf{Woman} & -0.051$^{***}$ & -0.063$^{***}$ & 0.006           & -0.067$^{***}$        & -0.016$^{**}$\\ 
\textbf{Nb. children} & -0.003         & 0.004          & -0.001          & 0.010$^{**}$          & 0.012$^{***}$\\ 
\textbf{Married} & 0.033$^{***}$  & 0.032$^{***}$  & -0.012$^{**}$   & 0.021$^{**}$          & -0.012$^{**}$\\

& & & & & \\ 
\textbf{Level of eductation (Ref : other)} & & & & & \\ 
High School & 0.012          & 0.003          & -0.001          & -0.015                & -0.027$^{**}$\\
HS + 2y College  & 0.025          & 0.003          & -0.013          & -0.016                & -0.042$^{***}$\\   
Bachelor degree & 0.054$^{***}$  & 0.034$^{*}$    & -0.026$^{***}$  & 0.010                 & -0.045$^{***}$\\
Master degree & 0.099$^{***}$  & 0.082$^{***}$  & 0.004           & 0.061$^{***}$         & -0.038$^{**}$\\    
Vocational degree & -0.007         & -0.005         & 0.003           & -0.007                & -0.0003\\
& & & & & \\ 
\textbf{Qualification} & & & & & \\ 
High skilled & 0.111$^{***}$  & 0.121$^{***}$  & -0.002          & 0.106$^{***}$         & -0.004\\ 
Qualified employees & -0.010         & -0.012         & 0.007           & -0.008                & 0.002\\
Non qualified employees & -0.011         & -0.026$^{*}$   & 0.015$^{**}$    & -0.025$^{*}$          & -0.015\\   
& & & & & \\ 
Experience (nb. of year) & 0.002$^{***}$  & 0.002$^{***}$  & -0.0002         & 0.002$^{***}$         & 0.0006$^{*}$\\
& & & & & \\ 
\textbf{Spell duration (Ref : $<$ 1 month)} & & & & & \\
Duration 1-3 months & -0.009         & -0.004         & 0.004           & -0.010                & -0.0009\\  
Duration 3-6 months & -0.020$^{*}$   & -0.023$^{**}$  & 0.003           & -0.031$^{***}$        & -0.011\\ 
Duration 6-12 months  & -0.044$^{***}$ & -0.036$^{***}$ & 0.005           & -0.034$^{***}$        & 0.009\\   
Duration 12-24 months & -0.050$^{***}$ & -0.043$^{***}$ & 0.014$^{**}$    & -0.041$^{***}$        & 0.008\\  
Duration > 24 months  & -0.036$^{**}$  & -0.030$^{**}$  & 0.011           & -0.026$^{*}$          & 0.010\\   
& & & & & \\
\textbf{Benefits duration (Ref : Exhausted)} & & & & & \\
Benefits < 6 months  & 0.003          & 0.005          & -0.0009         & 0.006                 & 0.004\\  
Benefits 6-12 months & -0.007         & -0.011         & 0.014$^{*}$     & -0.005                & 0.002\\ 
Benefits 12-24 months & 0.007          & 0.008          & -0.005          & 0.014                 & 0.007\\   
Benefits > 24 months & -0.005         & 0.0003         & -0.005          & 0.007                 & 0.012\\    
& & & & & \\ 
\textbf{One application}& -0.004         & -0.006         & -0.013$^{***}$  & -0.012                & -0.008\\  
\hline \\[-1.8ex] 
Region fixed effect & \checkmark & \checkmark & \checkmark & \checkmark & \checkmark \\ 
Target occupation fixed effect & \checkmark & \checkmark & \checkmark & \checkmark & \checkmark \\ 
Panel fixed effect & \checkmark & \checkmark & \checkmark & \checkmark & \checkmark \\ 
\midrule
Num. Obs. & 16,219         & 16,217         & 16,217          & 16,217                & 16,213\\  
$R^{2}$ & 0.141        & 0.139        & 0.101         & 0.138               & 0.108\\ 
\bottomrule
\end{tabular}

\begin{tablenotes}[flushleft]
\small
\item \textit{Notes:} This table presents the results of the linear regressions of individual biases on the wage distribution $B_i(\alpha)$ for different quantiles of the distribution: relatively low wages $\alpha=0.25$, median $\alpha=0.5$, and high wages $\alpha=0.75$, as well as for the absolute bias on the median $|B(0.5)|$ and the spread $B(0.75)-B(0.25)$. One application is a dummy indicating whether the job seekers made at least one application through the PES in the past. The sample pools Waves 1--8 of the panel survey (October 2021--July 2023) and is restricted to the 25,846 individuals who reported their perceived 3-month reemployment probability and a main occupation. Significance levels: $^{*}p<0.1$, $^{**}p<0.05$, $^{***}p<0.01$.
\end{tablenotes}

\end{threeparttable}
}
\end{table}

\begin{table}[!htbp]
\centering
\footnotesize
\setlength{\tabcolsep}{3pt}
\scalebox{0.85}{
\begin{threeparttable}

\caption{Determinants of subjective and objective offers' arrival rate}
\label{tab:determinants_arrival}

\begin{tabular}{lcccccc}
\toprule
\toprule
& \multicolumn{3}{c}{Probability of receiving at least 1 offer}
& \multicolumn{3}{c}{Probability of receiving at least 3 offers} \\
\cmidrule(lr){2-4}\cmidrule(lr){5-7}
& $\indic\{N \geq 1\}$ & $\indic\{N \geq 1\}$ & $\tilde{p}_N(1)$
& $\indic\{N \geq 3\}$ & $\indic\{N \geq 3\}$ & $\tilde{p}_N(3)$ \\
& (1) & (2) & (3) & (4) & (5) & (6) \\
\midrule

\textbf{Perceived arrival rate} \\
$\tilde{p}_N(1)$ &               & 0.344$^{***}$ &                &               &               &   \\ 
$\tilde{p}_N(3)$ &               &               &                &               & 0.332$^{***}$ &   \\

\addlinespace
Level of search effort & -0.0010       & -0.002$^{**}$ & 0.004$^{***}$  & 0.004$^{***}$ & 0.003$^{**}$  & 0.004$^{***}$\\

\addlinespace
\textbf{Age bin (Ref : $<$ 25)} \\
Age 25--35 & 0.038         & 0.042         & -0.011         & 0.079         & 0.086         & -0.011\\
Age 35--45 & 0.096         & 0.103         & -0.022         & 0.103$^{*}$   & 0.114$^{**}$  & -0.022\\   
Age 45--55 & 0.151$^{**}$  & 0.159$^{**}$  & -0.022         & 0.188$^{***}$ & 0.197$^{***}$ & -0.022\\ 
Age $>$ 55 & 0.015         & 0.060         & -0.131$^{***}$ & 0.093         & 0.139$^{**}$  & -0.131$^{***}$\\

\addlinespace
\textbf{Woman} & -0.010        & 0.004         & -0.041$^{**}$  & -0.042        & -0.023        & -0.041$^{**}$\\
\textbf{Nb. children} & 0.009         & 0.008         & 0.003          & 0.004         & 0.0003        & 0.003\\
\textbf{Married} & -0.058$^{**}$ & -0.048$^{**}$ & -0.029$^{**}$  & -0.057$^{**}$ & -0.048$^{**}$ & -0.029$^{**}$\\

\addlinespace
\textbf{Level of education (Ref : Other)} \\
High School & 0.042         & 0.036         & 0.019          & 0.019         & 0.020         & 0.019\\
HS + 2y College & -0.005        & -0.011        & 0.018          & -0.040        & -0.035        & 0.018\\ 
Bachelor degree & 0.010         & 0.006         & 0.013          & -0.029        & -0.029        & 0.013\\  
Master degree & 0.032         & 0.029         & 0.009          & -0.049        & -0.037        & 0.009\\ 
Professional degree  & -0.008        & -0.011        & 0.007          & -0.016        & -0.016        & 0.007\\ 

\addlinespace
\textbf{Qualification} \\
High skilled & -0.057        & -0.061        & 0.013          & -0.063        & -0.067$^{*}$  & 0.013\\  
Qualified employees & -0.010        & -0.004        & -0.017         & -0.035        & -0.034        & -0.017\\   
Non qualified employees & -0.042        & -0.019        & -0.067$^{**}$  & -0.016        & -0.003        & -0.067$^{**}$\\  

\addlinespace
\textbf{Experience (years)} & -0.003$^{**}$ & -0.003$^{**}$ & 0.0005         & -0.0009       & -0.0007       & 0.0005\\

\addlinespace
\textbf{Spell duration (Ref : $<$ 1 month)} \\
Duration 1--3 months & 0.011         & 0.028         & -0.049$^{***}$ & 0.043         & 0.055$^{*}$   & -0.049$^{***}$\\ 
Duration 3--6 months & 0.021         & 0.031         & -0.028         & -0.021        & -0.008        & -0.028\\   
Duration 6--12 months & 0.004         & 0.021         & -0.050$^{***}$ & -0.016        & 0.002         & -0.050$^{***}$\\
Duration 12--24 months & -0.039        & -0.003        & -0.103$^{***}$ & -0.083$^{**}$ & -0.047        & -0.103$^{***}$\\   
Duration $>$ 24 months & 0.002         & 0.045         & -0.125$^{***}$ & -0.041        & -0.005        & -0.125$^{***}$\\  

\addlinespace
\textbf{One application} & 0.135$^{***}$ & 0.129$^{***}$ & 0.017          & 0.130$^{***}$ & 0.121$^{***}$ & 0.017\\

\midrule
Remaining benefit duration & \checkmark & \checkmark & \checkmark & \checkmark & \checkmark & \checkmark \\
Region fixed effect & \checkmark & \checkmark & \checkmark & \checkmark & \checkmark & \checkmark \\
Target occupation fixed effect & \checkmark & \checkmark & \checkmark & \checkmark & \checkmark & \checkmark \\
Panel fixed effect & \checkmark & \checkmark & \checkmark & \checkmark & \checkmark & \checkmark \\
\midrule
Num. Obs. & 4,218         & 4,218         & 4,218          & 4,218         & 4,218         & 4,218\\
$R^2$ & 0.252       & 0.285       & 0.261        & 0.205       & 0.239       & 0.261\\ 
\bottomrule

\end{tabular}

\begin{tablenotes}[flushleft]
\small
\item \textit{Notes:} Columns (1) and (4) report linear probability models for receiving at least 1 and 3 offers. Columns (2) and (5) additionally include subjective expectations to test rational expectations. Columns (3) and (6) regress subjective expectations on observables. Location, occupation and panel fixed effects are included. One application is a dummy indicating whether the job seekers made at least one application through the PES in the past. The sample pools Waves 1--8 of the panel survey (October 2021--July 2023) and is restricted to the 25,846 individuals who reported their perceived 3-month reemployment probability and a main occupation and who took the survey at least twice. Significance levels: $^{*}p<0.1$, $^{**}p<0.05$, $^{***}p<0.01$.
\end{tablenotes}

\end{threeparttable}
}
\end{table}

\begin{table}[!htbp]
\centering
 \caption{Within individual beliefs updating with unemployment duration}\label{table_update_duration} 
\begin{adjustbox}{width = \textwidth}
\begin{threeparttable}
\begin{tabular}[t]{lcccccccccc}
\toprule
  & \multicolumn{2}{c}{Beliefs on arrival rate} &  \multicolumn{3}{c}{Beliefs on wages}  &  \\ 
  \cmidrule{2-3} 	\cmidrule{4-6}
  &  $ \widetilde{P}(N\geq 3) $, (A) & Bias on (A) & $B_i(0.5)$ & Bias spread & Bias at $w^*$ & Marg. 20h\\
   \cmidrule{2-3} 	\cmidrule{4-6}
   & (1) & (2)    & (3) & (4)    & (5) & (6)  \\
\midrule
Q2 &-0.004  & -0.008 & 0.003 & -0.023 & -0.042 & -1.589\\
 & (0.011)  & (0.007) & (0.014) & (0.011)* & (0.019)* & (1.536)\\
Q3 & 0.000  & 0.024 & 0.033 & -0.027 & -0.032 & -2.193\\
 & (0.013)  & (0.009)** & (0.016)* & (0.013)* & (0.023) & (1.778)\\
Q4&-0.003 & 0.021 & 0.012 & -0.059 & -0.082 & -3.082\\
  & (0.014)  & (0.010)* & (0.018) & (0.015)*** & (0.026)** & (1.889)\\
Q5 & -0.002  & 0.027 & 0.024 & -0.051 & -0.068 & -3.429\\
  & (0.016)  & (0.011)* & (0.020) & (0.017)** & (0.029)* & (1.994)+\\
Q6 & 0.001  & 0.031 & -0.006 & -0.040 & -0.160 & -4.139\\
 & (0.018)  & (0.012)* & (0.022) & (0.019)* & (0.032)*** & (2.131)+\\
Q7 & 0.002  & 0.032 & 0.010 & -0.047 & -0.210 & -2.609\\
 & (0.020)  & (0.014)* & (0.025) & (0.021)* & (0.036)*** & (2.314)\\
Q8 & -0.012  & 0.023 & -0.017 & -0.007 & -0.207 & -1.437\\
  & (0.023)  & (0.015) & (0.028) & (0.023) & (0.040)*** & (2.572)\\
Q8+ & -0.004 & -0.015 & -0.017 & -0.033 & -0.239 & 0.236\\
 & (0.024) & (0.016) & (0.029) & (0.025) & (0.042)*** & (2.606)\\
\midrule
Num. Obs.& 11,636 & 11,269 & 10,107 & 10,099 & 8,384 & 6,237\\
$R^2$ & 0.536 & 0.629 & 0.484 & 0.281 & 0.670 & 0.566\\
\bottomrule
\end{tabular}
\begin{tablenotes}[flushleft]
    \small
    \item \textit{Notes:} Estimates of model \eqref{eq:update} for various outcomes $Y$: ``$ \widetilde{P}(N\geq 3) $, (A)'' represents the subjective probability of receiving at least 3 offers, ``Bias on (A)'' represents the bias wrt the variable of the first column, ``$B_i(0.5)$'' is the bias on the median of the distribution of the wage of job offers, ``Bias spread'' the bias on the spread of the latter distribution, ``Bias at $w^*$'' the bias at the declared reservation wage, and ``Marg. 20h'' the marginal returns when searching 20h/week. $Q1$ to $Q8+$ correspond to the quarters of unemployment. Significance levels: $^{*}$p$<$0.1; $^{**}$p$<$0.05; $^{***}$p$<$0.01.
\end{tablenotes}
\end{threeparttable}
\end{adjustbox}
\end{table}

\FloatBarrier

\begin{table}[!htbp]\centering
\caption{Informational content of job-finding expectations}
\label{tab:predictive}

\begingroup
\setlength{\tabcolsep}{3pt}    

\begin{adjustbox}{width = \textwidth} 
\begin{threeparttable}
\begin{tabular}{l*{9}{c}}
\hline
\hline
& \multicolumn{9}{c}{\textbf{Dependent variable: return to work before}} \\
\cline{2-10}
& \multicolumn{3}{c}{3 months} & \multicolumn{3}{c}{6 months} & \multicolumn{3}{c}{12 months} \\
& (1) & (2) & (3) & (4) & (5) & (6) & (7) & (8) & (9) \\
\hline

\textbf{Subjective job-finding prob. ($\psi_t$)} \\
3 months  & & 0.414$^{***}$ & 0.370$^{***}$ & & & & & &   \\   
                                   & & (0.015) & (0.016) & & &                &                &               &   \\  
6 months  & & & & & 0.410$^{***}$ & 0.341$^{***}$  & & &   \\   
 & & & & & (0.015) & (0.017) & & &   \\
12 months & & & & & & & & 0.395$^{***}$ & 0.284$^{***}$\\   
 & & & & & & & & (0.015)       & (0.017)\\
\addlinespace

\midrule
Job seekers' controls & \checkmark & & \checkmark & \checkmark & & \checkmark & \checkmark & & \checkmark \\
Region FE & \checkmark & & \checkmark & \checkmark & & \checkmark & \checkmark & & \checkmark \\
Occupation FE  & \checkmark & & \checkmark & \checkmark & & \checkmark & \checkmark & & \checkmark \\
Panel FE    & \checkmark & & \checkmark & \checkmark & & \checkmark & \checkmark & & \checkmark \\
\midrule
Num. Obs. & 25,811 & 25,846 & 25,811 & 25,811 & 25,846 & 25,811 & 25,811 & 25,594 & 25,558\\    
   Adjusted R$^2$ & 0.061 & 0.071       & 0.109      & 0.076        & 0.065       & 0.112        & 0.104        & 0.063       & 0.129\\  
\hline
\hline
\end{tabular}
\begin{tablenotes}[flushleft]
    \small
    \item \textit{Notes:} Linear probability models of returning to work before each horizon on subjective expectations and controls. Fixed effects for region, occupation, and panel wave are included. Job seekers' controls include age, gender, education level, qualification, work experience, spell duration, remaining benefit duration and passed applications. The sample pools Waves 1--8 of the panel survey (October 2021--July 2023) and is restricted to the 25,846 individuals who reported their perceived 3-month reemployment probability and a main occupation. $^{*}p<0.1$, $^{**}p<0.05$, $^{***}p<0.01$.
\end{tablenotes}
\end{threeparttable}
\end{adjustbox}

\endgroup
\end{table}

\begin{table}[!ht]
\centering
\caption{Within individual variation in expectations with unemployment duration}\label{table_update_duration_psi}
\scalebox{0.85}{
\begin{tabular}[t]{lccccc}
\toprule
  & 1m prob. & 3m prob. & 6m prob. & 12m prob. \\
  & (1) & (2) & (3) & (4) \\
\midrule
Q2 & 0.011 & 0.009 & -0.012 & -0.020 \\
 & (0.008) & (0.008) & (0.008) & (0.008)* \\
Q3 & 0.038 & 0.012 & -0.018 & -0.030 \\
 & (0.009)*** & (0.009) & (0.009)+ & (0.010)** \\
Q4 & 0.069 & 0.042 & 0.001 & -0.030 \\
 & (0.010)*** & (0.011)*** & (0.010) & (0.011)** \\
Q5 & 0.078 & 0.030 & -0.012 & -0.040 \\
 & (0.011)*** & (0.011)** & (0.011) & (0.012)*** \\
Q6 & 0.077 & 0.026 & -0.025 & -0.051 \\
 & (0.012)*** & (0.013)* & (0.012)* & (0.013)*** \\
Q7& 0.074 & 0.023 & -0.038 & -0.070 \\
 & (0.013)*** & (0.014)+ & (0.014)** & (0.014)*** \\
Q8 & 0.096 & 0.047 & -0.029 & -0.077 \\
 & (0.015)*** & (0.015)** & (0.015)+ & (0.015)*** \\
Q8+ & 0.091 & 0.035 & -0.056 & -0.112 \\
 & (0.015)*** & (0.016)* & (0.015)*** & (0.016)*** \\
\midrule
Num. Obs. & 14,814 & 14,897 & 15,073 & 14,928 \\
$R^2$ & 0.537 & 0.599 & 0.674 & 0.712 \\
\bottomrule
\end{tabular}
}
\caption*{\scriptsize Notes: Estimates of model \eqref{eq:update} for various outcomes $Y$: ``Xm prob.'' represents the subjective probability of reemployment after X months. $Q1$ to $Q8+$ correspond to the quarters of unemployment. Significance levels: $^{*}$p$<$0.1; $^{**}$p$<$0.05; $^{***}$p$<$0.01.}
\end{table}

\FloatBarrier
\newpage

\begin{figure}
\centering

\caption{Binned scatter plot between biases on the arrival rate of offers and the wage offer distribution}

\includegraphics[width=\textwidth]{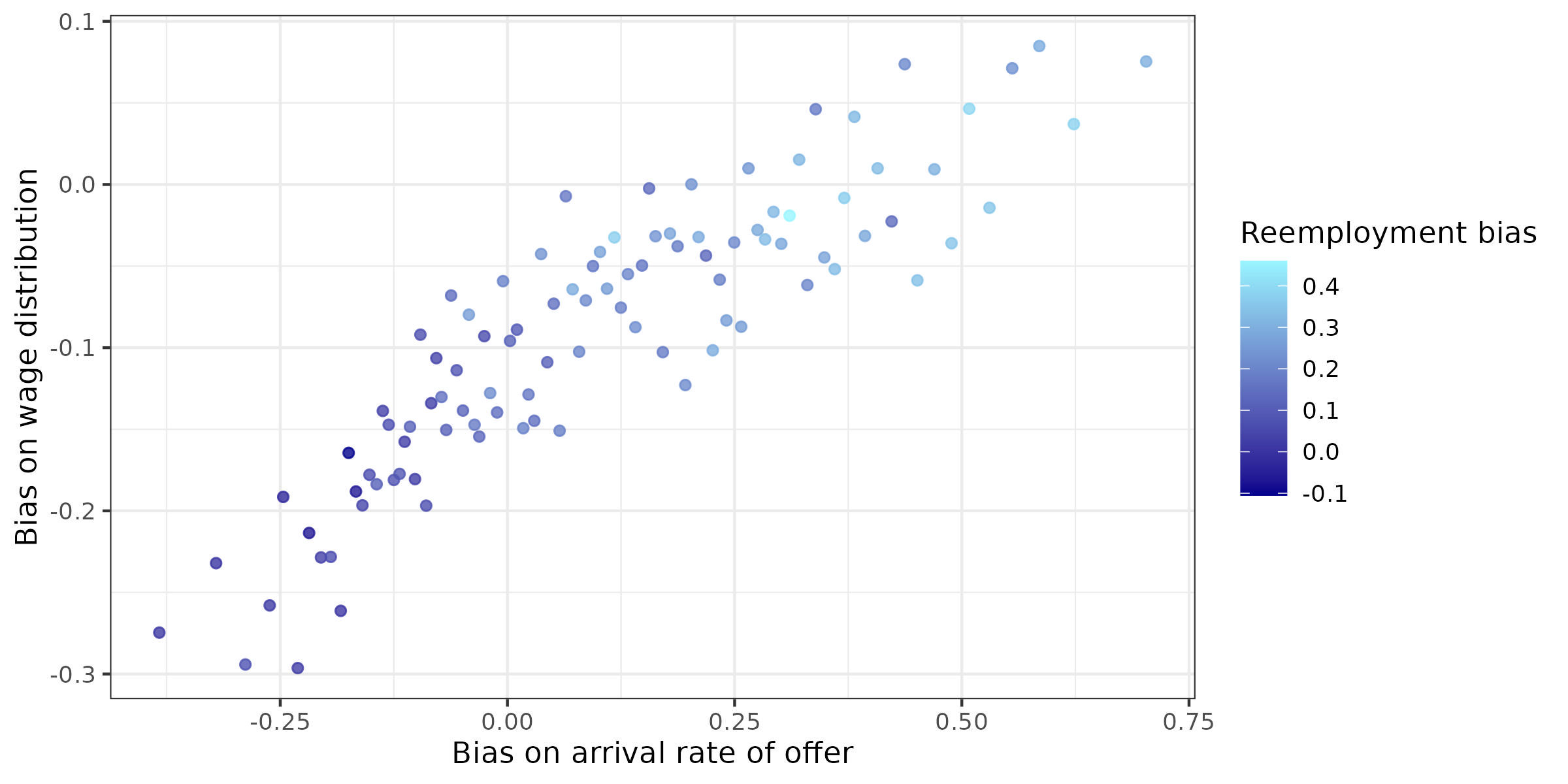}

\caption*{\footnotesize \textit{Notes:} “Bias on arrival rate of offers” refers to the raw difference between beliefs regarding the probability of receiving at least three job offers within the next three months and the corresponding objective prediction. We bin 15,343 observations into 100 cells defined by percentiles of the arrival-rate bias. Within each cell, we average both these biases and the biases on the median of the wage offer distribution and on the 3-month reemployment probability. The sample includes panel wave 1--8 restricting to individuals who answer questions about perceived 3-month reemployment probability and beliefs about wages and arrival rate of job offers. Weights are used.}

\label{fig:correlation_wage_arrival_biases}
\end{figure}

\begin{table}[!htpb]
    \centering
    \caption{Regression of reemployment biases on informational ones}
    \label{tab:reduced_form_jobfindingbias_perceptionbias}
\begin{adjustbox}{width = \textwidth}
\begin{threeparttable}

\begin{tabular}{@{\extracolsep{5pt}}lcccc} 
\\[-1.8ex]\hline 
\hline \\[-1.8ex] 
\\[-1.8ex] & \multicolumn{4}{c}{3 months reemployment bias : $\psi_3 - \indic\{T <3 \}$} \\ 

\\[-1.8ex] & (1) & (2) & (3) & (4) \\ \hline
Bias on arrival rate of offers : $\widetilde{P}(N\geq3) - P(N\geq3)$ & & 0.399$^{***}$  & 0.397$^{***}$  & 0.394$^{***}$\\   
                                   & & (0.029)        & (0.029)        & (0.029)\\   
Wage bias at the 1st quartile : $B_{.25}$ & & 0.050$^{**}$   &                &   \\   
                                   &                & (0.021)        &                &   \\   
Wage bias at the median : $B_{.5}$ &                &                & 0.052$^{**}$   &   \\   
                                   &                &                & (0.021)        &   \\   
Wage bias at the 3rd quartile : $B_{.75}$ &                &                &                & 0.065$^{***}$\\   
                                   &                &                &                & (0.020)\\   
\hline
Demographics & \checkmark & \checkmark & \checkmark & \checkmark \\
Target occupation fixed effects & \checkmark & \checkmark & \checkmark & \checkmark\\
Region fixed effects  & \checkmark & \checkmark & \checkmark & \checkmark\\
\hline
Adjusted R$^2$& 0.049        & 0.082        & 0.082        & 0.082\\  
   Num. Obs.                    & 15,293         & 15,293         & 15,293         & 15,293\\ 
\hline
\hline
\end{tabular}
\begin{tablenotes}[flushleft]
    \small 
    \item  \textit{Notes:} We regress an unbiased signal of 3-months reemployement bias on several key characteristics of job seekers (1). Then adding measures of misperception of labor market parameters improves the $R^2$. We test several measures of wage biases in columns (2-4). Each one corresponds to a different quartile level of the wage distribution. The sample includes panel wave 1--8 restricting to individuals who answer questions about perceived 3-month reemployment probability and beliefs about wages and arrival rate of job offers. Significance levels: $^{*}$p$<$0.1; $^{**}$p$<$0.05; $^{***}$p$<$0.01.
    
\end{tablenotes}
\end{threeparttable}
\end{adjustbox}
\caption*{\scriptsize}
\end{table}

\begin{table}[!htbp] \centering 
  \caption{Regressions of search effort on labor market beliefs} 
  \label{tab:search_behaviors_effort} 
\begin{adjustbox}{width = \textwidth}
\begin{threeparttable}
\begin{tabular}{l*{6}{c}}
\hline
\hline
\\[-1.8ex] & \multicolumn{5}{c}{Exerted search effort (nb. hours/week)} \\ 
\cline{2-7}
Sample & Waves 1--8 & \multicolumn{3}{c}{Waves 6--8} & \multicolumn{2}{c}{Waves 6--8} \\
Specification & \multicolumn{4}{c}{Cross section} & \multicolumn{2}{c}{Panel} \\
& (1) & (2) & (3) & (4) & (5) & (6) \\
\hline

$B_{.5}$ 
& 2.799$^{***}$ & 3.297$^{***}$ & 3.334$^{***}$  & 3.343$^{**}$ & 2.974$^{**}$ & 3.116$^{***}$ \\
& (0.5684)      & (1.018)       & (0.9979)       & (1.325) & (1.225) & (1.170)\\

$B_{.75}-B_{.25}$ 
& 2.016$^{**}$  & 4.478$^{**}$  & 4.423$^{**}$   & 2.285 & 1.248 & $-$0.227 \\ 
& (0.8878)      & (1.748)       & (1.762)        & (2.371) & (2.052) & (1.550) \\ 

$\widetilde{p_N}(3)-p_N(3)$
& 1.971$^{**}$  &  &  &    &  &  \\
& (0.7956) &  &  &  &  &  \\

$p_{N|e=20}(1)$
 & & 1.580 & & 0.194 & $-$2.553  &  \\
& & (1.065) & & (1.504) & (2.093) &  \\

$p_{N|e=30}(1)$ - $p_{N|e=10}(1)$
& & & 5.890$^{***}$  &   &  & 9.374$^{***}$ \\ 
&  &  & (1.467) &  &  & (2.687) \\

Spell duration (months)
&  &  &  &  & 0.130$^{**}$ & 0.098$^{*}$ \\ 
&  &  &  &  & (0.061) & (0.057) \\ 

\hline

Region fixed effects & \checkmark & \checkmark & \checkmark & \checkmark &  &  \\
Target occupation fixed effects & \checkmark & \checkmark & \checkmark & \checkmark &  &  \\
Panel fixed effects & \checkmark & \checkmark & \checkmark & \checkmark &  &  \\
Psychological traits &  &  &  & \checkmark &  &  \\
Individual fixed effects &  &  &  &  & \checkmark & \checkmark \\

\hline

Num. Obs. & 7,078 & 2,064 & 2,021 & 1,018 & 1,248 & 1,251 \\
Job seekers  & 7,078 & 2,064 & 2,021 & 1,018 & 494 & 494 \\

\hline
\hline
\end{tabular}
\begin{tablenotes}[flushleft]
    \small 
    \item \textit{Notes:} This table presents estimate of the influence of subjective beliefs on search effort. The first four columns are cross-sectional regressions. Beliefs about the wage distribution are measured by the bias at the median ($B_{.5}$) and the bias at the interquartile range of the distribution ($B_{.75} - B_{.25}$). Measures of perceived offer arrival rates include the bias in the probability of receiving at least three job offers within three months, the perceived arrival rate of job offers conditional on 20 hours weekly effort on average and the increase of the arrival rate of offers with additional search effort. Column (4) additionally controls for psychological traits. Columns (5)--(6) use panel data with individual fixed effects. Weights are used. Standard errors in parentheses are either heteroskedastic robust (columns (1)--(4)) or clustered at the individual level (columns (5)--(6)). $^{*}p<0.1$, $^{**}p<0.05$, $^{***}p<0.01$.
\end{tablenotes}
\end{threeparttable}
\end{adjustbox}
\end{table}

\begin{table}[!htbp] 
\centering 
\caption{Regression of labor market beliefs on psychological traits} 
\label{tab:psycho_perceptions} 
\begin{adjustbox}{width = .9\textwidth}
\begin{threeparttable}
\begin{tabular}{lcccccc} \\
[-1.8ex]\hline 
\hline \\
[-1.8ex] & \multicolumn{6}{c}{\textit{Beliefs}} \\ 
\cline{2-7} \\[-1.8ex] 
& \multicolumn{3}{c}{Conditional arrival rate} & \multicolumn{3}{c}{Return to effort} \\ 
\\[-1.8ex] & (1) & (2) & (3) & (4) & (5) & (6)\\
\midrule \\[-1.8ex] Internal locus & & 0.037$^{***}$ & & & 0.015$^{**}$ & \\ 
& & (0.009) & & & (0.007) & \\ 
External locus & & -0.024$^{**}$ & & & -0.004 & \\ 
& & (0.010) & & & (0.007) & \\
Conscientiousness & & & 0.021$^{**}$ & & & 0.006\\ 
& & & (0.009) & & & (0.007)\\
Neuroticism & & & -0.008 & & & -0.002\\
& & & (0.009) & & & (0.007)\\ 
\midrule \\
Demographic controls & \checkmark & \checkmark & \checkmark & \checkmark & \checkmark & \checkmark  \\ 
Region fixed effect & \checkmark & \checkmark & \checkmark & \checkmark & \checkmark & \checkmark  \\ 
Target occupation fixed effect & \checkmark & \checkmark & \checkmark & \checkmark & \checkmark & \checkmark\\ 
Panel fixed effect & \checkmark & \checkmark & \checkmark & \checkmark & \checkmark & \checkmark \\
\midrule
Sample mean & \multicolumn{3}{c}{0.47} & \multicolumn{3}{c}{0.18} \\ 
Num.Obs. & 2,098 & 2,098 & 2,098 & 2,112 & 2,112 & 2,112\\ 
Adjusted R$^{2}$ & 0.094 & 0.125 & 0.103 & 0.051 & 0.058 & 0.052\\ 
\hline 
\hline
\end{tabular}
\begin{tablenotes}[flushleft]
    \small
    \item \textit{Notes:} This table reports the regression of two different measures of beliefs regarding the arrival rate of offers on psychological traits: arrival rate of offers conditional on 20 hours search per week (elicited by $p_{N |e = 20}(1)$)  in columns (1)--(3), perceived probability for an application to turn into a job offer from the recruiters in columns (4)--(6) and perceived return to search effort on the arrival rate of offer (elicited by $p_{N |e = 30}(1) - p_{N |e = 10}(1)$) in columns (7)--(9). Psychological measures are built thanks to specific items in the survey, each of them is normalized and standardized. Significance levels: $^{*}$p$<$0.1; $^{**}$p$<$0.05; $^{***}$p$<$0.01.
\end{tablenotes}
\end{threeparttable}
\end{adjustbox}
\end{table}

\FloatBarrier

\section{Structural Model}

\renewcommand{\thetable}{\thesection.\arabic{table}}
\setcounter{table}{0}

\renewcommand{\thefigure}{\thesection.\arabic{figure}}
\setcounter{figure}{0}

\begin{table}[!htbp]
\centering
\caption{Empirical moments and simulated counterparts}
\label{tab:moments_matching}

\begin{adjustbox}{width=\textwidth}
\begin{threeparttable}

\begin{tabular}{lc|c|c}
\hline
\textbf{Moment} & & \textbf{Empirical value} & \textbf{Simulated value} \\
\hline

\multicolumn{4}{l}{\textbf{Average 3-month job-finding rate}} \\

1. High perceived employability, $D<6$ months &
$\mathbb{E}\!\left[\indic\{T\leq D+3\}\mid \widetilde{\lambda}_0 > m(\widetilde{\lambda}_0),\, D<6\right]$
& 0.437 & 0.427 \\

2. High perceived employability, $D\geq 6$ months &
$\mathbb{E}\!\left[\indic\{T\leq D+3\}\mid \widetilde{\lambda}_0 > m(\widetilde{\lambda}_0),\, D\geq 6\right]$
& 0.371 & 0.362 \\

3. Low perceived employability, $D<6$ months &
$\mathbb{E}\!\left[\indic\{T\leq D+3\}\mid \widetilde{\lambda}_0 \leq m(\widetilde{\lambda}_0),\, D<6\right]$
& 0.301 & 0.306 \\

4. Low perceived employability, $D\geq 6$ months &
$\mathbb{E}\!\left[\indic\{T\leq D+3\}\mid \widetilde{\lambda}_0 \leq m(\widetilde{\lambda}_0),\, D\geq 6\right]$
& 0.179 & 0.176 \\

\addlinespace

5. \textbf{Average 3-month reemployment bias} &
$\mathbb{E}\!\left[\psi_3-\indic\{T\leq D+3\}\right]$
& 0.215 & 0.180 \\

\addlinespace

\multicolumn{4}{l}{\textbf{Average weekly search effort}} \\

6. High perceived employability &
$\mathbb{E}\!\left[e^* \mid \widetilde{\lambda}_0 > m(\widetilde{\lambda}_0)\right]$
& 16.3 & 16.3 \\

7. Low perceived employability &
$\mathbb{E}\!\left[e^* \mid \widetilde{\lambda}_0 \leq m(\widetilde{\lambda}_0)\right]$
& 13.4 & 13.5 \\

\addlinespace

8. \textbf{Wage acceptance rate} &
$\mathbb{E}\!\left[1-F_W(w^*)\right]$
& 0.547 & 0.878 \\

\hline
\end{tabular}

\begin{tablenotes}[flushleft]
\small
\item \textit{Notes:} The table compares empirical moments with their simulated counterparts generated by the estimated model. The variable $T$ denotes unemployment duration (in months) until reemployment, $D$ is elapsed unemployment duration at the time of the survey, $\widetilde{\lambda}_0$ is the perceived job-offer arrival rate at baseline, $m(\widetilde{\lambda}_0)$ denotes its sample median, $\psi_3$ is the individual's perceived probability of finding a job within the next three months, $e^*$ denotes weekly search effort, and $w^*$ is the reservation wage. Empirical moments are computed using respondents in Panels 6--8 whose reported beliefs satisfy basic consistency restrictions: perceived job-finding probabilities are weakly increasing in the prediction horizon, perceived offer arrival rates are weakly increasing in search effort, and perceived wage offer distributions are valid cumulative distribution functions. Sampling weights are not used.
\end{tablenotes}

\end{threeparttable}
\end{adjustbox}

\end{table}

\begin{table}[!htbp]
\centering
\caption{Estimated parameters and 95\% confidence intervals}
\label{tab:model_parameters}

\begin{threeparttable}

\begin{tabular}{lccc}
\hline
\textbf{Parameter} & \textbf{Notation} & \textbf{Estimate} & \textbf{95\% CI} \\
\hline

Share of low types &
$\phi$
& 0.17501
& [0.17501, 0.17501] \\

Low-type employability parameter &
$\lambda^{0,l}$
& 0.00183
& [0.00180, 0.00185] \\

High-type employability parameter &
$\lambda^{0,h}$
& 0.01396
& [0.01392, 0.01400] \\

Classical search-cost parameter &
$c_1$
& 0.03839
& [0.03795, 0.03883] \\

Behavioral search-cost parameter ($\times 10^{5}$) &
$c_{\lambda}$
& 2.50
& [2.47, 2.52] \\

Arrival-rate depreciation parameter &
$\theta$
& 1.0000
& [0.9991, 1.0009] \\

Unemployment benefit parameter &
$b_u$
& 0.08204
& [0.08136, 0.08271] \\

\hline
\end{tabular}

\begin{tablenotes}[flushleft]
\small
\item \textit{Notes:} The table reports parameter estimates obtained by generalized method of moments (GMM). Confidence intervals are computed using the estimated asymptotic covariance matrix of the parameter estimates. For presentation purposes, the estimate of $c_{\lambda}$ is multiplied by $10^{5}$.
\end{tablenotes}

\end{threeparttable}
\end{table}

\begin{figure}[!htbp]
    \centering
    \caption{Informational biases and 3-month reemployment bias in the simulated sample}
    \includegraphics[width=\textwidth]{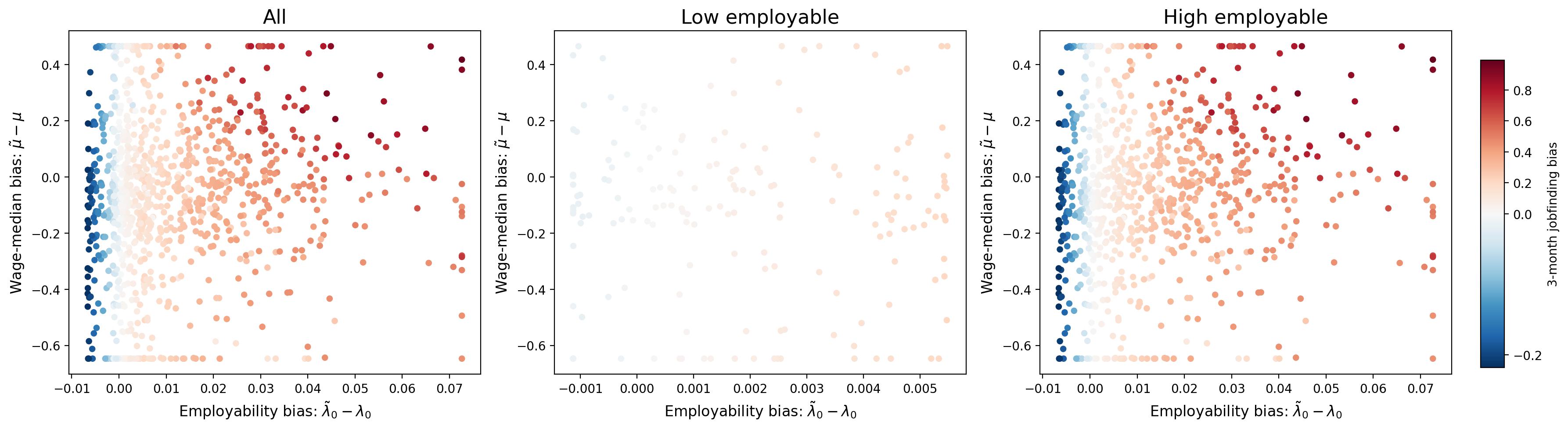}
    \label{fig:pepsi}
    
    \caption*{\footnotesize
    \textit{Notes:} This figure displays individual informational biases and reemployment biases in the simulated sample. Wage-offer biases are measured as the difference between the perceived and true median of the wage-offer distribution. Employability biases are measured as the difference between the perceived and true job-offer arrival rate (employability parameter). The 3-month reemployment bias is defined as the difference between the perceived probability of finding a job within three months and the corresponding true probability implied by the model.}
\end{figure}

\begin{table}[!htbp]
\centering
\caption{Average labor market outcomes under counterfactual reductions in informational biases}
\label{tab:avg_counterfactual}

\begin{adjustbox}{width=\textwidth}
\begin{threeparttable}

\begin{tabular}{lccc}
\hline
\hline
& \textbf{Search effort} & \textbf{Acceptance rate} & \textbf{3-month job-finding probability} \\
\hline

No treatment (A) & 15.170 & 0.896 & 0.398 \\

Reduce employability bias by 50\% (B) & 15.013 & 0.947 & 0.419 \\

Reduce wage-offer bias by 50\% (C) & 15.181 & 0.910 & 0.406 \\

Reduce both biases by 50\% (D) & 15.019 & 0.961 & 0.426 \\

\hline
\end{tabular}

\begin{tablenotes}[flushleft]
\small
\item \textit{Notes:} The table reports average simulated labor market outcomes under alternative counterfactual scenarios. In the baseline scenario (A), agents retain their estimated informational biases. In scenarios (B) and (C), employability biases and wage-offer biases are reduced by 50\%, respectively. Scenario (D) simultaneously reduces both sources of bias by 50\%. Reported outcomes correspond to the first period of the unemployment spell and include average search effort, the wage-offer acceptance rate, and the probability of finding a job within three months.
\end{tablenotes}

\end{threeparttable}
\end{adjustbox}
\end{table}

\begin{table}[!htbp]
\centering
\caption{Average labor market outcomes under counterfactual reductions of 50\% in informational biases, by employability type and reemployment-bias quartile}
\label{tab:conditional_counterfactual}

\begin{adjustbox}{width=\textwidth}
\begin{threeparttable}

\begin{tabular}{lllrrr}
\toprule
\textbf{Bias quartile} &
\textbf{Type} &
\textbf{Counterfactual} &
\textbf{Search effort} &
\textbf{Acceptance rate} &
\textbf{3-month job-finding probability} \\
\midrule

\multirow{8}{*}{Quartile 1}
& \multirow{4}{*}{High}
& No treatment (A) & 15.010 & 0.997 & 0.508 \\
& & Reduce employability bias (B) & 15.086 & 0.998 & 0.509 \\
& & Reduce wage-offer bias (C) & 15.013 & 1.000 & 0.509 \\
& & Reduce both biases (D) & 15.085 & 1.000 & 0.510 \\

& \multirow{4}{*}{Low}
& No treatment (A) & 7.019 & 1.000 & 0.037 \\
& & Reduce employability bias (B) & 9.387 & 1.000 & 0.050 \\
& & Reduce wage-offer bias (C) & 7.017 & 1.000 & 0.037 \\
& & Reduce both biases (D) & 9.383 & 1.000 & 0.050 \\

\midrule

\multirow{8}{*}{Quartile 2}
& \multirow{4}{*}{High}
& No treatment (A) & 15.501 & 0.975 & 0.510 \\
& & Reduce employability bias (B) & 15.341 & 0.990 & 0.512 \\
& & Reduce wage-offer bias (C) & 15.509 & 0.980 & 0.512 \\
& & Reduce both biases (D) & 15.348 & 0.994 & 0.515 \\

& \multirow{4}{*}{Low}
& No treatment (A) & 13.672 & 1.000 & 0.073 \\
& & Reduce employability bias (B) & 12.892 & 1.000 & 0.069 \\
& & Reduce wage-offer bias (C) & 13.677 & 1.000 & 0.073 \\
& & Reduce both biases (D) & 12.894 & 1.000 & 0.069 \\

\midrule

\multirow{8}{*}{Quartile 3}
& \multirow{4}{*}{High}
& No treatment (A) & 15.990 & 0.916 & 0.497 \\
& & Reduce employability bias (B) & 15.643 & 0.973 & 0.513 \\
& & Reduce wage-offer bias (C) & 16.015 & 0.917 & 0.498 \\
& & Reduce both biases (D) & 15.666 & 0.977 & 0.515 \\

& \multirow{4}{*}{Low}
& No treatment (A) & 14.766 & 1.000 & 0.079 \\
& & Reduce employability bias (B) & 14.359 & 1.000 & 0.077 \\
& & Reduce wage-offer bias (C) & 14.770 & 1.000 & 0.079 \\
& & Reduce both biases (D) & 14.364 & 1.000 & 0.077 \\

\midrule

\multirow{4}{*}{Quartile 4}
& \multirow{4}{*}{High}
& No treatment (A) & 16.496 & 0.673 & 0.387 \\
& & Reduce employability bias (B) & 16.098 & 0.820 & 0.453 \\
& & Reduce wage-offer bias (C) & 16.510 & 0.722 & 0.413 \\
& & Reduce both biases (D) & 16.098 & 0.866 & 0.476 \\

\bottomrule
\end{tabular}

\begin{tablenotes}[flushleft]
\small
\item \textit{Notes:} The table reports average simulated labor market outcomes under counterfactual scenarios in which informational biases are reduced by 50\%. Quartiles are defined using the distribution of ex-ante reemployment bias in the simulated sample. Within each quartile and employability type, outcomes are averaged using equal weights. Scenario (A) corresponds to the baseline model. Scenarios (B) and (C) reduce employability biases and wage-offer biases by 50\%, respectively, while scenario (D) reduces both biases by 50\%. The acceptance rate is defined as the probability that an arriving wage offer exceeds the reservation wage. The reported job-finding probability corresponds to the probability of reemployment within three months from the start of the unemployment spell. Quartile 4 contains only high-employability individuals.
\end{tablenotes}

\end{threeparttable}
\end{adjustbox}
\end{table}

\begin{table}[!htbp]
\centering
\caption{Average labor market outcomes under full correction of informational biases, by employability type and reemployment-bias quartile}
\label{tab:conditional_counterfactual_100}

\begin{adjustbox}{width=\textwidth}
\begin{threeparttable}

\begin{tabular}{lllrrr}
\toprule
\textbf{Bias quartile} &
\textbf{Type} &
\textbf{Counterfactual} &
\textbf{Search effort} &
\textbf{Acceptance rate} &
\textbf{3-month job-finding probability} \\
\midrule

\multirow{8}{*}{Quartile 1}
& \multirow{4}{*}{High}
& No treatment (A) & 15.010 & 0.997 & 0.508 \\
& & Correct employability bias (B) & 15.186 & 0.995 & 0.510 \\
& & Correct wage-offer bias (C) & 15.022 & 1.000 & 0.509 \\
& & Correct both biases (D) & 15.190 & 1.000 & 0.512 \\

& \multirow{4}{*}{Low}
& No treatment (A) & 7.019 & 1.000 & 0.037 \\
& & Correct employability bias (B) & 10.928 & 1.000 & 0.058 \\
& & Correct wage-offer bias (C) & 7.015 & 1.000 & 0.037 \\
& & Correct both biases (D) & 10.934 & 1.000 & 0.058 \\

\midrule

\multirow{8}{*}{Quartile 2}
& \multirow{4}{*}{High}
& No treatment (A) & 15.501 & 0.975 & 0.510 \\
& & Correct employability bias (B) & 15.181 & 0.996 & 0.511 \\
& & Correct wage-offer bias (C) & 15.525 & 0.979 & 0.512 \\
& & Correct both biases (D) & 15.201 & 1.000 & 0.513 \\

& \multirow{4}{*}{Low}
& No treatment (A) & 13.672 & 1.000 & 0.073 \\
& & Correct employability bias (B) & 10.926 & 1.000 & 0.058 \\
& & Correct wage-offer bias (C) & 13.691 & 1.000 & 0.073 \\
& & Correct both biases (D) & 10.936 & 1.000 & 0.058 \\

\midrule

\multirow{8}{*}{Quartile 3}
& \multirow{4}{*}{High}
& No treatment (A) & 15.990 & 0.916 & 0.497 \\
& & Correct employability bias (B) & 15.178 & 0.997 & 0.511 \\
& & Correct wage-offer bias (C) & 16.025 & 0.902 & 0.492 \\
& & Correct both biases (D) & 15.194 & 1.000 & 0.513 \\

& \multirow{4}{*}{Low}
& No treatment (A) & 14.766 & 1.000 & 0.079 \\
& & Correct employability bias (B) & 10.927 & 1.000 & 0.058 \\
& & Correct wage-offer bias (C) & 14.780 & 1.000 & 0.080 \\
& & Correct both biases (D) & 10.936 & 1.000 & 0.058 \\

\midrule

\multirow{4}{*}{Quartile 4}
& \multirow{4}{*}{High}
& No treatment (A) & 16.496 & 0.673 & 0.387 \\
& & Correct employability bias (B) & 15.202 & 0.993 & 0.510 \\
& & Correct wage-offer bias (C) & 16.522 & 0.737 & 0.422 \\
& & Correct both biases (D) & 15.179 & 1.000 & 0.512 \\

\bottomrule
\end{tabular}

\begin{tablenotes}[flushleft]
\small
\item \textit{Notes:} The table reports average simulated labor market outcomes under counterfactual scenarios in which informational biases are fully corrected. Quartiles are defined using the distribution of ex-ante reemployment bias in the simulated sample. Within each quartile and employability type, outcomes are averaged using equal weights. Scenario (A) corresponds to the baseline model. Scenarios (B) and (C) fully correct employability biases and wage-offer biases, respectively, while scenario (D) fully corrects both sources of bias. The acceptance rate is defined as the probability that an arriving wage offer exceeds the reservation wage. The reported job-finding probability corresponds to the probability of reemployment within three months from the start of the unemployment spell. Quartile 4 contains only high-employability individuals.
\end{tablenotes}

\end{threeparttable}
\end{adjustbox}
\end{table}

\FloatBarrier

\section{Classification results}
\label{app:glm_xgb}

\renewcommand{\thetable}{\thesection.\arabic{table}}
\setcounter{table}{0}

\renewcommand{\thefigure}{\thesection.\arabic{figure}}
\setcounter{figure}{0}

\begin{table}[!htbp]
\centering
\caption{Comparison of causal machine-learning methods}
\label{tab:ML_comparison}

\begin{adjustbox}{width=0.8\textwidth}
\begin{threeparttable}

\begin{tabular}{lcccc}
\hline
\hline
& Elastic Net & XGBoost & Neural Network & Random Forest \\
\hline

$\widehat{\beta}_2$
& 1.0469
& 1.0414
& 0.8018
& 0.9380 \\
& [0.9690, 1.1311]
& [0.9645, 1.1084]
& [0.7564, 0.8551]
& [0.8470, 1.1015] \\

\cmidrule{2-5}

BLP indicator
& 0.0313
& 0.0329
& 0.0278
& 0.0316 \\
& [0.0293, 0.0332]
& [0.0304, 0.0344]
& [0.0261, 0.0297]
& [0.0293, 0.0332] \\

GATES indicator
& 0.251
& 0.253
& 0.237
& 0.248 \\
& [0.237, 0.264]
& [0.239, 0.267]
& [0.226, 0.249]
& [0.238, 0.261] \\

\hline
\hline
\end{tabular}

\begin{tablenotes}[flushleft]
\small
\item \textit{Notes:} The table compares four machine-learning methods used to estimate heterogeneous treatment effects. For each method, the Best Linear Predictor (BLP) and Group Average Treatment Effects (GATES) statistics are computed on each sample split. $\widehat{\beta}_2$ corresponds to estimated heterogeneity coefficient in the BLP. Then ``BLP indicator'' and  ``GATES indicator'' are the two indicators used to select the best performing ML method. The BLP based indicator  is defined as
\[
|\widehat{\beta}_2|^2 \operatorname{Var}\!\bigl(\delta(X)\bigr),
\]
where $\delta(X)$ denotes the predicted treatment effect. The GATES based indicator is defined as
\[
\sum_{k=1}^{K}\widehat{\overline{\gamma}}(G_{k,M}^{*}),
\]
where the predicted treatment effects are partitioned into $K=4$ quartile groups. Reported point estimates correspond to the median across sample splits. Values in brackets correspond to the 2.5th and 97.5th percentiles of the distribution across sample splits.
\end{tablenotes}

\end{threeparttable}
\end{adjustbox}
\end{table}

\begin{figure}[!htbp]
\centering
\caption{Illustration of the classification methodology (XGBoost)}
\label{fig:steps_xgb}

\begin{subfigure}{0.45\linewidth}
    \centering
    \includegraphics[width=\linewidth]{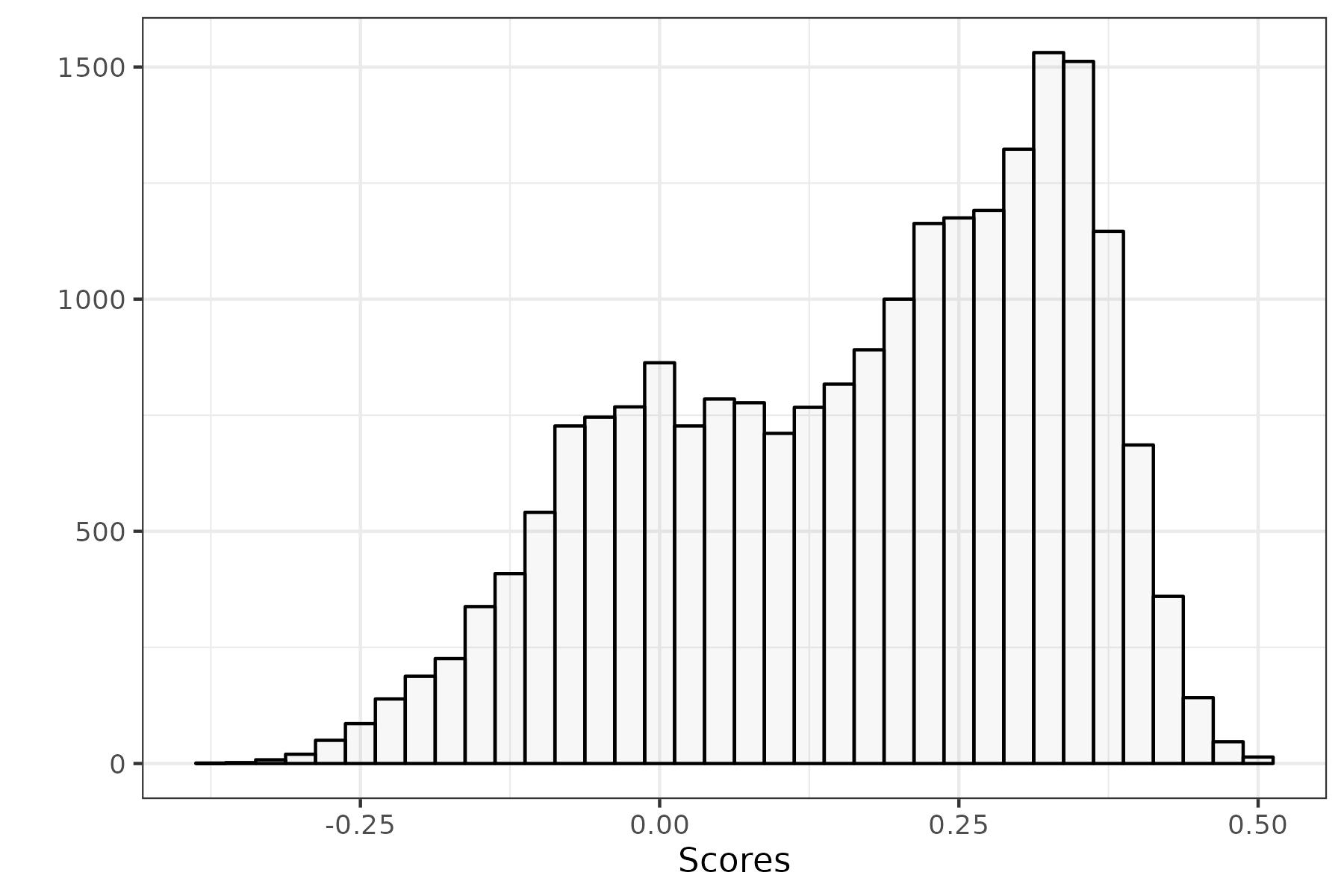}
    \caption{Distribution of predicted scores}
    \label{fig:steps_a_xgb}
\end{subfigure}
\hfill
\begin{subfigure}{0.45\linewidth}
    \centering
    \includegraphics[width=\linewidth]{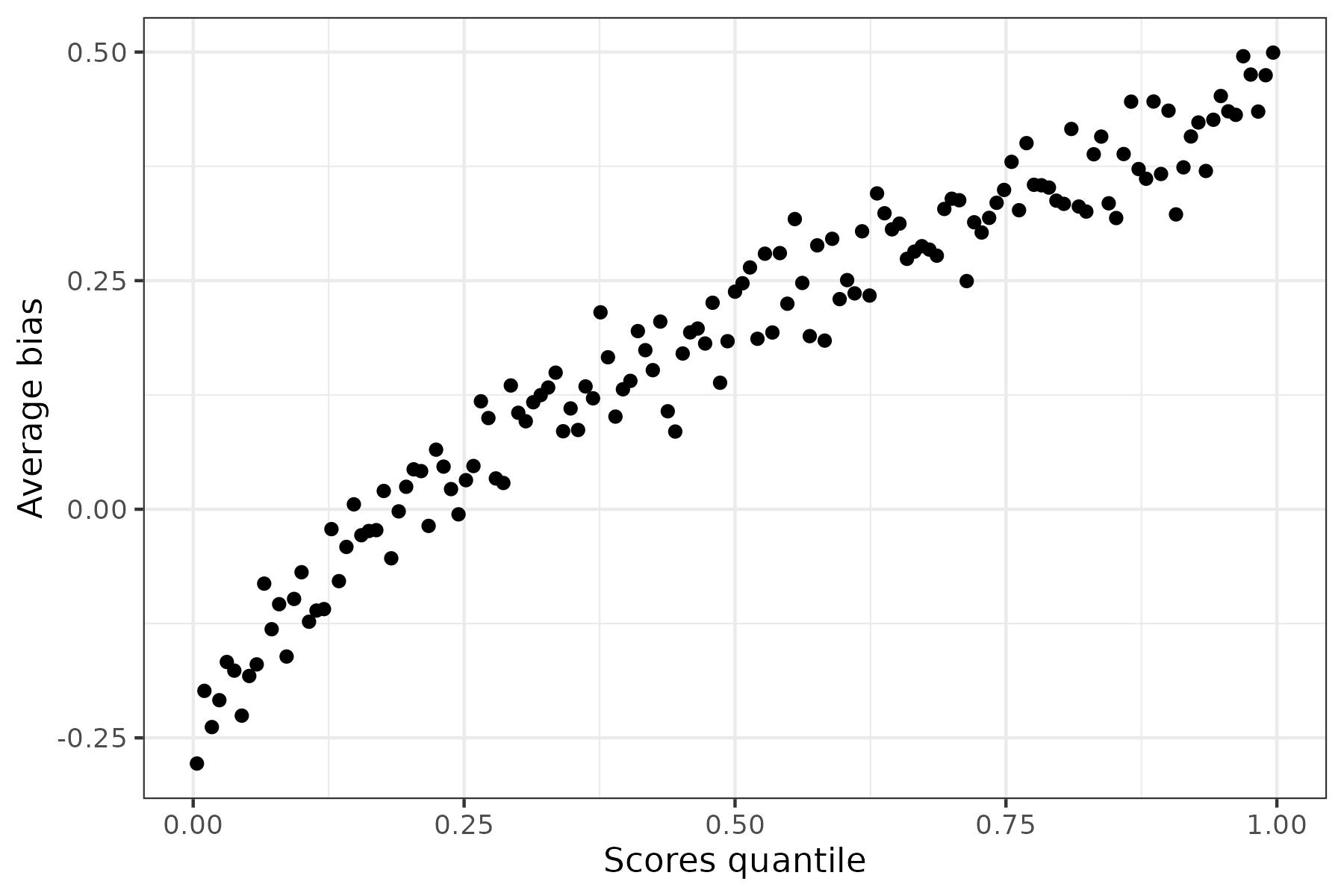}
    \caption{Average bias by quantile, $\widetilde{\gamma}(I_{j,M})$}
    \label{fig:steps_b_xgb}
\end{subfigure}

\vspace{0.3cm}

\begin{subfigure}{0.45\linewidth}
    \centering
    \includegraphics[width=\linewidth]{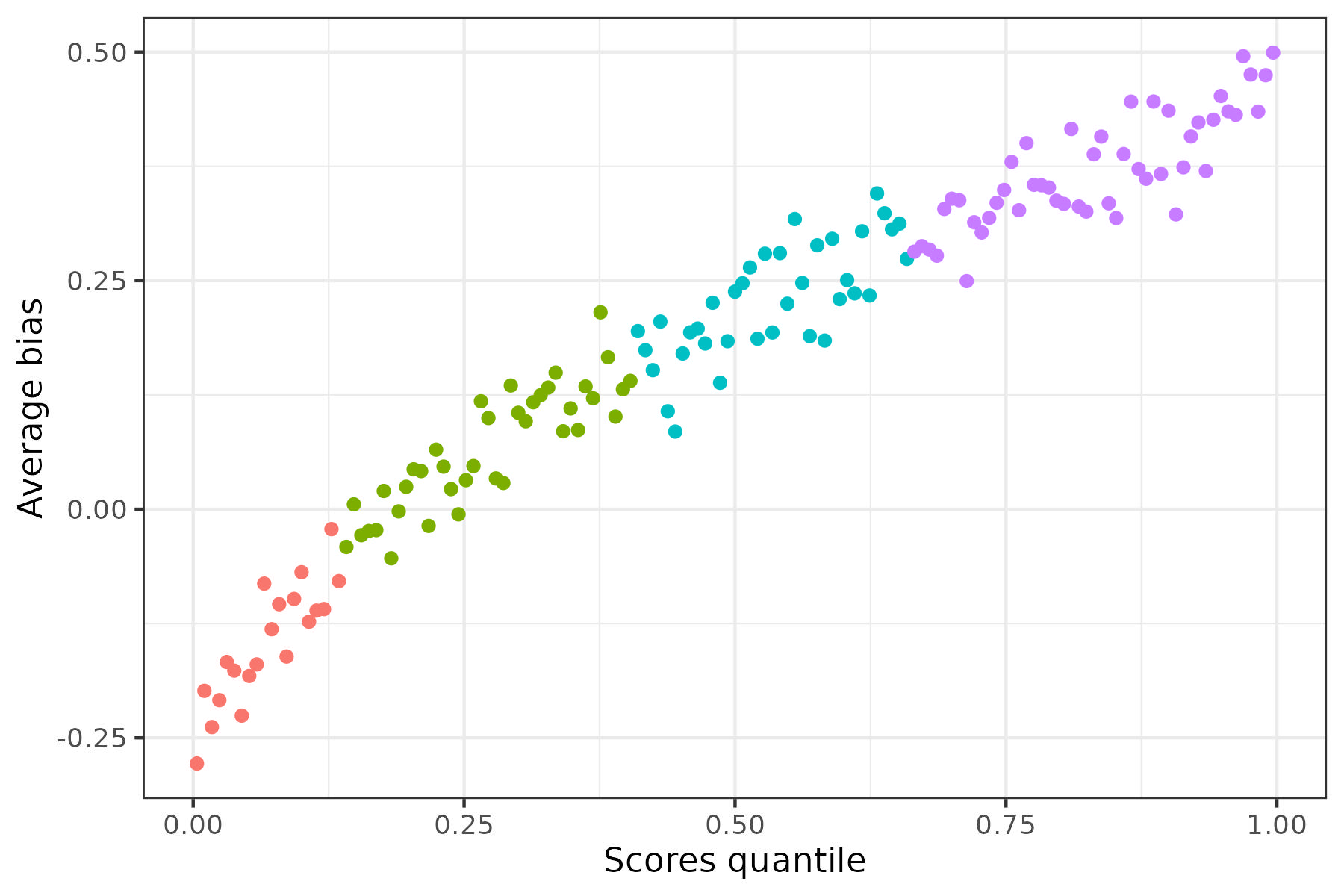}
    \caption{Clustering of average biases}
    \label{fig:steps_c_xgb}
\end{subfigure}
\hfill
\begin{subfigure}{0.45\linewidth}
    \centering
    \includegraphics[width=\linewidth]{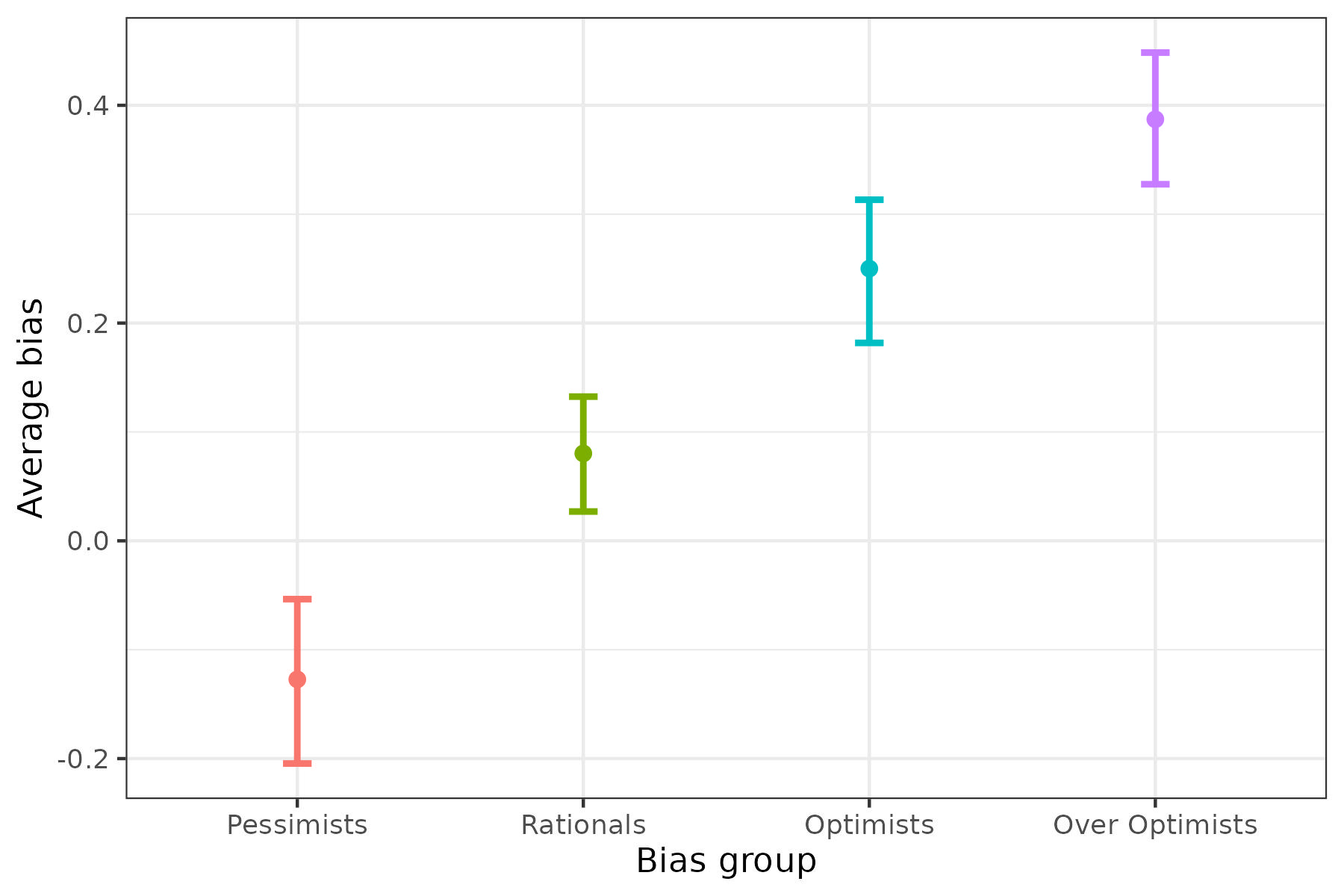}
    \caption{Average bias by group, $G_k$}
    \label{fig:steps_d_xgb}
\end{subfigure}

\caption*{\footnotesize
\textit{Notes:} This figure illustrates the successive steps of the classification procedure. Panel (a) displays the distribution of the machine-learning score used to partition the sample. Panel (b) reports the estimated average 3-month reemployment bias, $\widetilde{\gamma}(I_{j,M})$, for each cell of the partition $\mathcal{I}^M$. Panel (c) shows the same estimates after grouping adjacent cells into four clusters according to their average reemployment bias. Panel (d) reports the resulting average 3-month reemployment bias within each group $G_k$. Point estimates and confidence intervals are obtained by repeating the procedure over 100 sample splits; the median estimate across splits is reported. Panels (a)--(c) are illustrative and are constructed using the median predicted score for each individual across sample splits. The XGBoost algorithm is used to predict bias groups.
}
\end{figure}

\begin{figure}[!htbp]
    \centering
    \caption{Perceived and realized job-finding probabilities by GML group (Random Forest)}
    \includegraphics[width=\textwidth]{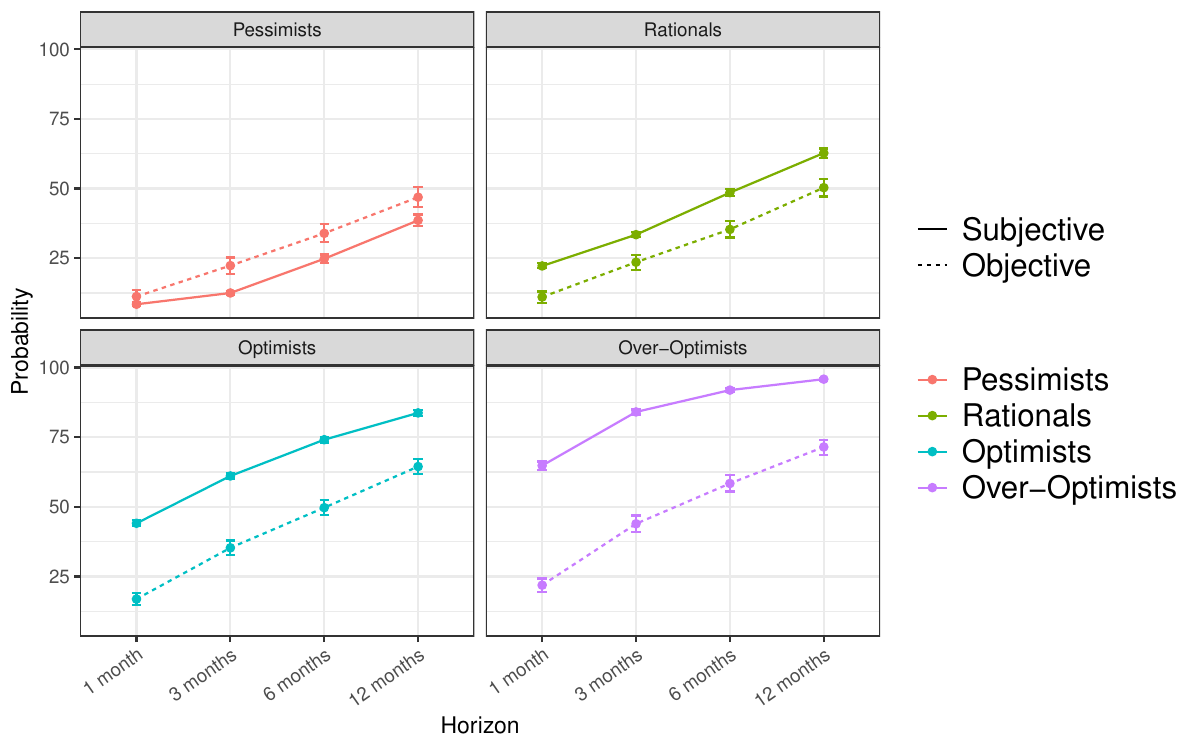}
    \label{fig:GLM_bias_grps_horizon}

    \caption*{\footnotesize
    \textit{Notes:} This figure reports average perceived and realized job-finding probabilities by reemployment-bias group for different prediction horizons. Realized job-finding probabilities are computed as the proportion of individuals who are reemployed by the corresponding horizon after the survey date. Group membership is predicted using the Random Forest algorithm and is estimated separately across 100 sample splits. For each split, group-specific means and confidence intervals are computed; the reported estimates correspond to the median across splits. The sample is restricted to individuals with consistent beliefs regarding reemployment probabilities.
    }
\end{figure}

\begin{table}[!htbp]
\centering
\footnotesize
\caption{Demographic characteristics and unemployment histories by reemployment-bias group (XGBoost)}
\label{tab:bias_group_stat_desc_xgb}

\begin{adjustbox}{width=\textwidth}
\begin{threeparttable}

\begin{tabular}{lcccccccc}
\hline
\hline
& \multicolumn{2}{c}{Pessimist} & \multicolumn{2}{c}{Rational} & \multicolumn{2}{c}{Optimist} & \multicolumn{2}{c}{Over-optimist} \\
\midrule

\textbf{Women (\%)} & 54.60 & [50.7, 58.4] & 52.70 & [49.7, 55.7] & 52.10 & [49.0, 55.2] & 47.60 & [45.0, 50.3] \\

\addlinespace
\textbf{Age} & 41.60 & [40.6, 42.6] & 43.10 & [42.3, 43.9] & 38.80 & [38.1, 39.5] & 37.80 & [37.2, 38.4] \\
$\leq 25$ & 11.90 & [9.1, 14.8] & 9.40 & [7.5, 11.4] & 12.30 & [10.0, 14.5] & 11.80 & [9.8, 13.7] \\
25--35 & 24.60 & [21.2, 28.1] & 24.40 & [21.8, 27.0] & 32.70 & [29.6, 35.7] & 35.60 & [32.9, 38.2] \\
35--45 & 21.00 & [17.8, 24.3] & 21.60 & [19.1, 24.0] & 23.60 & [21.0, 26.2] & 27.00 & [24.7, 29.3] \\
45--55 & 23.00 & [19.9, 26.2] & 18.70 & [16.6, 20.8] & 19.70 & [17.4, 22.1] & 17.60 & [15.7, 19.4] \\
$> 55$ & 19.10 & [17.1, 21.0] & 25.70 & [24.1, 27.2] & 11.20 & [10.1, 12.4] & 8.00 & [7.1, 8.9] \\

\addlinespace
\textbf{Number of children} & 0.79 & [0.70, 0.88] & 0.81 & [0.75, 0.88] & 0.82 & [0.75, 0.89] & 0.82 & [0.76, 0.88] \\
\textbf{Married (\%)} & 40.70 & [37.0, 44.5] & 47.80 & [44.9, 50.7] & 44.00 & [41.0, 47.1] & 43.80 & [41.3, 46.5] \\

\addlinespace
\textbf{Education} & & & & & & & & \\
High school & 24.50 & [21.0, 27.8] & 22.40 & [20.0, 24.9] & 24.10 & [21.5, 26.9] & 23.90 & [21.6, 26.2] \\
High school + 2 years of college & 16.60 & [13.7, 19.5] & 15.40 & [13.3, 17.5] & 17.10 & [14.7, 19.4] & 17.60 & [15.5, 19.6] \\
Bachelor's degree & 9.60 & [7.6, 11.8] & 11.60 & [9.8, 13.4] & 11.10 & [9.2, 12.9] & 14.60 & [12.7, 16.4] \\
Master's degree & 8.60 & [6.6, 10.5] & 12.50 & [10.7, 14.2] & 12.60 & [10.7, 14.5] & 14.10 & [12.4, 15.8] \\
Vocational degree & 30.10 & [26.5, 33.8] & 24.90 & [22.4, 27.5] & 24.30 & [21.6, 27.1] & 20.80 & [18.7, 23.0] \\
Other degree & 10.60 & [8.2, 13.2] & 13.10 & [11.0, 15.2] & 10.70 & [8.7, 12.7] & 9.10 & [7.5, 10.7] \\

\addlinespace
\textbf{Occupation} & & & & & & & & \\
High-skilled worker & 8.70 & [6.9, 10.5] & 15.30 & [13.6, 17.1] & 12.80 & [11.1, 14.5] & 14.50 & [12.9, 16.2] \\
Qualified employee & 46.20 & [42.3, 50.2] & 44.40 & [41.5, 47.4] & 45.90 & [42.7, 49.0] & 46.60 & [44.0, 49.3] \\
Non-qualified employee & 21.80 & [18.5, 25.2] & 20.00 & [17.5, 22.5] & 19.40 & [16.8, 22.1] & 16.10 & [14.0, 18.2] \\

\addlinespace
\textbf{Experience (years)} & 6.78 & [6.16, 7.37] & 7.94 & [7.44, 8.45] & 6.54 & [6.10, 6.99] & 6.42 & [6.07, 6.77] \\

\addlinespace
\textbf{Current unemployment duration} & & & & & & & & \\
$\leq 1$ month & 8.3 & [7.0, 9.5] & 8.7 & [7.8, 9.6] & 12.3 & [11.1, 13.5] & 13.3 & [12.2, 14.4] \\
1--3 months & 8.5 & [6.9, 10.2] & 10.8 & [9.3, 12.4] & 13.9 & [11.9, 15.8] & 12.8 & [11.3, 14.4] \\
3--6 months & 13.9 & [11.0, 16.8] & 15.3 & [13.2, 17.6] & 17.1 & [14.6, 19.6] & 17.6 & [15.4, 19.7] \\
6--12 months & 21.1 & [18.0, 24.2] & 23.5 & [21.1, 26.0] & 22.2 & [19.6, 24.8] & 21.8 & [19.6, 24.0] \\
12--24 months & 21.0 & [17.5, 24.5] & 19.5 & [17.0, 22.0] & 17.7 & [15.0, 20.3] & 18.0 & [15.7, 20.2] \\
$> 24$ months & 26.9 & [23.2, 30.4] & 22.0 & [19.4, 24.6] & 16.6 & [14.1, 19.2] & 16.8 & [14.5, 19.0] \\

\addlinespace
\textbf{Remaining benefit duration} & & & & & & & & \\
Exhausted & 43.5 & [39.5, 47.5] & 32.4 & [29.5, 35.3] & 30.0 & [26.9, 33.1] & 30.4 & [27.7, 33.1] \\
$< 6$ months & 13.3 & [10.5, 15.9] & 11.3 & [9.3, 13.3] & 12.6 & [10.5, 14.8] & 13.7 & [11.8, 15.5] \\
6--12 months & 12.6 & [10.1, 15.1] & 11.9 & [10.1, 13.7] & 14.2 & [12.0, 16.3] & 14.5 & [12.6, 16.3] \\
12--24 months & 22.7 & [19.8, 25.8] & 33.9 & [31.2, 36.6] & 36.6 & [33.7, 39.5] & 37.1 & [34.6, 39.5] \\
$> 24$ months & 7.8 & [5.8, 9.8] & 10.4 & [8.9, 12.0] & 6.4 & [5.0, 7.8] & 4.4 & [3.3, 5.4] \\

\addlinespace
\textbf{Cumulative unemployment} & & & & & & & & \\
\textbf{(months)} & 95.3 & [88.7, 101.9] & 71.0 & [66.4, 75.6] & 73.6 & [68.7, 78.4] & 66.8 & [62.9, 70.8] \\

\hline
\hline
\end{tabular}

\begin{tablenotes}[flushleft]
\small
\item \textit{Notes:} This table reports descriptive statistics by 3-month reemployment-bias group. Group membership is estimated using the XGBoost classification procedure and is recomputed across 100 sample splits. Reported means correspond to the median estimate across splits, while the values in brackets correspond to the 2.5th and 97.5th percentiles across splits. Estimated on panels 1--8 restricted to individuals with consistent beliefs regarding reemployment probabilities at different time horizons. Sampling weights are used throughout.
\end{tablenotes}

\end{threeparttable}
\end{adjustbox}
\end{table}

\begin{table}[!htbp]
\centering
\footnotesize
\caption{Demographic characteristics and unemployment histories by reemployment-bias group (Random Forest)}
\label{tab:bias_group_stat_desc}

\begin{adjustbox}{width=\textwidth}
\begin{threeparttable}

\begin{tabular}{lcccccccc}
\hline
\hline
& \multicolumn{2}{c}{Pessimist} & \multicolumn{2}{c}{Rational} & \multicolumn{2}{c}{Optimist} & \multicolumn{2}{c}{Over-optimist} \\
\midrule

\textbf{Women (\%)} & 52.8 & [49.3, 56.3] & 53.4 & [50.2, 56.7] & 52.6 & [49.8, 55.4] & 46.7 & [43.8, 49.6] \\

\addlinespace
\textbf{Age} & 42.9 & [42.0, 43.8] & 42.3 & [41.5, 43.2] & 39.4 & [38.8, 40.0] & 37.1 & [36.5, 37.7] \\
$\leq 25$ & 11.7 & [9.0, 14.2] & 9.8 & [7.7, 11.9] & 10.9 & [9.0, 12.9] & 12.5 & [10.3, 14.6] \\
25--35 & 22.3 & [19.3, 25.2] & 25.8 & [22.9, 28.8] & 32.3 & [29.5, 35.0] & 37.0 & [34.1, 39.9] \\
35--45 & 20.5 & [17.6, 23.4] & 21.8 & [19.2, 24.5] & 23.7 & [21.4, 26.1] & 27.9 & [25.3, 30.5] \\
45--55 & 21.6 & [18.9, 24.3] & 18.9 & [16.6, 21.3] & 20.9 & [18.8, 23.0] & 16.2 & [14.2, 18.2] \\
$> 55$ & 23.8 & [20.9, 26.6] & 23.1 & [20.5, 25.8] & 12.1 & [10.5, 13.8] & 6.7 & [5.3, 8.0] \\

\addlinespace
\textbf{Number of children} & 0.79 & [0.71, 0.87] & 0.81 & [0.73, 0.88] & 0.84 & [0.78, 0.91] & 0.81 & [0.75, 0.87] \\
\textbf{Married (\%)} & 42.5 & [39.1, 46.0] & 46.6 & [43.4, 49.8] & 45.3 & [42.5, 48.1] & 43.4 & [40.6, 46.3] \\

\addlinespace
\textbf{Education} & & & & & & & & \\
High school & 23.3 & [20.3, 26.4] & 22.8 & [19.9, 25.5] & 23.9 & [21.5, 26.4] & 24.2 & [21.6, 26.7] \\
High school + 2 years of college & 16.1 & [13.5, 18.7] & 15.9 & [13.6, 18.2] & 17.0 & [15.0, 19.1] & 17.2 & [15.1, 19.4] \\
Bachelor's degree & 9.5 & [7.7, 11.5] & 11.3 & [9.3, 13.3] & 11.9 & [10.1, 13.5] & 15.0 & [13.0, 17.1] \\
Master's degree & 9.3 & [7.5, 11.2] & 12.3 & [10.4, 14.2] & 13.0 & [11.3, 14.7] & 13.8 & [11.8, 15.7] \\
Vocational degree & 29.3 & [27.2, 31.4] & 25.4 & [23.7, 27.2] & 23.5 & [22.1, 24.9] & 20.7 & [19.2, 22.1] \\
No degree & 12.3 & [9.9, 14.7] & 12.2 & [9.9, 14.5] & 10.5 & [8.8, 12.3] & 8.9 & [7.2, 10.6] \\

\addlinespace
\textbf{Occupation} & & & & & & & & \\
High-skilled worker & 10.1 & [8.3, 11.9] & 14.8 & [12.8, 16.7] & 13.4 & [11.9, 15.0] & 14.8 & [13.0, 16.6] \\
Qualified employee & 44.1 & [40.6, 47.7] & 44.8 & [41.6, 48.1] & 46.5 & [43.7, 49.4] & 47.0 & [44.1, 49.9] \\
Non-qualified employee & 21.6 & [18.6, 24.6] & 20.4 & [17.6, 23.2] & 18.9 & [16.5, 21.2] & 15.9 & [13.6, 18.1] \\

\addlinespace
\textbf{Experience (years)} & 7.40 & [6.81, 8.00] & 7.71 & [7.18, 8.24] & 6.79 & [6.39, 7.20] & 6.08 & [5.72, 6.43] \\

\addlinespace
\textbf{Current unemployment duration} & & & & & & & & \\
$\leq 1$ month & 8.1 & [6.9, 9.3] & 8.8 & [7.7, 9.9] & 11.9 & [10.8, 13.0] & 14.0 & [12.8, 15.2] \\
1--3 months & 8.2 & [6.7, 9.7] & 11.1 & [9.4, 12.7] & 13.4 & [11.7, 15.1] & 13.2 & [11.4, 15.0] \\
3--6 months & 13.6 & [11.0, 16.2] & 15.6 & [13.2, 18.0] & 16.9 & [14.7, 19.1] & 18.0 & [15.7, 20.4] \\
6--12 months & 20.3 & [17.6, 23.1] & 23.7 & [21.0, 26.5] & 22.1 & [19.8, 24.4] & 22.6 & [20.0, 25.1] \\
12--24 months & 21.3 & [18.1, 24.3] & 18.6 & [15.9, 21.5] & 18.3 & [16.0, 20.6] & 17.8 & [15.4, 20.2] \\
$> 24$ months & 28.3 & [25.1, 31.5] & 21.8 & [19.0, 24.7] & 17.7 & [15.3, 20.0] & 14.3 & [12.0, 16.6] \\

\addlinespace
\textbf{Remaining benefit duration} & & & & & & & & \\
Exhausted & 43.4 & [39.9, 46.9] & 32.5 & [29.4, 35.7] & 30.9 & [28.1, 33.7] & 27.7 & [24.9, 30.6] \\
$< 6$ months & 12.2 & [9.8, 14.5] & 11.9 & [9.7, 14.0] & 12.5 & [10.7, 14.4] & 14.2 & [12.1, 16.2] \\
6--12 months & 12.1 & [9.9, 14.4] & 11.8 & [9.8, 13.8] & 13.4 & [11.5, 15.3] & 15.6 & [13.5, 17.7] \\
12--24 months & 23.0 & [20.3, 25.8] & 33.6 & [30.7, 36.5] & 36.3 & [33.7, 38.8] & 39.1 & [36.3, 41.9] \\
$> 24$ months & 9.2 & [7.3, 11.0] & 10.0 & [8.1, 11.7] & 6.7 & [5.4, 7.9] & 3.5 & [2.5, 4.5] \\

\addlinespace
\textbf{Cumulative unemployment} & & & & & & & & \\
\textbf{(months)} & 87.9 & [82.1, 93.8] & 72.4 & [67.2, 77.4] & 73.9 & [69.6, 78.2] & 66.0 & [62.1, 70.1] \\

\hline
\hline
\end{tabular}

\begin{tablenotes}[flushleft]
\footnotesize
\item \textit{Notes:} This table reports descriptive statistics by 3-month reemployment-bias group. Group membership is estimated using the Random Forest classification procedure and is recomputed across 100 sample splits. Reported means correspond to the median estimate across splits, while the values in brackets correspond to the 2.5th and 97.5th percentiles across splits. Estimated on panels 1--8 restricted to individuals with consistent beliefs regarding reemployment probabilities at different time horizons. Sampling weights are used throughout.
\end{tablenotes}

\end{threeparttable}
\end{adjustbox}
\end{table}

\begin{table}[!htbp]
\centering
\caption{Average informational biases by reemployment-bias group (XGBoost)}
\label{tab:informational_biases_by_group_xgb}

\begin{adjustbox}{width=\textwidth}
\begin{threeparttable}

\begin{tabular}{lcccccccc}
\hline
\hline
& \multicolumn{2}{c}{Pessimists}
& \multicolumn{2}{c}{Rational}
& \multicolumn{2}{c}{Optimists}
& \multicolumn{2}{c}{Over-optimists} \\
\midrule

\multicolumn{9}{l}{\textit{Panel A. Biases in the wage-offer distribution}} \\
Median wage bias & -21.4 & [-24.4, -18.3] & -13.2 & [-15.3, -11.0] & -6.0 & [-8.3, -3.6] & 0.1 & [-2.0, 2.3] \\
25th-percentile wage bias & -41.4 & [-44.7, -38.1] & -32.8 & [-35.1, -30.5] & -25.6 & [-27.8, -23.3] & -19.2 & [-21.2, -17.1] \\
75th-percentile wage bias & -1.7 & [-4.6, 1.3] & 6.1 & [3.9, 8.3] & 12.2 & [9.9, 14.5] & 17.5 & [15.3, 19.7] \\

\addlinespace
\multicolumn{9}{l}{\textit{Panel B. Biases in the probability of receiving job offers: subjective minus predicted}} \\
$\widetilde{P}(N\geq 1)$ & 22.9 & [20.4, 25.3] & 32.4 & [30.6, 34.3] & 49.2 & [47.3, 51.1] & 64.9 & [63.0, 66.7] \\
$P(N\geq 1)$ & 61.8 & [60.3, 63.3] & 61.3 & [60.2, 62.5] & 68.3 & [67.2, 69.4] & 72.4 & [71.5, 73.3] \\
$\widetilde{P}(N\geq 1)-P(N\geq 1)$ & -39.1 & [-41.0, -37.1] & -28.8 & [-30.1, -27.4] & -19.2 & [-20.7, -17.7] & -7.5 & [-8.9, -6.0] \\
$\widetilde{P}(N\geq 3)$ & 20.4 & [18.0, 22.8] & 30.0 & [28.2, 31.7] & 46.0 & [44.0, 47.9] & 58.2 & [56.3, 60.2] \\
$P(N\geq 3)$ & 25.8 & [24.7, 26.9] & 25.8 & [25.0, 26.7] & 31.5 & [30.4, 32.5] & 34.6 & [33.6, 35.6] \\
$\widetilde{P}(N\geq 3)-P(N\geq 3)$ & -5.3 & [-7.3, -3.5] & 4.1 & [2.8, 5.5] & 14.4 & [12.9, 15.9] & 23.7 & [22.3, 25.1] \\

\addlinespace
\multicolumn{9}{l}{\textit{Panel C. Biases in the probability of receiving job offers: subjective minus realized}} \\
$\widetilde{P}(N\geq 1)$ & 24.5 & [19.4, 29.6] & 31.4 & [28.1, 34.8] & 47.7 & [43.7, 51.7] & 63.5 & [59.5, 67.3] \\
$P(N\geq 1)$ & 54.1 & [44.3, 63.8] & 58.4 & [52.1, 64.6] & 68.7 & [61.5, 75.8] & 75.3 & [69.0, 81.7] \\
$\widetilde{P}(N\geq 1)-P(N\geq 1)$ & -29.7 & [-39.4, -19.6] & -27.2 & [-33.8, -20.5] & -20.5 & [-27.9, -12.9] & -12.2 & [-18.9, -5.6] \\
$\widetilde{P}(N\geq 3)$ & 22.2 & [17.5, 26.9] & 28.9 & [25.7, 32.0] & 45.4 & [41.2, 49.7] & 59.5 & [55.3, 63.7] \\
$P(N\geq 3)$ & 25.7 & [17.3, 34.3] & 23.5 & [18.1, 28.8] & 34.0 & [26.7, 41.2] & 35.1 & [28.0, 42.1] \\
$\widetilde{P}(N\geq 3)-P(N\geq 3)$ & -3.6 & [-12.4, 5.3] & 5.5 & [0.1, 10.8] & 11.4 & [3.7, 19.3] & 24.7 & [17.5, 31.9] \\

\hline
\hline
\end{tabular}

\begin{tablenotes}[flushleft]
\small
\item \textit{Notes:} This table reports average informational biases by 3-month reemployment-bias group. Panel A reports biases at the 25th percentile, median, and 75th percentile of the wage-offer distribution. Panel B reports subjective probabilities, predicted probabilities, and the corresponding biases in the probability of receiving at least one or at least three job offers within the next three months. Panel C reports the same quantities using realized job-offer receipt and restricts the sample to individuals who responded to the survey at least twice. Bias group membership is estimated using the XGBoost classification procedure and is recomputed across 100 sample splits. Reported means correspond to the median estimate across splits, while the values in brackets correspond to the 2.5th and 97.5th percentiles across splits.  Estimated on panels 1--8 restricted to individuals with consistent beliefs regarding reemployment probabilities at different time horizons.   Sampling weights are used throughout.
\end{tablenotes}

\end{threeparttable}
\end{adjustbox}
\end{table}

\begin{table}[!htbp] \centering 
\caption{Average informational biases by group (Random Forest)} 
\label{tab:informational_biases_by_group} 
\begin{adjustbox}{width = \textwidth}
\begin{threeparttable}
\begin{tabular}{lrcrcrcrc} 
\hline
\hline
 & \multicolumn{2}{c}{Pessimists} & \multicolumn{2}{c}{Rationals} & \multicolumn{2}{c}{Optimists} & \multicolumn{2}{c}{Over-Optimists} \\ 
 \midrule
 \multicolumn{9}{l}{\textit{Panel A. Bias on the wage distribution}} \\
 Bias q0.5 & -20.6 & [-23.3;-17.8] & -12.5 & [-14.9;-10.1] &  -5.4 & [-7.6;-3.3] &   0.5 & [-1.8;2.8] \\ 
Bias q0.25 & -40.4 & [-43.3;-37.6] & -32.0 & [-34.5;-29.6] & -25.0 & [-27.1;-22.9] & -18.9 & [-21.2;-16.6] \\ 
Bias q0.75 &  -0.9 & [-3.4;1.9] &   6.5 & [4.1;8.8] &  12.7 & [10.6;14.8] &  17.9 & [15.5;20.3] \\ 
\addlinespace
 \multicolumn{9}{l}{\textit{Panel B. Bias on the arrival rate of job offers (subjective - predicted)}} \\
$\widetilde{P}(N\geq 1)$ &  22.8 & [20.5;24.9] &  35.3 & [33.2;37.3] &  50.9 & [49.1;52.7] & 64.5 & [62.5;66.6] \\  
$P(N\geq 1)$ &  60.9 & [59.6;62.2] &  62.7 & [61.5;64] &  68.5 & [67.6;69.5] & 72.0 & [71;72.9] \\ 
$\widetilde{P}(N\geq 1) - P(N\geq 1)$ & -38.0 & [-39.8;-36.2] & -27.6 & [-29.1;-25.9] & -17.6 & [-19;-16.2] & -7.5 & [-9;-5.9] \\ 
$\widetilde{P}(N\geq 3)$ &  20.6 & [18.5;22.6] &  32.4 & [30.4;34.3] &  47.4 & [45.5;49.3] & 57.7 & [55.7;59.8] \\  
$P(N\geq 3)$  &  25.2 & [24.3;26.2] &  27.0 & [26;28] &  32.0 & [31.1;32.9] & 33.8 & [32.7;34.9] \\ 
$\widetilde{P}(N\geq 3) - P(N\geq 3)$ &  -4.7 & [-6.4;-3] &   5.4 & [3.9;6.8] &  15.5 & [14.1;16.8] & 23.9 & [22.4;25.4] \\ 
\addlinespace
\multicolumn{9}{l}{\textit{Panel C. Bias on the arrival rate of job offers (subjective - observed)}} \\
$\widetilde{P}(N\geq 1)$ &  24.2 & [19.4;28.9] &  33.9 & [30.1;37.7] &  48.0 & [44.5;51.6] &  63.5 & [59.3;67.7] \\ 
$P(N\geq 1)$ &  53.2 & [44.4;61.7] &  60.7 & [53.8;67.4] &  68.6 & [62.3;75] &  75.0 & [67.5;81.9] \\  
$\widetilde{P}(N\geq 1) - P(N\geq 1)$ & -29.2 & [-37.8;-20.4] & -27.0 & [-34.4;-19.6] & -20.7 & [-27.3;-14.1] & -11.7 & [-18.7;-4.5] \\
$\widetilde{P}(N\geq 3)$ &  21.6 & [17.3;25.9] &  31.3 & [27.7;34.8] &  46.1 & [42.4;49.9] &  59.3 & [54.7;64] \\ 
$P(N\geq 3)$ &  23.6 & [16.5;30.8] &  25.0 & [18.8;31] &  33.2 & [26.9;39.5] &  35.4 & [27.4;43.4] \\ 
$\widetilde{P}(N\geq 3) - P(N\geq 3)$ &  -1.9 & [-9.6;5.5] &   6.2 & [-0.1;12.6] &  13.2 & [6.2;20.4] &  24.4 & [16.5;31.9] \\ 
\hline 
\end{tabular}
\begin{tablenotes}[flushleft]
\small
\item \textit{Notes:}Average informational biases across job-finding bias groups. Panel~A 
reports biases at three quantiles of the wage distribution. Panel~B reports the true probability, the predicted probability, and the associated bias on the probability of receiving at least 1 or 3 job offers within the next three months. Panel~C reports the same quantities, restricting the sample to individuals who responded to the survey at least twice, in order to observe the realized number 
of job offers received. Bias group assignments are estimated over 100 sample splits. For each group, the reported mean is the median of the 100 sample-split-specific means. Panels 1--8 restricted to individuals with consistent beliefs regarding reemployment probabilities at different time horizons. Weights are used. The Random Forest algorithm is used to predict bias groups.
\end{tablenotes}
\end{threeparttable}
\end{adjustbox}
\end{table}

\begin{figure}
\caption{Informational biases and predicted jobfinding bias group indicators}
\label{fig:individual_info_biases_and_group}
     \centering
    \begin{subfigure}{0.45\linewidth}
        \centering
        \includegraphics[width=\linewidth]{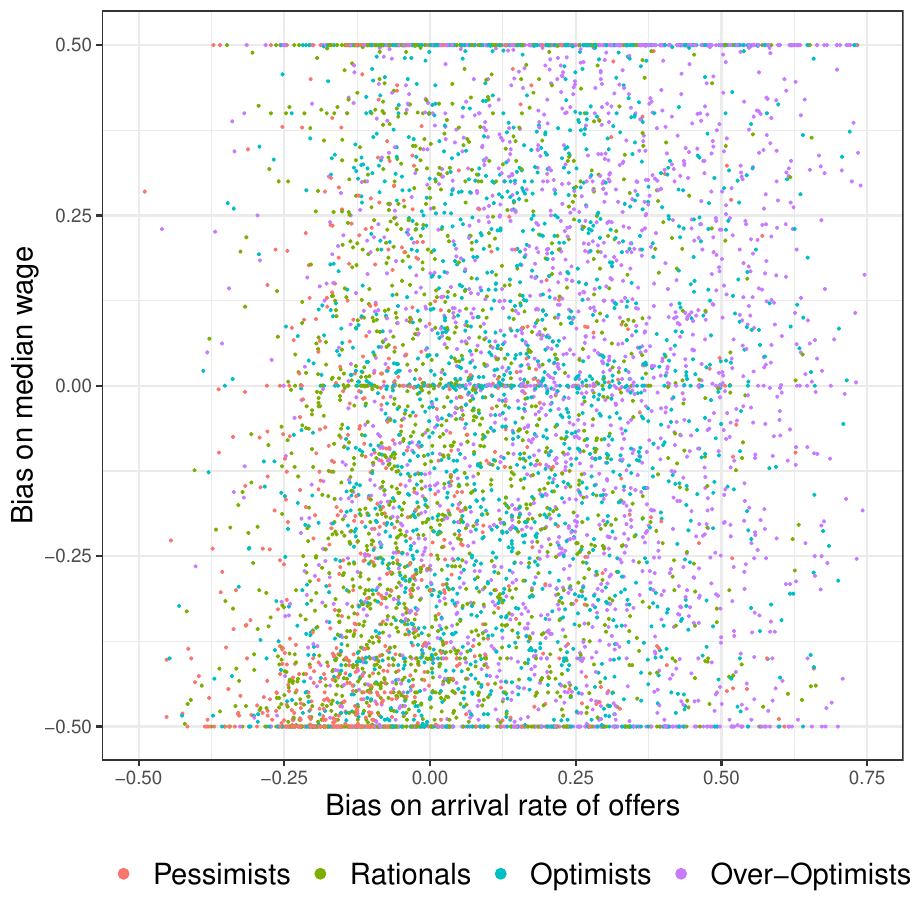}
        \caption{XGBoost}
    \end{subfigure}
\hfill  
    \begin{subfigure}{0.45\linewidth}
        \centering
        \includegraphics[width=\linewidth]{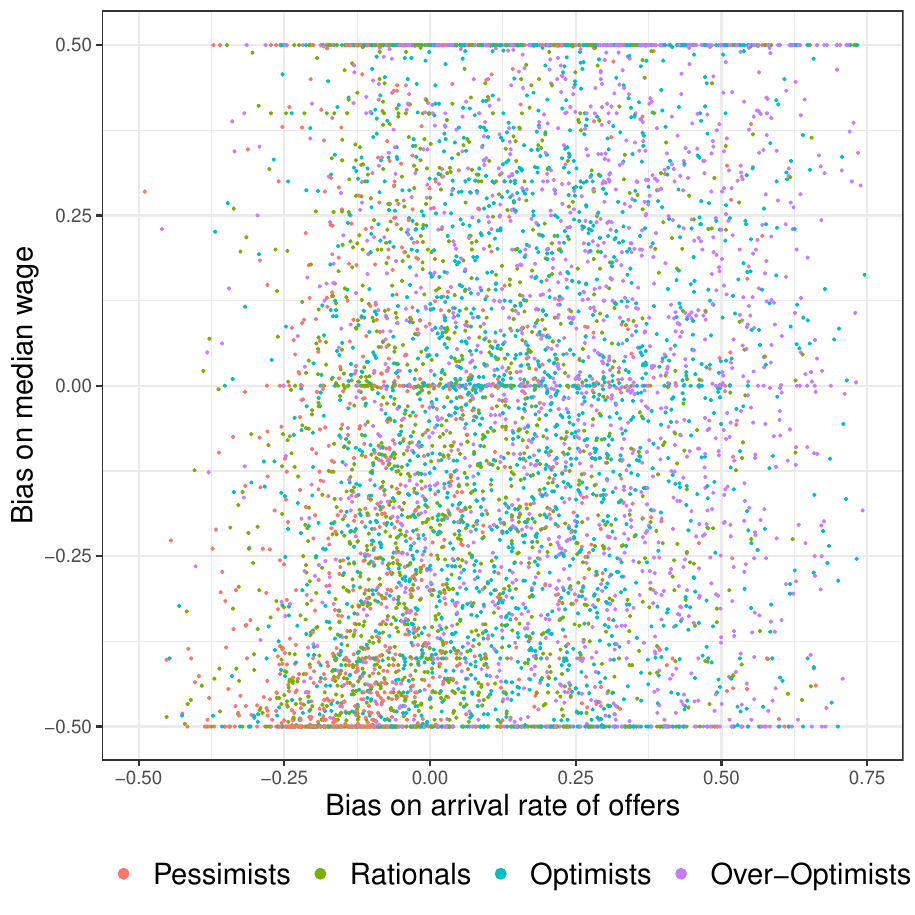}
        \caption{Random Forest}
    \end{subfigure}
    \caption*{\footnotesize \textit{Notes:} The figure plots observations from a single sample split according to their informational biases. The arrival rate of offers bias is computed as the raw difference between the perceived probability of receiving at least 3 job offers within the next three months and its objective predicted counterpart. The bias on the median wage is the raw difference between the perceived probability that a job offer proposes a wage above the median of the wage distribution and its true counterpart (equal to 50\% by definition of the median). 
Bias group assignments are estimated over 100 sample splits. The XGBoost algorithm is used to predict bias groups in panel~(a) and the Random Forest algorithm in panel~(b). Panels 1--8 restricted to individuals with consistent beliefs regarding reemployment probabilities at different time horizons. Weights are used.}
\end{figure}

\begin{figure}[!htbp]
    \centering
    \caption{True and perceived wage distributions by reemployment-bias group (XGBoost)}
    \label{fig:true_perceived_cdf_by_group_xgb}

    \includegraphics[width=\textwidth]{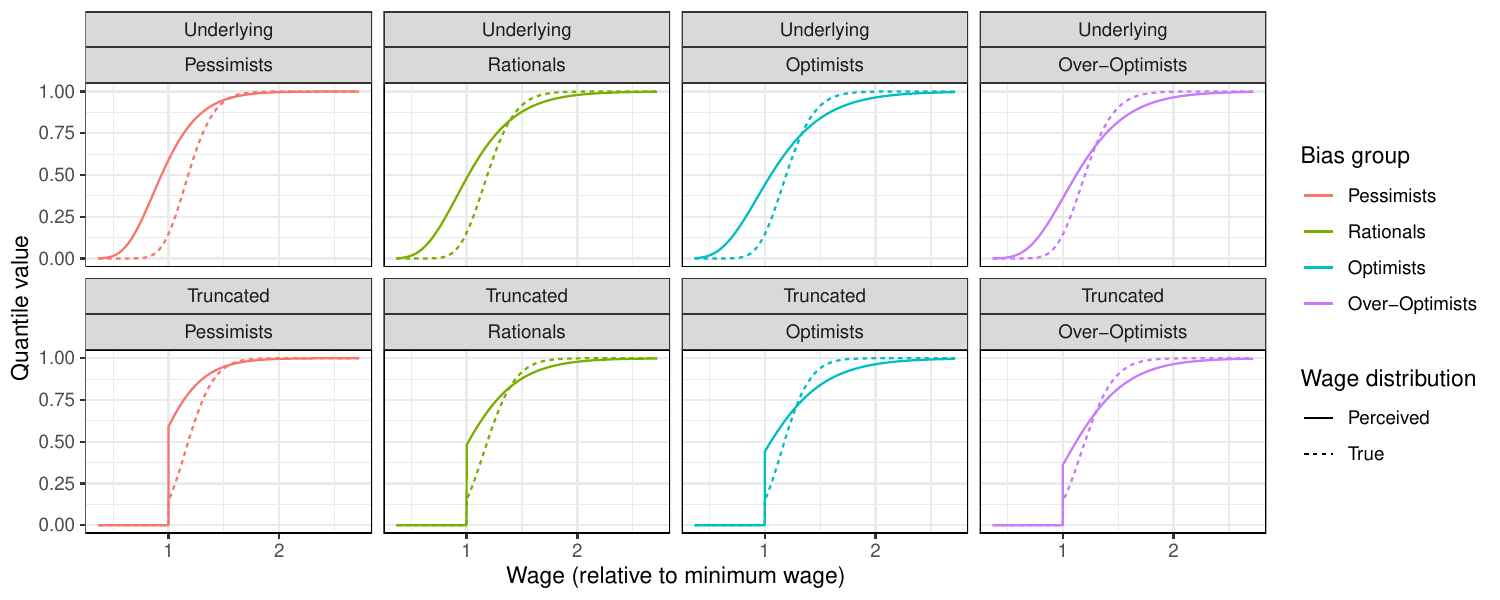}

    \caption*{\footnotesize \textit{Notes:} This figure compares the average true and perceived wage distributions across 3-month reemployment-bias groups. For each individual, lognormal distributions are fitted to both the perceived and the true wage distributions. Reemployment-bias group membership is estimated using the XGBoost classification procedure and is recomputed across 100 sample splits. Within each split, average distribution parameters are computed by group, and the reported distributions are obtained using the median parameter estimates across splits. The sample is restricted to individuals with strictly increasing perceived wage cumulative distribution functions (CDFs) and whose reported reemployment beliefs satisfy the consistency restrictions. Sampling weights are used throughout.}
\end{figure}

\begin{figure}[!htbp]
\centering

\caption{Perceived job-offer arrival rates by search effort and reemployment-bias group}
\label{fig:perceived_return}

\begin{subfigure}{0.48\linewidth}
    \centering
    \includegraphics[width=\linewidth]{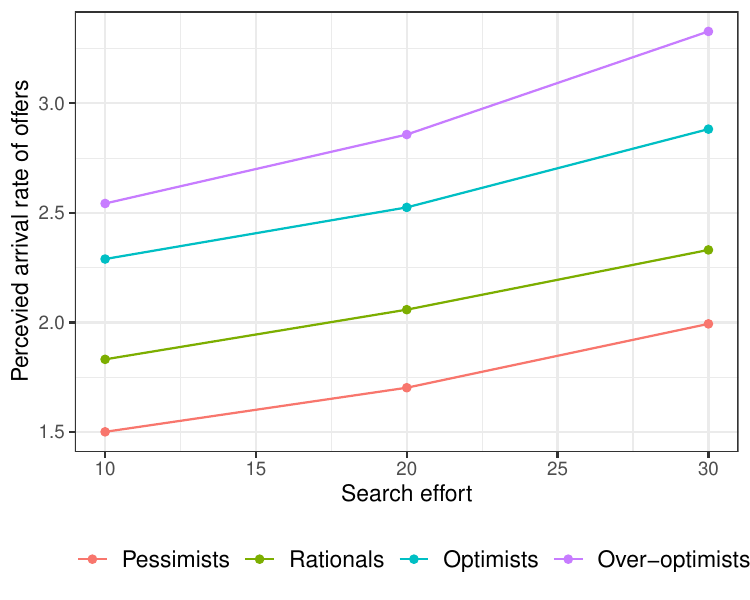}
    \caption{XGBoost}
\end{subfigure}
\hfill
\begin{subfigure}{0.48\linewidth}
    \centering
    \includegraphics[width=\linewidth]{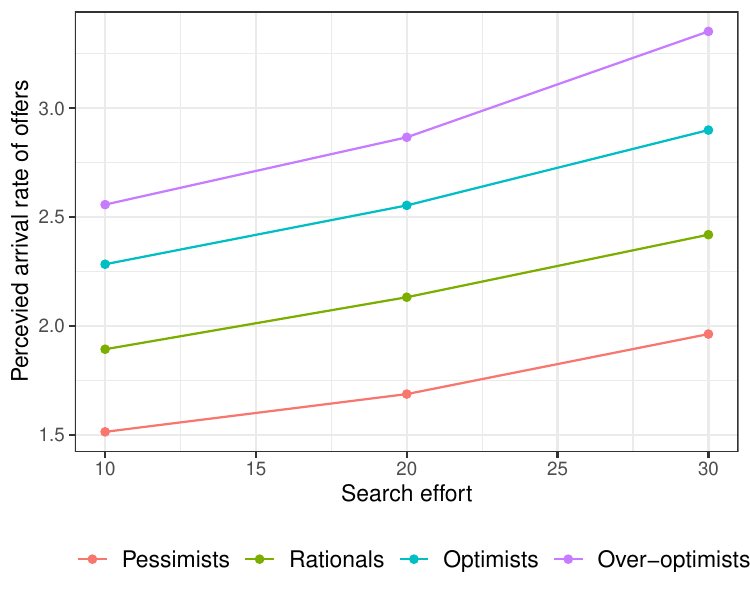}
    \caption{Random Forest}
\end{subfigure}

\caption*{\small \textit{Notes:} This figure reports the average perceived arrival rate of job offers for weekly search efforts of 10, 20, and 30 hours, by 3-month reemployment-bias group. Reemployment-bias group membership is estimated across 100 sample splits. For each individual, the perceived arrival rate of job offers is recovered under the assumption that job offers arrive according to a Poisson process. Average arrival rates are computed within each bias group and sample split, and the reported values correspond to the median across splits. Panel (a) uses the XGBoost classification procedure, while Panel (b) uses the Random Forest classification procedure. The sample is restricted to individuals whose reported reemployment beliefs satisfy the consistency restrictions and whose perceived probability of receiving at least one job offer lies strictly between 0 and 1 for all considered levels of search effort. Sampling weights are used throughout.}
\end{figure}

\begin{table}[!htbp]
\centering
\caption{Reemployment-bias groups and search behaviors (Random Forest)}
\label{tab:behaviors_gml}

\begin{adjustbox}{width=\textwidth}
\begin{threeparttable}

\begin{tabular}{lcccccc}
\hline
\hline
& Reservation wage & Reservation wage & Search effort & Reservation & Number of applications & Number of applications \\
& (log, administrative) & (log, self-reported) & (hours/week) & mobility & (PES website) & (other occupations) \\
& (1) & (2) & (3) & (4) & (5) & (6) \\
\midrule

Pessimists
& -0.024*** & -0.030** & -0.939** & 0.364 & 0.431** & 0.318* \\
& (0.011) & (0.014) & (0.450) & (0.636) & (0.209) & (0.174) \\

Rationals
& -- & -- & -- & -- & -- & -- \\
& & & & & & \\

Optimists
& 0.004 & -0.007 & 1.267*** & 1.122* & 0.058 & 0.014 \\
& (0.010) & (0.013) & (0.394) & (0.552) & (0.176) & (0.147) \\

Over-optimists
& 0.012 & 0.030** & 1.176*** & 0.787 & -0.406** & -0.347** \\
& (0.010) & (0.014) & (0.421) & (0.586) & (0.171) & (0.143) \\

\midrule

Demographics & \checkmark & \checkmark & \checkmark & \checkmark & \checkmark & \checkmark \\
Target occupation & \checkmark & \checkmark & \checkmark & \checkmark & \checkmark & \checkmark \\
Region fixed effects & \checkmark & \checkmark & \checkmark & \checkmark & \checkmark & \checkmark \\

\hline

Observations & 9,571 & 6,748 & 9,350 & 10,930 & 10,938 & 10,938 \\
Mean (Rationals) & \euro 2,367 & \euro 2,196 & 12.5 & 23.3 km & 1.09 & 0.846 \\
$p$-value (joint test) & 0.000 & 0.000 & 0.000 & 0.112 & 0.000 & 0.000 \\

\hline
\hline
\end{tabular}

\begin{tablenotes}[flushleft]
\small
\item \textit{Notes:} This table reports the relationship between reemployment-bias groups and search behaviors. Each outcome is regressed on reemployment-bias group indicators, demographic characteristics, target-occupation fixed effects, and region fixed effects, with the Rational group serving as the reference category. Group membership is estimated using the Random Forest classification procedure and is recomputed across 100 sample splits. Regressions are estimated separately for each split, and the reported coefficients, standard errors, sample sizes, and test statistics correspond to the median across splits.

The administrative reservation wage is trimmed at \euro 10,000, while the self-reported reservation wage is winsorized at the 2.5th and 97.5th percentiles. The number of applications is winsorized at the 98th percentile among positive values. The reported $p$-values correspond to joint tests of whether all reemployment-bias group coefficients are equal to zero and are computed separately for each split before taking the median across splits.

The sample is restricted to individuals whose reported reemployment beliefs satisfy the consistency restrictions described in Section~\ref{ssec:panel_data}. Sampling weights are used throughout.

Significance levels: $^{*}p<0.10$, $^{**}p<0.05$, and $^{***}p<0.01$.
\end{tablenotes}

\end{threeparttable}
\end{adjustbox}
\end{table}

\FloatBarrier

\newpage

\section{Perceived arrival rate of job offers}\label{app:perceived_arrival_rate}

The survey includes several items designed to elicit job seekers’ beliefs about the arrival of job offers. Two baseline questions ask respondents to report the subjective probability of receiving at least one job offer and at least three job offers within the next three months. Under the benchmark assumption that the number of offers follows a Poisson distribution, and that respondents’ answers are internally consistent, these two probabilities should imply the same underlying arrival-rate parameter.

\medskip

Empirically, however, the two reported probabilities are strikingly similar. The average reported probability is 45.2\% for receiving at least one job offer and 41.1\% for receiving at least three job offers within the next three months (see Table~\ref{tab:subj_exp}). Under a Poisson specification, these responses imply substantially different arrival rates: approximately $\lambda \approx 0.6$ when inferred from the first question and $\lambda \approx 2.3$ when inferred from the second.

More generally, the two probabilities imply that respondents assign only 4.1\% probability to receiving exactly one or two offers. This pattern is not logically inconsistent with every possible count process. For example, a sufficiently flexible model such as a hurdle or zero-inflated specification, allowing for strong heterogeneity or excess mass at zero could generate substantial probability mass at both zero and relatively high offer counts. It is, however, difficult to reconcile with the parsimonious Poisson process used as our benchmark and suggests either that respondents perceive a highly dispersed offer distribution or that they do not interpret the two questions literally.

\medskip

Additional inconsistencies favor the latter interpretation. If the perceived probability of receiving at least one job offer within three months is 45\%, then the perceived probability of reemployment over the same horizon cannot exceed this value, even under the extreme assumption that job seekers accept every offer they receive. Yet the average perceived probability of reemployment is close to 50\%. Moreover, beliefs about receiving at least one job offer exhibit a strong pessimistic bias relative to realized outcomes, whereas beliefs about receiving at least three job offers are, on average, optimistic and more consistent with reported reemployment expectations. Taken together, these patterns suggest that the ``at least one offer'' framing may generate substantial measurement error.

\medskip

From wave~6 onward, we introduce additional questions designed to elicit perceived offer-arrival probabilities conditional on specific levels of weekly search effort (10, 20, and 30 hours). These items also use the ``at least one offer'' framing. Given the concerns discussed above, we assess whether this wording affects respondents’ answers.

\medskip

\begin{figure}[!htbp]
    \centering
    \caption{Perceived job-offer arrival rates by search effort and framing}
    \label{fig:framing_arrival}

    \includegraphics[width=\linewidth]{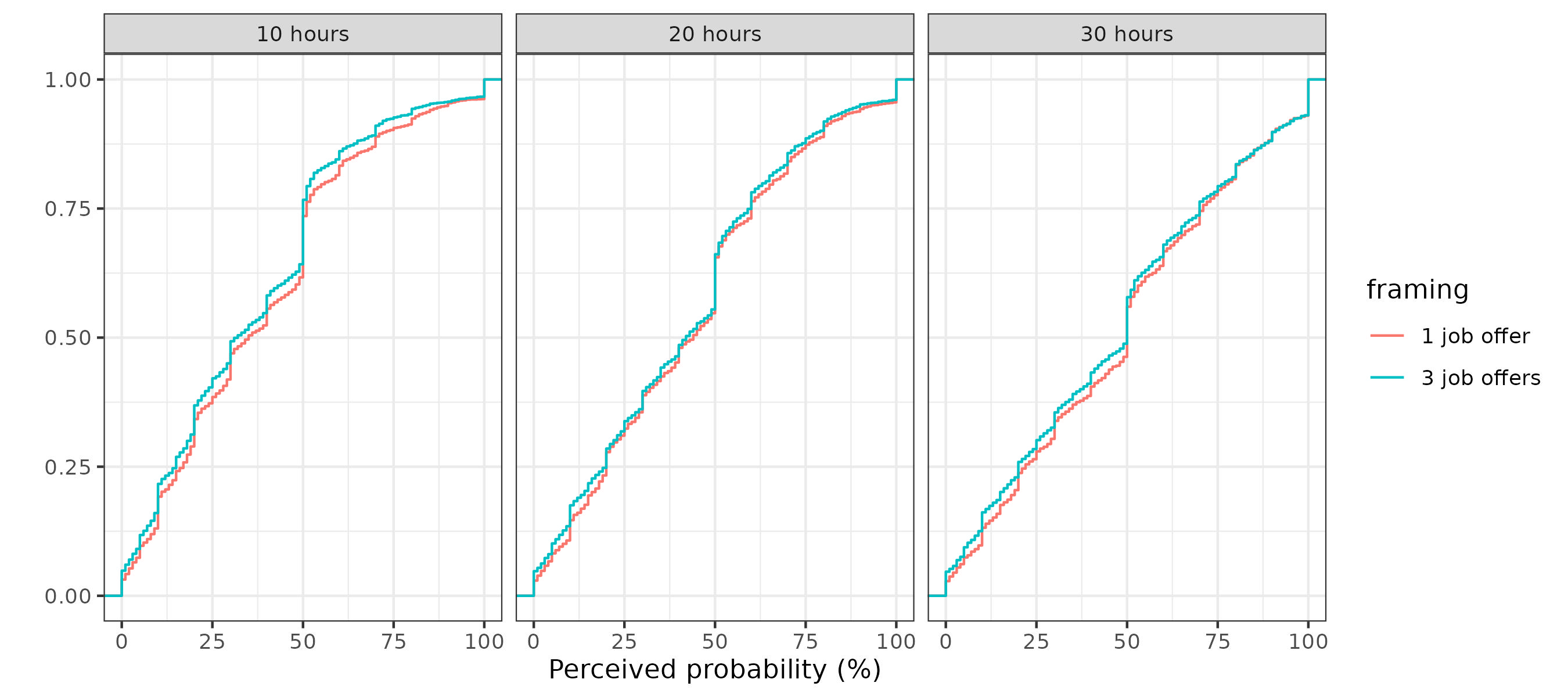}

    \begin{minipage}{\textwidth}
    \small
    \textit{Notes:} This figure reports average subjective job-offer arrival rates as a function of self-reported search effort under the different survey framing conditions. Search effort is measured using the survey question described in Section~\ref{sec:data}. 
    \end{minipage}
\end{figure}

To investigate this issue, we exploit an experimental variation implemented in a later survey wave (July 2025), in which the numerical threshold is randomized. Half of respondents are asked about the probability of receiving at least \textit{one offer} conditional on a given level of search effort, while the other half are asked about the probability of receiving at least \textit{three offers}. The empirical distributions of reported probabilities are nearly identical across the two groups (see Figure~\ref{fig:framing_arrival}). This absence of a framing effect suggests that respondents may not interpret the ``at least one offer'' question literally and may instead map both formulations to a similar, less precisely defined event.

\medskip

Based on this evidence, our main analysis relies primarily on the items that explicitly refer to receiving at least three job offers, which appear to provide more reliable information. For the effort-conditional questions framed in terms of receiving at least one offer, we interpret the responses as capturing beliefs about a comparable higher-threshold event and use them to measure perceived returns to search effort. Although this interpretation should be viewed with caution, the experimental evidence suggests that it provides a reasonable approximation.

\section{Proof of Proposition \ref{prop:k_means_lit}.}

We proceed in four steps.

\medskip
\noindent
\textbf{Step 1: Population problem as best step-function approximation.}

Let \(\mathcal G=\{G_1,\dots,G_K\}\) be a partition of the support of \(D\), and define
\[
B_{\mathcal G}(d)
=
\sum_{k=1}^K \gamma(G_k)\mathbf 1\{d\in G_k\},
\qquad
\gamma(G_k)=\E[\varepsilon^*\mid D\in G_k].
\]
By the law of total variance,
\[
\E[B(D)^2]
=
\sum_{k=1}^K P(D\in G_k)\gamma(G_k)^2
+
\E\!\left[(B(D)-B_{\mathcal G}(D))^2\right].
\]
Hence maximizing \eqref{eq:maxhet} is equivalent to minimizing $\|B-B_{\mathcal G}\|_{L^2(P_D)}^2$. Thus, the population problem corresponds to the best \(K\)-step approximation of \(B\) in \(L^2(P_D)\).

\medskip
\noindent

\textbf{Step 2: Approximation by quantile partitions (grid discretization of the optimal quantizer).}

Pass to quantile coordinates. Let \(U=F_D(D)\); by assumption (iii), \(U\sim\mathrm{Unif}(0,1)\)
and the change of variables \(d=F_D^{-1}(u)\) is an isometry for the \(L^2(P_D)\) norm.
Set \(g(u):=B\!\left(F_D^{-1}(u)\right)=\E[\varepsilon^*\mid U=u]\). By (i) and (iii), \(g\) is
bounded, nondecreasing, and càdlàg on \([0,1]\), hence \(g\in L^2([0,1])\); and, because \(g\)
is monotone, every best \(K\)-step \(L^2\)-approximation of \(B\) corresponds to an \emph{interval}
partition of \([0,1]\), i.e. to a threshold vector \(\mathbf t=(t_1,\dots,t_{K-1})\) with
\(0=t_0\le t_1\le\dots\le t_{K-1}\le t_K=1\). Writing \(\Pi_{\mathbf t}g\) for the \(L^2\)-projection
of \(g\) onto step functions constant on the cells \([t_{k-1},t_k)\), Step~1 identifies \(B_K^*\)
with \(\Pi_{\mathbf t^*}g\) for a minimizer \(\mathbf t^*\) of the approximation error
\[
R(\mathbf t):=\bigl\|g-\Pi_{\mathbf t}g\bigr\|_{L^2([0,1])}^2 .
\]
The existence of $\mathbf t^*$ is the existence of an optimal $K$-level
quantizer of the law of $g(U)$ (see \cite{pollard1982quantization} for the $k$-means formulation, or Theorem 4.12 in \cite{graf2000foundations}).

\medskip

The quantile partition \(\mathcal I^M\) is the \emph{uniform} grid \(\{0,1/M,\dots,1\}\) in these coordinates, and \(\mathcal S_{K,M}\) is the sub-class of \(K\)-step functions whose thresholds lie on this grid. Restricting the cell boundaries to the grid is asymptotically without loss. First, \(R\) is continuous on the compact simplex
\(\{0\le t_1\le\dots\le t_{K-1}\le1\}\): since \(g\) is bounded and \(U\) has no
atoms, \(t\mapsto\mathbf 1\{u\le t\}\) is continuous in \(L^2\), and \(R\) is a
continuous function of these indicators. Second, the uniform grid is
\(1/M\)-dense, so snapping the optimal thresholds \(\mathbf t^{*}\) to their
nearest grid nodes \(\mathbf t^{(M)}\) (with \(\|\mathbf t^{(M)}-\mathbf t^{*}\|_\infty\le 1/M\)) gives \(R(B_{K,M}^*)\le R(\Pi_{\mathbf t^{(M)}}g)\to R(\mathbf t^{*})=R(B_K^*)\); combined with \(R(B_{K,M}^*)\ge R(B_K^*)\), this yields \(R(B^*_{K,M})\to R(B_K^*)\). If in addition the optimal \(K\)-partition \(\mathbf t^\star\) is unique (assumption~(vi)), argmin consistency plus \(L^2\)-continuity of
\(\mathbf t\mapsto\Pi_{\mathbf t}g\) yield \(B^*_{K,M}\to B_K^*\) in \(L^2(P_D)\);
otherwise convergence is to the set of optimal \(K\)-step approximations.

\medskip
\noindent
\textbf{Step 3: Estimation and empirical optimization.}

For each bin \(I_{j,M}\) set
\(\gamma_{j,M}=\E[\varepsilon^*\mid D\in I_{j,M}]=\E[\varepsilon\mid D\in I_{j,M}]\),
the second equality using that \(\varepsilon\) is an unbiased signal of \(\varepsilon^*\).

\medskip
\noindent\emph{Uniform control of the bin means.}
Under (iv) the bins carry equal mass \(1/M\), so
\(N_{j,M}=|\{i:D_i\in I_{j,M}\}|\) equals \(\lceil n/M\rceil\) exactly for empirical
quantile bins, and satisfies \(N_{j,M}\sim\mathrm{Bin}(n,1/M)\) for population
quantile bins. In the latter case a Chernoff bound together with
\(M^2/n\to 0\) (whence \(n/M\to\infty\) faster than \(\log M\)) gives
\(\min_j N_{j,M}\ge n/(2M)\) with probability tending to one. 

\medskip

Since \(\varepsilon\) is bounded (a difference of probabilities),
\(\bar\sigma^2:=\sup_j\operatorname{Var}(\varepsilon\mid D\in I_{j,M})<\infty\), and
Chebyshev's inequality with a union bound yields, for every \(t>0\),
\[
P\Big(\max_{1\le j\le M}\bigl|\widetilde\gamma(I_{j,M})-\gamma_{j,M}\bigr|>t\Big)
\;\le\;\frac{1}{t^2}\sum_{j=1}^{M}\frac{\bar\sigma^2}{N_{j,M}}
\;\lesssim\;\frac{\bar\sigma^2\,M^2}{n\,t^2}\;\xrightarrow[n\to\infty]{}\;0 .
\]
Hence \(\max_j|\widetilde\gamma(I_{j,M})-\gamma_{j,M}|\to0\) in probability precisely
under \(M^2/n\to0\). Because \((\gamma_{j,M})_j\) is nondecreasing by (i), the rearrangement inequality of \citet{chernozhukov2009improving} gives
\(\max_j|\widehat\gamma(I_{j,M})-\gamma_{j,M}|\le\max_j|\widetilde\gamma(I_{j,M})-\gamma_{j,M}|\), so the isotonic rearrangement \(\widehat\gamma\) inherits the uniform consistency of \(\widetilde\gamma\).

\medskip
\noindent\emph{Uniform convergence of the criterion.}
Each partition \(\mathcal G\in\mathcal G_{K,M}\) groups the fine bins into \(K\) blocks
of equal mass, so the population and empirical objectives
\(\overline\Lambda(\mathcal G)\) and \(\widehat{\overline\Lambda}(\mathcal G)\)
in \eqref{eq:maxhet_M} and \eqref{eq:gKM} are the same quadratic functional of
\((\gamma_{j,M})_j\) and \((\widehat\gamma(I_{j,M}))_j\). Writing
\(\bar\gamma:=\sup_j|\gamma_{j,M}|<\infty\), a block average of deviations is bounded
by the maximal deviation, so
\[
\sup_{\mathcal G\in\mathcal G_{K,M}}
\bigl|\widehat{\overline\Lambda}(\mathcal G)-\overline\Lambda(\mathcal G)\bigr|
\;\le\;2\bar\gamma\,\max_j\bigl|\widehat\gamma(I_{j,M})-\gamma_{j,M}\bigr|
\;\to\;0\quad\text{in probability.}
\]

\medskip
\noindent\emph{Asymptotic optimality.} We use the argmax theorem \citep[Thm.~5.7]{vandervaart1998asymptotic}. Let
\(\widehat{\mathcal G}_{n,M}\in\arg\max_{\mathcal G_{K,M}}\widehat{\overline\Lambda}\)
and \(\{G^*_{k,M}\}\in\arg\max_{\mathcal G_{K,M}}\overline\Lambda\). Since
\(\widehat{\mathcal G}_{n,M}\) maximizes \(\widehat{\overline\Lambda}\),
\[
0\;\le\;\overline\Lambda(\{G^*_{k,M}\})-\overline\Lambda(\widehat{\mathcal G}_{n,M})
\;\le\;2\sup_{\mathcal G\in\mathcal G_{K,M}}
\bigl|\widehat{\overline\Lambda}(\mathcal G)-\overline\Lambda(\mathcal G)\bigr|
\;\to\;0\quad\text{in probability.}
\]
By the law-of-total-variance identity of Step 1,
\(\overline\Lambda(\mathcal G)=\E[B(D)^2]-\|B-B_{\mathcal G}\|_{L^2(P_D)}^2\), so this is
equivalent to
\[
\bigl\|B-\widehat B_{n,M}\bigr\|_{L^2(P_D)}^2
-\bigl\|B-B_{K,M}^*\bigr\|_{L^2(P_D)}^2\;\to\;0
\quad\text{in probability:}
\]
the estimator \(\widehat B_{n,M}\) attains, asymptotically, the same \(L^2\) approximation
error as the grid optimum \(B_{K,M}^*\).
\medskip
\noindent

\textbf{Step 4: Conclusion.}
By Step 2, \(\|B-B_{K,M}^*\|_{L^2(P_D)}^2\to\|B-B_K^*\|_{L^2(P_D)}^2
=\inf_{\text{$K$-step }h}\|B-h\|_{L^2(P_D)}^2\). Combined with Step 3, this gives
\[
\bigl\|B-\widehat B_{n,M}\bigr\|_{L^2(P_D)}^2
\;\xrightarrow[n\to\infty]{}\;
\inf_{\text{$K$-step }h}\|B-h\|_{L^2(P_D)}^2
\quad\text{in probability,}
\]
i.e. \(\widehat B_{n,M}\) is an asymptotically minimizing sequence for the best
\(K\)-step approximation of \(B\). Under assumption~(vi) the minimizer \(B_K^*\) is
unique and well-separated, so \(\widehat B_{n,M}\to B_K^*\) in \(L^2(P_D)\). \qquad\(\square\)

\end{document}